\documentclass[onecolumn]{aastex631}

\usepackage{amsmath}
\usepackage{hyperref}
\usepackage{color,soul}
\usepackage{multirow}
\usepackage{longtable}
\usepackage[figuresright]{rotating}
\usepackage{colortbl}
\usepackage{hhline}
\usepackage{cellspace}
\usepackage{makecell}
\usepackage[version=4]{mhchem}

\defcitealias{madhusudhan2023}{M23}
\defcitealias{madhusudhan2025}{M25}
\defcitealias{horst2018}{H18}
\defcitealias{He24}{H24}
\defcitealias{khare1984}{K84}

\begin{document}

\title{Hydrocarbon Hazes on Temperate sub-Neptune K2-18b supported by data from the James Webb Space Telescope}
\author[0009-0006-1490-1098]{Ruohan Liu}
\affiliation{Department of Physics and Astronomy, University College London, Gower Street, WC1E 6BT London, United Kingdom}

\author[0000-0002-5360-3660]{Panayotis Lavvas}
\affiliation{Groupe de Spectrométrie Moléculaire et Atmosphérique, Universite de Reims Champagne Ardenne, Reims, France}
\author[0000-0001-6058-6654]{Giovanna Tinetti}
\affiliation{NMES Faculty, King’s College London, Strand Building, Strand, WC2R 2LS London, United Kingdom}
\author[0000-0002-2218-5689]{Jesus Maldonado}
\affiliation{Osservatorio Astronomico di Palermo, Piazza del Parlamento 1, 90134, Palermo Italy}
\author[0000-0001-9010-0539]{Sushuang Ma}
\affiliation{Department of Physics and Astronomy, University College London, Gower Street, WC1E 6BT London, United Kingdom}
\affiliation{NMES Faculty, King’s College London, Strand Building, Strand, WC2R 2LS London, United Kingdom}
\author[0000-0002-1437-4228]{Arianna Saba}
\affiliation{Blue Skies Space Science Foundation, 69 Wilson Street, London, EC2A 2BB, UK}

\begin{abstract}
K2-18b, a sub-Neptune orbiting in the habitable zone of an M dwarf, has attracted significant interest following observations with the Hubble Space Telescope (HST) and, more recently, with the James Webb Space Telescope (JWST), which reveal detectable atmospheric features across the near- and mid-infrared.
Using free-chemistry Bayesian retrievals, we investigate whether hydrocarbon hazes can explain the apparent mismatch of spectral feature amplitudes between the JWST NIRISS/NIRSpec and MIRI LRS datasets. We additionally assess the impact of stellar parameter uncertainties on the derived bulk properties of the planet and explore how planetary mass uncertainties affect atmospheric retrievals. We find that hazy scenarios can reproduce the combined JWST spectrum and provide a consistent explanation for the reduced NIRISS/NIRSpec feature amplitudes relative to the stronger MIRI features. Across all retrievals, the atmosphere remains consistent with an H$_2$-dominated sub-Neptune, with CH$_4$ and CO$_2$ as the dominant absorbers. Our hazy models retrieve systematically lower molecular abundances compared to haze-free models, reflecting the degeneracy between haze opacity and mean molecular weight. In addition, we identify strong degeneracies between planetary mass, temperature, and mean molecular weight. The retrieved planetary mass is particularly poorly constrained, with $2\sigma$ uncertainties reaching up to $\sim71\%$. We demonstrate that different mass assumptions can significantly bias the inferred atmospheric properties, with higher masses favouring warmer and lower mean molecular weight atmospheres. Breaking these degeneracies will require improved stellar characterisation to obtain more precise mass measurements. More laboratory-focused studies and future JWST observations are essential for interpreting these temperate sub-Neptune atmospheres.
\end{abstract}

\section{Introduction}
\label{sec:intro}

\subsection{A brief history of K2-18b}
\label{sec:intro-brief-history}
Super-Earths (\(1R_{\oplus} \lesssim R_p \lesssim 1.8 R_{\oplus}\))  and sub-Neptunes (\(1.8R_{\oplus} \lesssim R_p \lesssim 3.5 R_{\oplus}\))\footnote{The transition between super-Earths and sub-Neptunes is marked by the Radius Valley, between $1.7-1.9R_\oplus$ (see e.g., Cloutier 2024).} make up the most common types of exoplanets in our galaxy, yet their interior compositions remain unclear due to a wide range of possible bulk densities. Proposed interior structure models include rocky planets with thin atmospheres, gaseous``mini-Neptunes"\footnote{Here, "mini-Neptune" is used when describing the interior structure, whereas "sub-Neptune" refers to the planet in terms of size.} with thick H$_2$-dominated envelopes, and ``water-worlds" with volatile-rich interiors \citep[e.g.][]{grasset2009, valencia2007, valencia2013, ikoma2012,  miozzi2018, chachan2018,  zeng2019, ito2025}. They are particularly abundant around M dwarfs, which make up $\sim$75\% of the stellar population. Studies suggest that each M dwarf hosts on average 2.5 planets with radii between \(1\textrm{--}4 R_{\oplus}\) and orbital periods under 200 days \citep{dressing2015}. The advent of the James Webb Space Telescope (JWST) has revolutionized exoplanet studies by providing high-precision near-infrared (NIR) spectroscopy, allowing us to characterise their atmospheres in unprecedented detail. Among these sub-Neptunes, K2-18b is one that has captured significant attention. Discovered in 2015 during the Kepler K2 mission \citep{foreman-mackey2015, Montet15}, it orbits within the habitable zone of an M dwarf, making it a prime target for studying the atmospheric composition, interior structure, and habitability of sub-Neptunes. \par

Early atmospheric retrieval studies of K2-18b, based on Hubble Space Telescope (HST, spanning 1.1–1.7 $\mu$m) and Spitzer data, identified an H$_2$-rich atmosphere with strong atmospheric features attributed mainly to H$_2$O \citep{benneke2019,tsiaras2019,madhusudhan2020}. This made K2-18b the first non-hot Jupiter with a statistically significant ($>3\sigma$) spectroscopic detection of an atmosphere. However, similarities between CH$_4$ and H$_2$O absorption features in the WFC3 bandpass (1.1–1.7 µm) led to competing interpretations of K2-18b’s atmospheric chemistry \citep{bezard2022,blain2021}. A 1D climate-chemistry model by \citet{scheucher2020} found that an H$_2$-He atmosphere with limited H$_2$O and CH$_4$, up to 50× solar metallicity, could also be compatible with the HST and Spitzer observations. \citet{barclay2021} suggested that the H$_2$O signature could also be the result of contamination due to stellar surface inhomogeneities, such as star spots. The narrow spectral range and low resolution of the HST and Spitzer data made it difficult to distinguish between these scenarios, and highlights the need for a broader wavelength coverage to break the degeneracies between H$_2$O and CH$_4$.\par

\citet{madhusudhan2021} proposed K2-18b as a candidate ``Hycean'' world --- an exoplanet with a relatively thin H$_2$ atmosphere overlying a liquid water ocean with a thick layer of ice beneath it, making it a prime target for astrobiological studies. Using photochemical and radiative-convective models, \citet{hu2021} predicted that a ``Hycean'' atmosphere would primarily contain CO$_2$, N$_2$, CO, and CH$_4$, whereas a mini-Neptune with a massive H$_2$ envelope would be dominated by CH$_4$ and NH$_3$. \citet{tsai2021} further suggested that NH$_3$ and CH$_3$OH abundances could provide insight into the surface properties of sub-Neptunes.\par

The first set of JWST observations of K2-18b (NIRISS SOSS and NIRSpec G395H) revealed strong absorption features between 0.9–5.2 $\mu$m. \citet{madhusudhan2023} interpreted these as robust detections of CO$_2$ ($\sim1\%$ detected at $5\sigma$) and CH$_4$ ($\sim1\%$ at $3\sigma$) in an H$_2$-rich atmosphere, alongside non-detections of H$_2$O, CO, and NH$_3$. They also reported tentative ($\sim 1\sigma$) signs of dimethyl sulfide (DMS), a potential biosignature, though the statistical significance remains weak. \citet{madhusudhan2023} argued that the observed CO$_2$ and CH$_4$ abundances, coupled with the absence of NH$_3$, support a Hycean planet scenario, where a liquid water ocean enhances CH$_4$ and CO$_2$ production, increases the CO$_2$/CO ratio, while further depleting NH$_3$. According to \citet{hu2021}, such a atmospheric composition is incompatible with a mini-Neptune scenario under standard thermochemical models.\par

\citet{wogan2024} performed 1D photochemical and climate models on the JWST observations, and reported two compatible scenarios: (1) an inhabited Hycean world requiring a methane-producing biosphere or (2) a mini-Neptune with a massive H$_2$ atmosphere, where CH$_4$ and CO$_2$ are thermochemically produced in the deep atmosphere. They favoured the latter scenario as it does not rely on the existence of life for the explanation to work. Additionally, \citet{huang2024} showed that water condensation on sub-Neptunes depletes the atmospheric O$_2$, which significantly impacts the overall predicted composition. Taking this into account, they concluded that K2-18b is more consistent with a volatile-rich mini-Neptune with a deep surface and thick atmosphere, aligning with \citet{wogan2024}. \par

\citet{cooke2024} later challenged this conclusion using two independent photochemical models, arguing that \citet{wogan2024}’s preferred mini-Neptune scenario is inconsistent with their retrieved mixing ratios. They found that key parameters—such as photochemical cross sections, stellar spectrum, surface pressure, UV albedo, and metallicity—significantly influence the resulting atmospheric abundances. Their findings instead support a Hycean interpretation of K2-18b, where both inhabited and uninhabited cases remain viable depending on the assumed stellar UV flux and surface pressure.\par

Determining K2-18b’s atmospheric composition and lower atmospheric conditions is crucial for understanding sub-Neptune interiors. \citet{yang2024} explored a range of atmospheric scenarios from H$_2$- to H$_2$O-dominated compositions, with equilibrium temperatures between 250–400 K. They found that the CO$_2$/CH$_4$ ratio can be used to infer the planet’s deep-interior H$_2$O/H$_2$ ratio. Based on the JWST-derived CO$_2$ and CH$_4$ abundances \citep{madhusudhan2023}, they estimated that K2-18b’s interior is likely 50\% water-rich, exceeding the expectations for a scenario with 100$\times$ solar metallicity.\par

Some studies suggest that Hycean planets, particularly those in close-in orbits or in the runaway greenhouse state, may not sustain liquid-water oceans and could instead evolve to have a supercritical water interior \citep{mousis2020,innes2023,pierrehumbert2023,leconte2024}. \citet{luu2024} examined geochemical interactions between a hot supercritical water ocean and an H$_2$-rich atmosphere, resembling a global hydrothermal system. Their models showed that a supercritical ocean on K2-18b is viable if its temperature is between 710–1070 K and the pressure is between 1–10 kbar. \par

\citet{hu2021} argued that if K2-18b possessed a massive H$_2$-rich envelope, NH$_3$ would be dynamically transported to the observable regions of the atmosphere, and therefore be detectable in its transmission spectrum. \citet{shorttle2024} investigated whether a molten silicate surface on a mini-Neptune could explain the missing NH$_3$ in the JWST data, as nitrogen species are highly soluble under magma-reducing conditions. \citet{rigby2024} proposed that the abundances of CO$_2$, CO, NH$_3$, and sulfur-bearing species could help determine whether K2-18b’s atmosphere is consistent with a magma ocean scenario. However, they found that the observed composition does not support a magma ocean, contradicting \citet{shorttle2024}. \par

All preceding JWST NIRISS and NIRSpec studies of K2-18b used the transmission spectra as prepared by \citet{madhusudhan2023}. These datasets were extracted using a single, custom-built reduction pipeline that is not publicly available.  \citet{schmidt2025} performed a comprehensive reanalysis of the JWST observations from \citet{madhusudhan2023}, incorporating multiple independent data reduction pipelines, retrieval codes, and photochemical-climate and interior models to reassess the atmospheric composition of K2-18b. Across all explored data combinations, they reported a CH$_4$ volume mixing ratio of $7^{+11}_{-5}\%$ with a detection significance of $\approx4\sigma$, consistent with the findings of \citet{madhusudhan2023}. However, in contrast to the original study, they found no reliable evidence for CO$_2$ or DMS, as nearly all 60 data combinations they tested resulted in detection significances below $2.1\sigma$ for both molecules. Their interior structure models suggest that K2-18b is best characterized as a mini-Neptune with an oxygen-poor, nitrogen-depleted atmosphere and $100\times$ solar metallicity. Their findings emphasize that our interpretation of JWST data is highly sensitive to the choice of data reduction methods and spectral resolutions used \citep[e.g. see][]{mugnai2024}.\par

Similarly, as stressed by \citet{welbanks2025},  a preference for one retrieval model over another does not lead to the conclusion that the better-fitting model provides a more reliable or preferred interpretation of the data. Such is the case for K2-18b, as its existing observations yield low signal-to-noise ratios.  Moreover, retrieval models have intrinsic limitations as they heavily rely on the defined hypothesis space, alongside a limited subset of chemically plausible species \citep{ma2025}. It is computationally impossible to perform an exploration of the complete parameter space, especially when there are hundreds of plausible molecular species, with many lacking adequate opacity data. In essence, retrieval models that yield degenerate or inadequate fits to a spectrum should not be ruled out as a worse-performing model over those that offer a single, better-fitting solution. 

The most recent addition of observations from the JWST Mid-Infrared Instrument (MIRI) has extended our window of K2-18b's transmission spectrum up to $\sim$12 $\mu$m, making it the most frequently observed sub-Neptune in the JWST era. In theory, the extended coverage from MIRI could help us break degeneracies within existing models that so far have been limited to the HST, NIRISS, and NIRSpec datasets. \citet{madhusudhan2025} (hereafter M25), reported evidence for DMS and/or dimethyl disulfide (C$_2$H$_6$S$_2$ or DMDS) at 3$\sigma$ significance. The authors claim that the MIRI spectrum shows distinct features that cannot be explained by most of the expected molecules in K2-18b's atmosphere, with the exception of DMS and DMDS. In response to M25's claims, \citet{welbanks2025} and \citet{luque2025} demonstrated that numerous physically distinct models are compatible with the MIRI data without the requirement of DMS or DMDS. 

Notably, the amplitudes of spectral features observed in the MIRI data are significantly larger than those in the NIRISS, and NIRSpec (collectively, NIR) datasets. This implies larger atmospheric scale heights, requiring  higher temperatures and/or a lower mean molecular weights in retrieval models. Consequently, independent analyses of MIRI- and NIR-only datasets have led to conflicting interpretations of K2-18b's atmosphere (e.g. \citetalias{madhusudhan2023}; \citetalias{madhusudhan2025}; see also \citeauthor{luque2025} \citeyear{luque2025}).

There is no unanimous agreement on which model best explains the observed atmospheric composition of K2-18b. It could be a Hycean world, a mini-Neptune with a massive H$_2$ atmosphere, a planet with a magma ocean, or one with a hydrothermal system --- or something entirely unexplored. Amidst the undeniable degeneracies, the community's desire to search for a definitive answer has unlocked new physically-plausible scenarios, expanding the realm of what we currently know to be sub-Neptunes.

JWST observations have provided new insight into K2-18b's atmosphere, but it also marks a turning point in how we should interpret the data and report the results, particularly for small exoplanets with low SNRs. K2-18b has become an exoplanet laboratory that has helped facilitate ongoing open-ended discussions, interdisciplinary collaborations, and novel approaches within exoplanetary science. \par

\subsection{Aims of this work}
\label{sec:intro-aims}

Although exploring a wide range of solutions is crucial for understanding K2-18b, the growing number of degenerate solutions highlights the need to re-establish the boundaries of fundamental stellar and planetary parameters to ensure that a full scope of plausible scenarios is explored. Planetary properties are all measured as a function of the stellar quantities. Thus, it is important to check the robustness of our results against the uncertainties due to the methods used to derive the stellar parameters.

As discussed above, independent MIRI studies highlight its limited constraining power \citep{welbanks2025} and reveal discrepancies with NIR-based interpretations, driven by differences in spectral feature amplitudes. To address these concerns, in this work, we simultaneously analyse the entire range of JWST observations to obtain atmospheric constraints that satisfy the entire spectral range. In addition, we assess the impact of stellar parameter uncertainties on the inferred planetary properties, and carry out independent measurements of the host star, K2-18, using publicly available HARPS spectra. We then independently produce data-reduced transmission spectra using publicly available MIRI LRS observations and combine this with previously published NIRISS SOSS and NIRSpec G395H observations from \citetalias{madhusudhan2023} to perform atmospheric retrievals on the full JWST transmission spectrum, spanning 0.8-12 $\mu$m. To investigate the role of hazes, we include the scattering and absorption properties of laboratory-produced haze analogues from \citet{khare1984} and \citet{He24}. The inclusion of hazes is motivated by recent results of the self-consistent disequilibrium chemistry model presented in \citet{lavvas2026} \citep[see also][]{jaziri2025}.

In Section \ref{methods}, we describe our methodology for: (i) stellar characterisation; (ii) the MIRI data reduction and light curve extraction; and (iii) the atmospheric retrieval set-up. We present our planetary density estimates and atmospheric retrieval results in Section \ref{sec:results}. By considering the effects of uncertainties in the stellar mass and radius on derived planetary parameters in retrievals and models, as well as the addition of a haze continuum, we discuss the implications of our results and conclude in Sections \ref{sec:discussion} and \ref{sec:conclusions}, respectively.

\section{Methodology}
\label{methods}
\subsection{Stellar Properties}

The main stellar properties of K2-18 are listed in Table~\ref{physical_properties}.
Effective temperature and iron abundance were computed using a methodology based on ratios of spectral features\footnote{\url{https://github.com/jesusmaldonadoprado/mdslines}} \citep[][hereafter MA15]{2015A&A...577A.132M} measured on high-resolution optical spectra. A total of 107 HARPS spectra were downloaded from the ESO archive\footnote{\url{http://archive.eso.org/wdb/wdb/adp/phase3_spectral/form?phase3_collection=HARPS}} and co-added into one single spectrum.

The effective temperature provided by MA15 is calibrated using stars with interferometric estimates of their radii and it is in the revised scale by \cite{2013ApJ...779..188M}. On the other hand, metallicities provided by MA15 are based on the photometric relationship by \cite{2012A&A...538A..25N}. MA15 also provides empirical calibrations to derive masses and radii as a function of the effective temperature and the metallicity. 

Specifically, the mass scale in MA15 is based on the M$_{\rm Ks}$-mass relationship by \cite{1993AJ....106..773H} and has typical uncertainties of the order of 13\%.  Using stars with known interferometric radius as well as low-mass eclipsing binaries, MA15 derive their own stellar mass-radius relationship. Typical uncertainties in the radius are of the order of 12\%.

Finally, Galactic spatial-velocity components $(U,V,W)$ were computed by using {\it Gaia} parallaxes, proper motions and radial velocity \citep{2022yCat.1355....0G} using the methods described in \cite{2010A&A...521A..12M}. From the analysis of its velocity components, we identify K2-18 as an old thin disc star.

\begin{table}[!tb]
\centering
\caption{Physical properties of K2-18.}
\label{physical_properties}
\begin{tabular}{lrl}
\hline\noalign{\smallskip}
 Parameter                 &  Value                &  Notes \\
\hline
$\alpha$ (ICRS epoch 2016.0, deg)     &  172.5601297578 & a       \\
$\delta$ (ICRS epoch 2016.0, deg)     &  7.5878131221   & a       \\
\hline
 Spectral Type                 &   M3                 &  b  \\
 T$_{\rm eff}$  (K)            &  3500  $\pm$     68  &  b  \\
 ${\rm [Fe/H]}$ (dex)          &  +0.03 $\pm$   0.10  &  b  \\
 $M_{\star}$ ($M_{\odot}$)     &  0.46  $\pm$   0.07  &  b  \\
 $R_{\star}$ ($R_{\odot}$)     &  0.45  $\pm$   0.06  &  b  \\
 $\log g$ (cm s$^{\rm -2}$)    &  4.80  $\pm$   0.06  &  b  \\
 $\log (L_{\star}/L_{\odot})$  & -1.56  $\pm$   0.12  &  b  \\
\hline
 $\pi$ (mas)                     & 26.2469    $\pm$ 0.0266     &   a       \\
$\mu_{\alpha}$ (mas/yr)          & -80.4790   $\pm$ 0.0257    & a  \\
 $\mu_{\delta}$ (mas/yr)         & -133.0068  $\pm$ 0.0231    & a  \\
 v$_{\rm rad}$ (km s$^{\rm -1}$) & 0.02       $\pm$  0.52     & a   \\
 $U$ (km s$^{\rm -1}$)           & -1.36      $\pm$   0.06    & b   \\
 $V$ (km s$^{\rm -1}$)           & -25.01     $\pm$   0.23    & b   \\
 $W$ (km s$^{\rm -1}$)           & -12.68     $\pm$   0.46    & b   \\
 \hline
\end{tabular}
\tablecomments{(a) \cite{2022yCat.1355....0G}; (b) This work}
\end{table}
\subsection{MIRI Data Reduction}
\label{sec:miri data reduction}
A transit of K2-18b was observed in the mid-infrared using the JWST MIRI low-resolution spectrograph (LRS) in slitless prism mode with the P750L double prism filter and the FASTR1 readout pattern \citep{kendrew2015,bouwman2023}. The observations were carried out from 23:13:29 UTC on April 25 2024 to 05:04:37 UTC on April 26 2024, for a total of 5.85 hours, as part of JWST GO Program 2722 (PI: N. Madhusudhan). The exposure consists of a total of 5095 integrations (split into 7 segments), with 25 groups per integration. The exposures produced a spectrum with average R$\sim$100 between 5-12 $\mu$m, with a resolving power of R$\sim$40 at 5 $\mu$m increasing linearly to R$\sim$160 at 10 $\mu$m \citep{wright2015}. We perform data reduction using the official JWST Science Calibration Pipeline \citep{bushouse2020} --- hereafter \texttt{JWST-sci} --- and light curve fitting with the \texttt{PyLightcurve} package \citep{tsiaras2016-pylightcurve}, as described below.

\subsubsection{JWST Science Calibration Pipeline}
\label{sec:jwst-pipeline}
We employ the \texttt{JWST-sci} pipeline to perform our Stage 1-3 reductions on the K2-18b MIRI LRS observations. We start with the uncalibrated (\texttt{.uncal}) files taken from the MAST\footnote{\url{https://mast.stsci.edu}} archive and apply \texttt{Detector1Pipeline} in Stage 1 to all exposures for basic detector-level corrections. We use the same steps as described in \citetalias{madhusudhan2025}, which are: data quality initialization (\texttt{dq\_init}), electromagnetic interference correction (\texttt{emicorr}), saturation flagging (\texttt{saturation}), first and last frame flagging (\texttt{firstframe} and \texttt{lastframe}), linearity correction (\texttt{linearity}), reset switch charge decay correction (\texttt{rscd}), dark current subtraction (\texttt{dark\_current}), and ramp fitting (\texttt{ramp\_fitting}). Note that for MIRI LRS TSO data products, the \texttt{jump} and \texttt{reset} steps - skipped here - would not be skipped by default. 

In Stage 2, we assign the world coordinate system (\texttt{assign\_wcs}), set the source type to "POINT" (\texttt{srctype}), and apply the flat fielding correction to the data products. We then apply outlier detection (\texttt{OutlierDetectionStep}) on each frame to mask out cosmic rays and bad pixels, setting the signal-to-noise ratio threshold to \texttt{snr=('4.0 4.0')}, where the first value is for detecting the primary cosmic ray, and the second is for masking lower-level bad pixels adjacent to those found in the first pass. The outliers are replaced with "NaNs". We tested an additional data reduction method that replaces the \texttt{OutlierDetectionStep} with the \texttt{jump} step \citep[from \texttt{transitspectroscopy};][]{espinoza_nestor2022} in Stage 1, and we describe this in more detail in  Appendex \ref{appendix: including jump step}.

We then use the \texttt{PixelReplaceStep} to interpolate the flux values of the "NaN" pixels using adjacent profile approximation. We set the number of adjacent columns surrounding a bad pixel to 10, which is used in creation of the source profile. The error values and variance components for the interpolated pixels are similarly updated with estimated values. 

For Stage 3 (\texttt{calwebb\_tso3}), we correct for the background using the box extraction method. We define two background columns surrounding the trace with pixel numbers 11-30 and 44-63, and use an aperture of 9 pixels for the source extraction area (following \citetalias{madhusudhan2025}). For each integration frame, the median of the background values is calculated at each pixel within the source extraction region for that detector column and is subtracted from the source count rate. We then perform spectral extraction at native resolution and combine all segmented exposures to produce 1D, flux calibrated spectra.

\subsubsection{PyLightcurve}
\label{sec:pylightcurve}

\begin{table}[!tb]
    \centering
        \caption{Parameter estimates with uncertainties as a result of fitting the white light curve of our JWST MIRI LRS observation of K2-18b. The lower and upper uncertainties refer to the median of the 16th and 84th percentile of the MCMC chain, respectively.}
    \label{tab:pylightcurve_parameter_estimates}
    \begin{tabular}{lcc}
    \hline \hline
        Parameter& & Value\\ \hline
        Mid-transit time, $T_0$ $(\text{BJD} - 2400000.5)$& & $60426.12873^{+0.000098}_{-0.000098}$\\
        Inclination, $i$ $(^{\circ})$& &$89.536^{+0.026}_{-0.027}$\\
 Normalised semi-major axis, $a/R_*$& &$78.80^{+1.78}_{-1.82}$\\
        Planet-to-star radius ratio, $R_p/R_*$& &$0.05303^{+0.00030}_{-0.00031}$\\ 
 First limb-darkening coefficient, $u_1$& &$0.024^{+0.031}_{-0.017}$\\
 Second limb-darkening coefficient, $u_2$& &$0.049^{+0.043}_{-0.034}$\\
 \hline \hline
    \end{tabular}
\end{table}

\begin{figure}[!tb]
    \centering
    \includegraphics[width=1\linewidth]{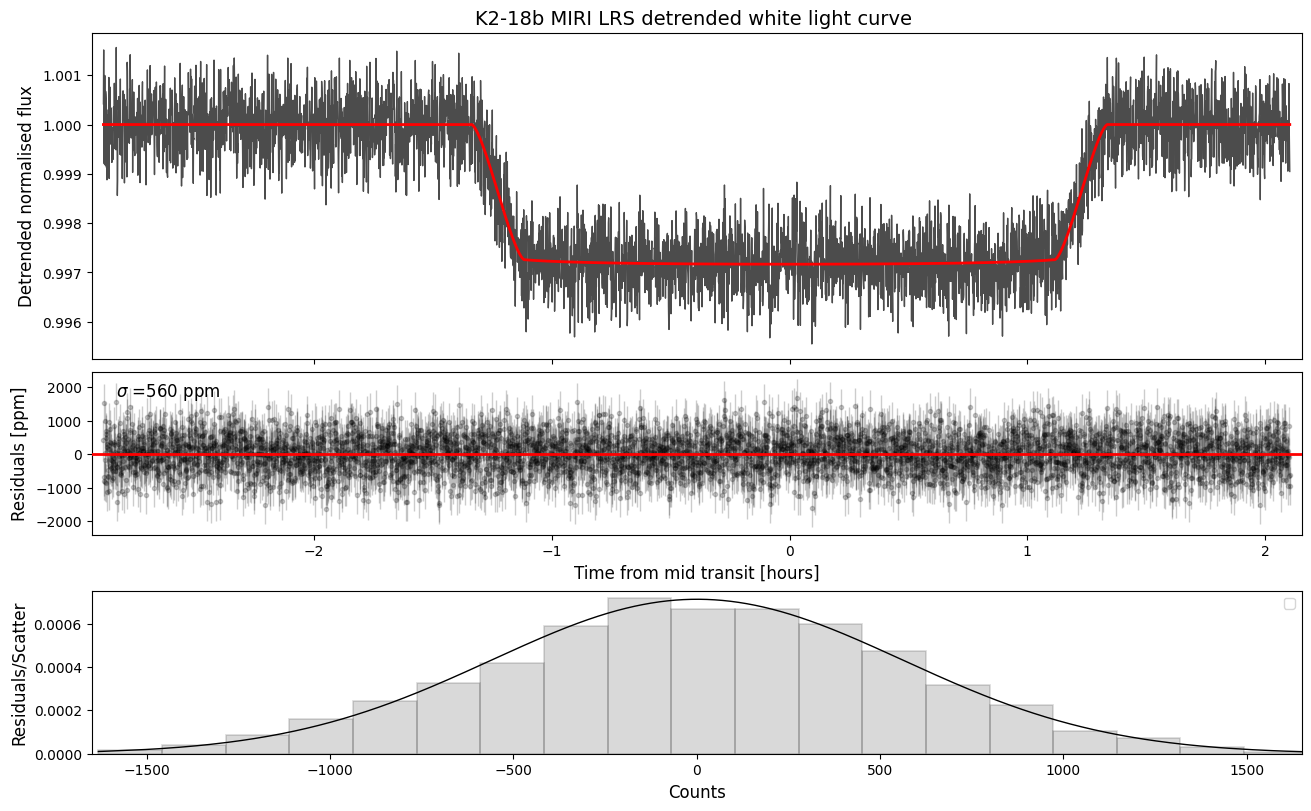}
    \caption{\textit{Top}: The detrended MIRI LRS white light curve of K2-18b, obtained from integrating spectral light curves between 5-12 $\mu$m, with the best-fitting transit model over-plotted in red. \textit{Middle}: Residuals after subtracting the model. The standard deviation of the residuals is $\sigma=560$ppm. \textit{Bottom}: Histogram of residuals, showing the probability density function, assuming a Gaussian distribution, given the mean and standard deviation of the residuals.}
    \label{fig:miri-white-lc}
\end{figure}

Using the \texttt{PyLightcurve} package, we fit spectral lightcurves spanning 5-12 $\mu$m, and construct a white light curve by summing the fluxes in this range. 

As described in \citet{feinstein2023} for their analysis of WASP-39b, for each light curve, \texttt{PyLightcurve}: (1) calculates the limb-darkening coefficients by calling the ExoTETHyS package \citep{morello2020}, using the wavelength range of the bin, the response curve of the MIRI LRS Slitless prism filter, and the stellar parameters; (2) finds the maximum-likelihood model for the data (an exposure-integrated transit model together with a quadratic trend model using the Nelder-Mead minimisation algorithm included in the SciPy package \citep{virtanen2020}; (3) removes outliers that deviate from the maximum-likelihood model by more than three times the standard deviation of the normalised residuals; (4) scales the uncertainties by the root mean square of the normalised residuals, to take into account any extra scatter; (5) performs an MCMC optimisation process using the emcee package \citep{foreman-mackey2013}.

We adopt host stellar parameters listed in Table \ref{physical_properties}, and assume initial planetary transit parameters from \citet{Benneke17}. The white light curve is modelled by fitting for the semi-major axis $a/R_*$, inclination $i$, mid-transit time $T_0$ and planet-to-star radius ratio $R_p/R_*$. We fix the period to 32.94171 days \citep{kruse2019} and apply a quadratic detrending function, and mask out the first 750 integrations to remove the strongest effect of the detector settling. We investigated the effect of varying the number of integrations masked in more detail in Appendix \ref{appendix: miri: no. of integrations masked}. We use the ATLAS stellar model grid \citep{claret2000} and adopt a quadratic limb-darkening model. 

\begin{figure}
    \centering
    \includegraphics[width=1\linewidth]{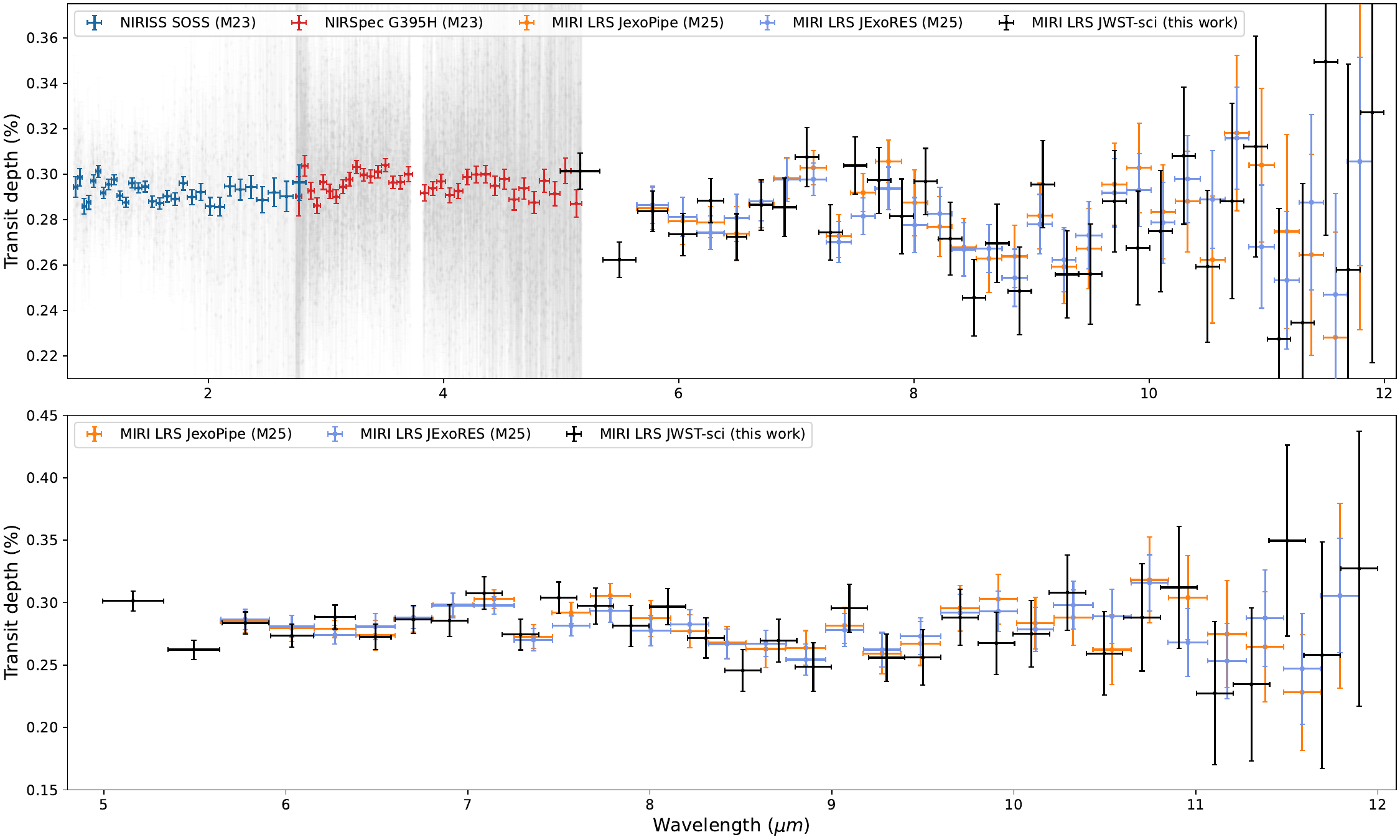}
    \caption{Comparison between our MIRI LRS transmission spectra and the reductions presented in \citetalias{madhusudhan2025}, JexoPipe (orange) and JExoRES (green), of K2-18b. Our nominal spectrum (black) is binned to match the resolution of JexoPipe and JExoRES, with a width of either a maximum of 5 pixels or 0.2$\mu$m (whichever contains the most pixels). \textit{Top}: The NIRISS SOSS (blue) and NIRSpec G395H (red) datasets from \citetalias{madhusudhan2023} are plotted to show the full span of observations, with their native resolution datasets under-plotted in grey. \textit{Bottom}: Zoomed-in spectrum spanning only the MIRI wavelength range.}
    \label{fig:nir+miri spectra comparison}
\end{figure}

Table \ref{tab:pylightcurve_parameter_estimates} shows the parameter estimates from fitting white light curve. Figure \ref{fig:miri-white-lc} shows the MIRI white light curve along with the fitted model. Figure \ref{fig:nir+miri spectra comparison} shows the resulting MIRI transmission spectrum of K2-18b, plotted with \citetalias{madhusudhan2025}'s JExoRES and JexoPipe spectra, for comparison, as well as the NIRISS and NIRSpec observations from \citetalias{madhusudhan2023}.

The spectral light curves were modelled by fitting for the $R_p/R_*$ in each wavelength bin, and fixing the values $a/R_*$, $i$, and $T_0$ to the values obtained from the white light curve fit. We fix the limb darkening coefficients for each spectral light curve to the derived values from the non-linear limb-darkening law by \citet{claret2000}. A quadratic detrending function is also applied here, and we mask out the first 750 integrations. 

\begin{figure}[!h]
    \centering
    \includegraphics[width=1\linewidth]{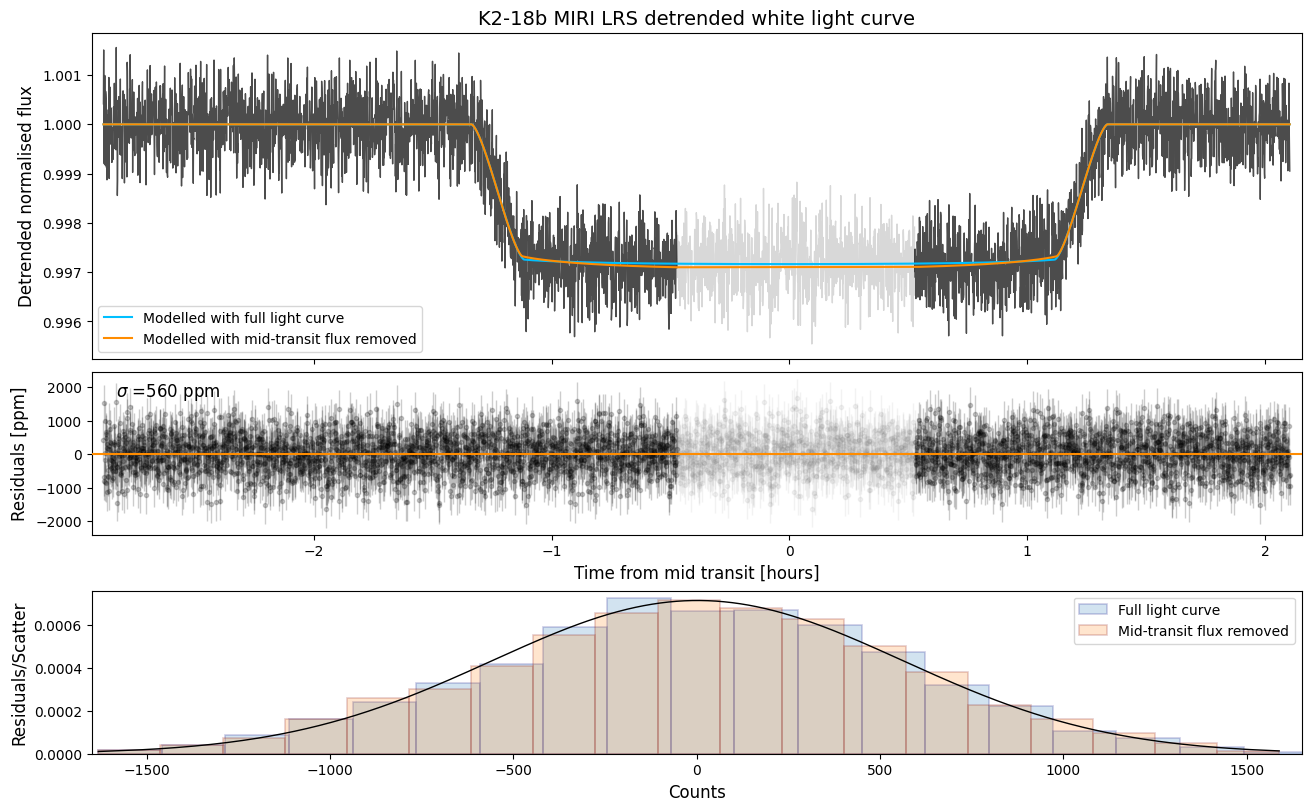}
    \caption{\textit{Top}: The detrended MIRI LRS white light curve of K2-18b. The dark grey lightcurve is identical to Figure \ref{fig:miri-white-lc}, except with the mid-transit region removed, shown in light grey ($\pm$1 hour about the mid-transit time). 
  The best-fitting transit model for the mid-transit-removed lightcurve is over-plotted in orange. The best-fitting model of the original lightcurve is shown in blue. \textit{Middle}: Residuals after subtracting the model. The standard deviation of the residuals is $\sigma=560$ppm. \textit{Bottom}: Histogram of residuals.}
    \label{fig:miri-white-lc-midtransit-masked}
\end{figure}

\begin{figure}[!h]
    \centering
    \includegraphics[width=1\linewidth]{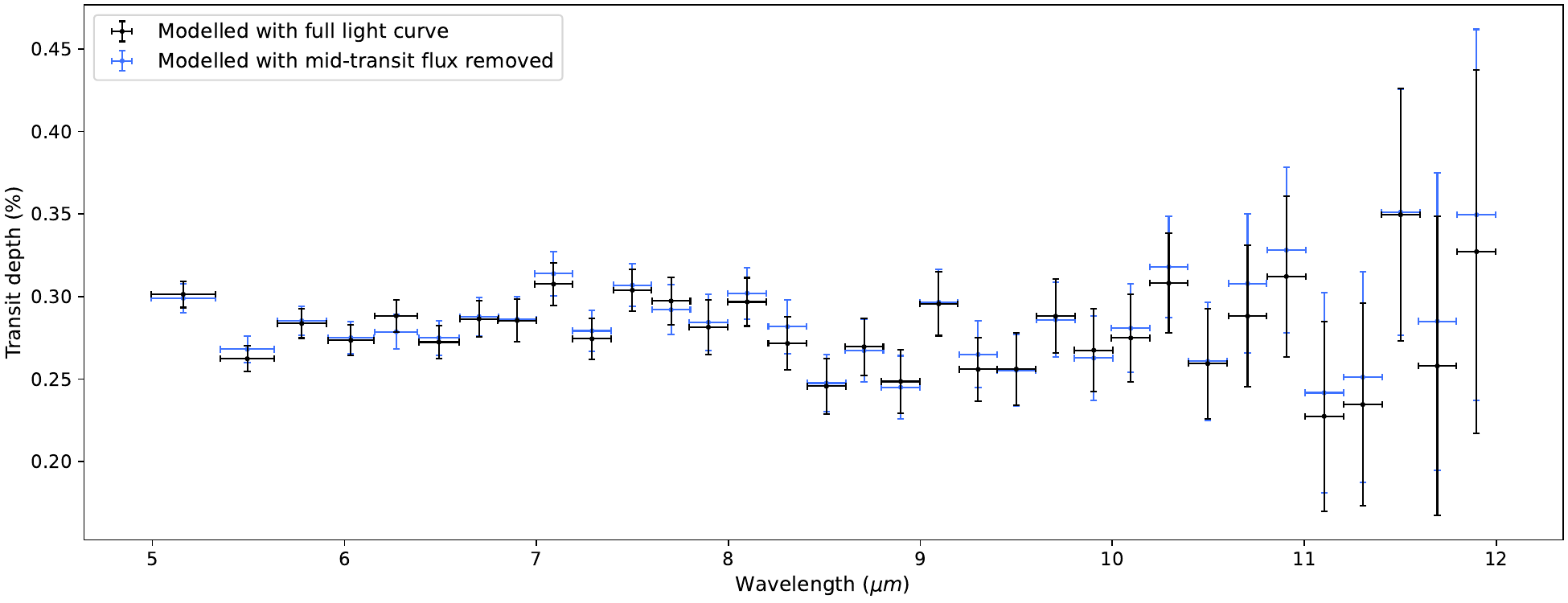}
    \caption{Comparison between the MIRI LRS spectra extracted from the lightcurve in Figure \ref{fig:miri-white-lc} (in black) and the mid-transit-removed lightcurve in Figure \ref{fig:miri spectrum comparison mid-transit masked} (in red).}
    \label{fig:miri spectrum comparison mid-transit masked}
\end{figure}

Prior to fitting, we bin the spectral lightcurves with a width of 0.2$\mu$m or 5 pixels, whichever contains the most pixels (as in \citetalias{madhusudhan2025}). We also explored binning with different widths in Appendix \ref{appendix: miri: spectral binning}. We additionally investigate the sensitivity of spectral transit depths to systematic noise during the light curve's mid-transit. We perform light curve fitting with a section of the mid-transit masked out (between 60426.1041 and 60426.1458 MJD UTC, or $\pm$1 hour about the mid-transit time; Figure \ref{fig:miri-white-lc-midtransit-masked}). We find that the extracted spectrum from the masked lightcurve is in good agreement with the original spectrum fitted using a full light curve (Figure \ref{fig:miri spectrum comparison mid-transit masked}). In general, the spectrum of the mid-transit-masked light curve is offset by $\sim50$ppm on average. The offset increases at longer wavelengths, with the exception of the transit depth at $\sim6.2$$\mu$m, which exhibits an offset of $-97$ppm. In both cases, the first 750 integrations were removed. We adopt the spectrum derived from the full light curve as our final dataset, while noting that the most variable points may be susceptible to unresolved instrumental systematics. 


\subsection{Atmospheric Modelling and Retrieval Setup}
\label{sec:atmospheric retrieval}

\texttt{TauREx 3}\footnote{\url{https://github.com/ucl-exoplanets/taurex3/}}  (hereafter \texttt{TauREx}) is retrieval framework designed to model and interpret exoplanet atmospheric data \citep{al-refaie2021,al-refaie2022}. \texttt{TauREx} includes a built-in free chemistry model and offers flexibility for incorporating new chemical codes in addition to supporting equilibrium chemistry models \citep{al-refaie2022}: \texttt{ACE} \citep{agundez2012,agundez2020}, \texttt{GGchem} \citep{Woitke18}, and \texttt{FastChem} \citep{stock2018}. Radiative transfer is computed from user-provided opacities of chemical species, collision-induced absorptions (CIA), Rayleigh and Mie scattering. Retrievals in \texttt{TauREx} use Bayesian sampling to statistically constrain atmospheric properties that best match the observations. \texttt{TauREx} has been widely applied to analyse the atmospheres of hot Jupiters, sub-Neptunes, and super-Earths \citep[e.g.][]{tsiaras2016,tsiaras2018,tsiaras2019,edwards2020, edwards2023, changeat2022}. \texttt{TauREx}’s forward model outputs have been cross-validated against the retrieval suites \texttt{NEMESIS} and \texttt{CHIMERA}, and demonstrates excellent agreement \citep{barstow2020}. 

\subsubsection{Free Chemistry Retrievals}
\label{sec:freechem retrieval}
We employ \texttt{TauREx} to perform our parametric 'free chemistry' atmospheric retrievals. In this framework, chemical abundances are treated as free parameters without enforcing chemical or physical self-consistency. The retrieval is driven by the spectral features associated with the molecular species defined in the model, such that the selected input priors directly influence the inferred atmospheric composition.

We define an equally spaced logarithmic pressure grid with 100 layers between $10^{-9} - 10^{2}$ bar, and assume an isothermal temperature profile to represent the average atmospheric temperature. We assume a fill gas mixture of H$_2$ and He, with a solar ratio of He/H$_2$ = 0.172 \citep{asplund2009}. \texttt{TauREx} relies exclusively on absorption cross sections of active molecules, which are stored as temperature–pressure–wavelength grids. 
We include the following absorbing species: H$_2$O, CH$_4$, CO$_2$, CO, NH$_3$, OCS, H$_2$S, C$_2$H$_2$, and  C$_2$H$_4$. Our selection is informed by self-consistent model results from \citet{lavvas2026}, which couples disequilibrium chemistry with haze/cloud microphysics. These species are also predicted in \citet{jaziri2025}'s disequilibrium model results. For consistency, we use absorption cross-sections computed by ExoMol\footnote{\url{https://www.exomol.com/}} sampled at R=15,000 between 0.3–50 $\mu$m \citep{polyansky2018,yurchenko2024,yurchenko2020,li2015,coles2019,owens2024,azzam2016,chubb2020,mant2018}. We acknowledge that our selection omits some heavier hydrocarbons predicted in these models (e.g. C$_2$H$_6$) due to their lack of available opacity data for hydrogen-dominated atmospheres. We also include Rayleigh scattering and collision-induced absorption (CIA) from H$_2$-H$_2$ and H$_2$-He interactions \citep[HITRAN;][]{karman2019}. \par

The stellar emission spectrum of K2-18 is modelled using a PHOENIX spectrum \citep{husser2013}, with the stellar mass, radius, and effective temperature set to values listed in Table \ref{physical_properties}. The planetary mass and radius are fitting parameters.\par

Photochemical hazes play a key role in shaping the thermal structure of K2-18b's atmosphere \citep{lavvas2026}. Their strong Rayleigh scattering slopes across the NIRISS and NIRSpec wavelength ranges, combined with absorption features at longer wavelengths, make them important for interpreting the JWST observations. Moreover, the formation environment and gas composition directly influence their optical properties. For consistency with \citet{lavvas2026}, we consider two haze analogues based on laboratory-derived refractive indices: (i) a Titan-like "tholin" \citep[][hereafter K84]{khare1984}, and (ii) a "super-solar" exoplantary haze representative of atmospheres at 1,000 times solar metallicity \citep[][hereafter H24]{horst2018,He24}. Table \ref{tab:tholin experiments} summarises the gas mixtures and the temperatures under which these hazes were synthesised. 

\begin{figure}[!h]
    \centering
    \includegraphics[width=0.75\linewidth]{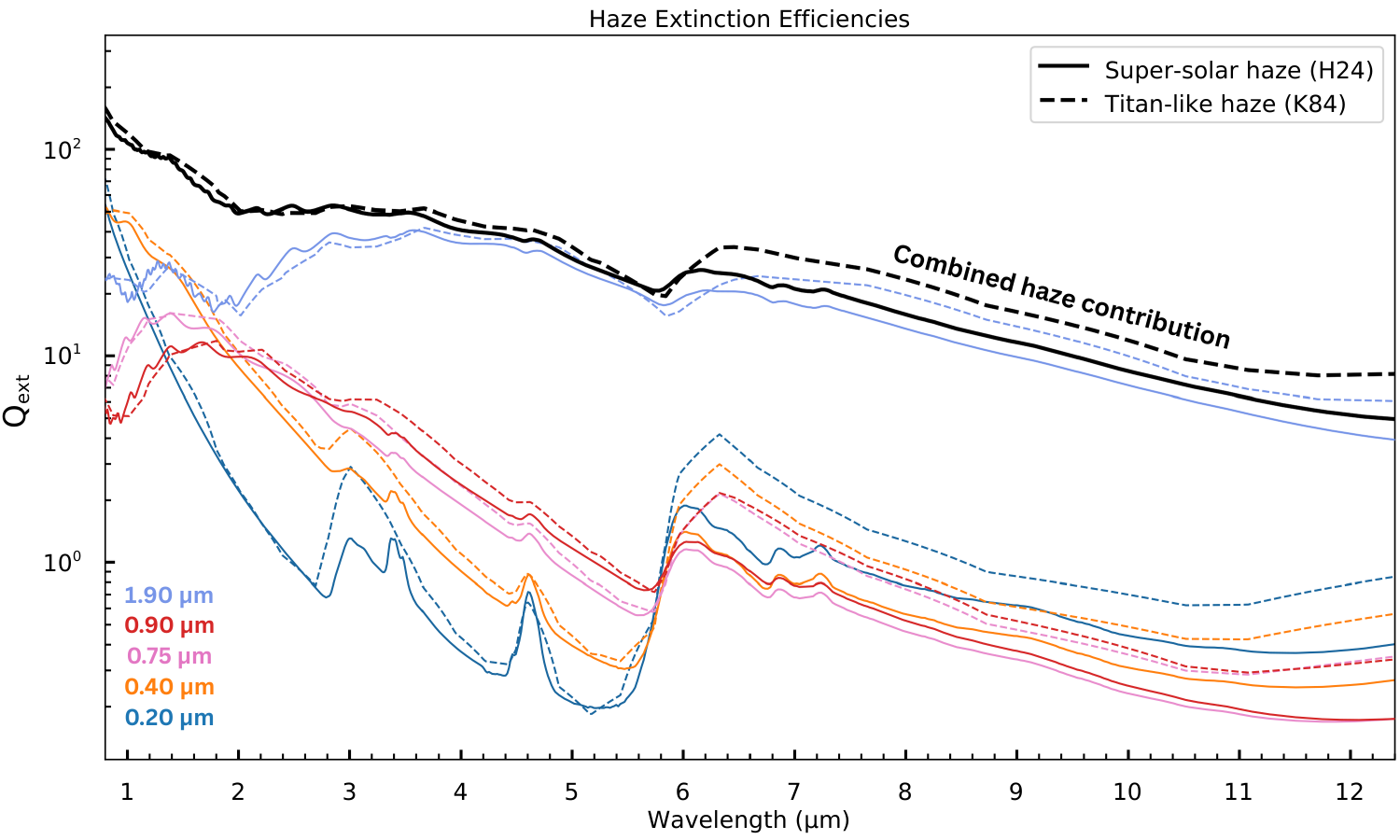}
    \caption{\textbf{Extinction efficiencies, $Q_{ext}$, of the haze particles from \citetalias{He24} (solid lines) and \citetalias{khare1984} (dashed lines) of the five particle radii used in this study: 0.20$\mu \text{m}$, 0.40$\mu \text{m}$, 0.75$\mu \text{m}$, 0.90$\mu \text{m}$, and 1.90$\mu \text{m}$, each normalised to their number densities shown in Table \ref{tab:haze particle radii and densities} . The combined haze contributions (black) are the summed $Q_{ext}$ for each individual particle size and their corresponding number density.}}
    \label{fig:tholin Qext}
\end{figure}

While both \citetalias{khare1984} and \citetalias{He24} hazes form under similar methane abundances, \citetalias{khare1984} haze contains significantly more N$_2$, whereas \citetalias{He24} haze forms in a warmer, water-rich environment. These differences lead to distinct absorption properties; for example, \citetalias{khare1984} exhibits stronger absorption near 6$\mu$m due to the enhanced nitrogen incorporation of the C$\equiv$N functional group. 
As shown in Figure \ref{fig:tholin Qext}, \citetalias{He24} hazes generally exhibit weaker absorption features, which in \citet{lavvas2026} results in a slightly cooler atmospheric structure. Other haze compositions, such as those formed in nitrogen-poor, methane-rich environments, have been explored in \citet{jaziri+drant2025}. 

\begin{table}[!tb]
\centering
\caption{The experimental conditions for the hazes used in this work}
    \label{tab:tholin experiments}
\begin{tabular}{lccc} \hline \hline
Haze analogue&Gas mixture& Temperature& Reference\\
\hline
 Titan-like&90\% N$_2$, 10\% CH$_4$& 300 K& \citetalias{khare1984}\\
 &&&\\
\multirow{2}{*}{Super-solar}&56\% H$_2$O, 11\% CH$_4$, 10\% CO$_2$, & \multirow{2}{*}{400 K}& \multirow{2}{*}{\citetalias{He24}}\\
                                                                    &6.4\% N$_2$, 1.9\% H$_2$, 14.7\% He & & \\
\hline \hline   
\end{tabular}
\end{table}

The haze opacities are calculated using Mie scattering theory  \citep{bohren2008}, assuming spherical particles and accounting for both absorption and scattering. These are computed with \texttt{YunMa} \citep{ma2023} to produce a semi-opaque haze layer in \texttt{TauREx}, given a uniform particle radius, number density, and the pressure bounds of the haze layer. Following \citet{lavvas2026}, typical haze particle sizes are expected to be $\sim$0.1$\mu$m, with growth up to $\sim$2$\mu$m in the pressure range probed by JWST. Larger particles produce stronger broadband absorption, while smaller particles enhance scattering. We therefore adopt five representative particle sizes, with corresponding radii and relative number densities listed in Table \ref{tab:haze particle radii and densities}, broadly consistent with \citet{lavvas2026}. Figure \ref{fig:tholin Qext} shows the extinction efficiencies of each size component, scaled by their relative number densities, along with the total combined contribution for both Titan-like and super-solar hazes. 

In our retrievals, the pressure bounds of the haze layer and the overall haze abundance are treated as free parameters, while the relative number densities between particle sizes are fixed at a ratio of 300:20:2:1:0.8 for radii of 0.20, 0.40, 0.75, 0.9, and 1.9 $\mu$m, normalised to the density of the 0.9 $\mu$m particles. The corresponding particle number densities are scaled by a retrieved factor, $X_{haze}$, which sets the total haze abundance. This setup allows the haze opacity to vary with pressure and total abundance while preserving a fixed particle size distribution. We note that this parametrisation is not unique, and alternative combinations of particle sizes and densities could yield similar opacity profiles while remaining consistent with \citet{lavvas2026}. In this study, we therefore adopt a single representative distribution to explore its impact on the retrievals.

\begin{table}[!tb]
    \centering
    \caption{Uniform particle radii and initial number densities used in this study to construct a haze continuum.}
    \label{tab:haze particle radii and densities}
    \begin{tabular}{ccc} \hline \hline
         Particle radius ($\mu\text{m}$)&&Initial particle number density ($\text{m}^{-3}$)\\
         \hline
         0.20&&  $3\times10^{5}$\\
         0.40&&  $2\times10^{4}$\\
         0.75&&  $2\times10^{3}$\\
         0.9&&  $1\times10^{3}$\\
         1.9&& $8\times10^{2}$\\ 
\hline \hline
    \end{tabular}
\end{table}

We perform \texttt{TauREx} retrievals on the combined NIRISS, NIRSpec, and MIRI transmission spectrum spanning 0.85-12 $\mu$m. For the NIR range, we use the JWST NIRISS and NIRSpec G395H data taken from \citet{madhusudhan2023}, spanning 0.85–2.8 $\mu$m ($R\approx930$) and 2.8–5.2 $\mu$m ($R\approx5900$), respectively, at their native resolutions. For the MIRI range, we use the \texttt{JWST-sci} spectrum extracted in Section \ref{sec:miri data reduction}, sampled at $R\approx21$, corresponding to a bin width of 0.4 $\mu$m (see Section \ref{appendix: miri: spectral binning}). As the spectra from each instrument are derived using different pipelines, systematic offsets may arise due to variations in data reduction and assumed physical parameters. In particular, NIRSpec is known to exhibit systematic offsets in transit observations \citep[e.g.,][]{moran2023,may2023,gressier2024,alderson2025}. These studies have generally introduced additional free parameters to account for these offsets. Previous studies have addressed this by introducing additional free parameters; for example, \citetalias{madhusudhan2023} reported a NIRSpec offset of –41 ppm relative to the NIRISS. In this work, we include free offset parameters for NIRSpec (NRS1 and NRS2) and MIRI LRS, relative to the NIRISS dataset.

Bayesian inference and parameter estimation are performed using the MultiNest sampling algorithm \citep{feroz2009}, implemented via PyMultiNest \citep{buchner2014}, with 1,000 MultiNest live points to ensure thorough sampling of the parameter space. Table \ref{tab:free-chem-priors} summarises the fitting parameters and prior bounds used in our baseline retrievals. 
We initially include a comprehensive set of molecules (H$_2$O, CH$_4$, CO$_2$, CO, NH$_3$, OCS, H$_2$S, C$_2$H$_2$, and  C$_2$H$_4$) to identify the dominant absorbers, and subsequently perform reduced retrievals excluding species with negligible opacity contributions. For each case, we also run retrievals without hazes to assess their impact and statistical significance. 

To explore the effect of planetary mass uncertainties, we perform additional retrievals with fixed masses spanning the $\pm0.5\sigma$ and $\pm1\sigma$ bounds derived from K2-18b’s density estimates  (MA15; Table \ref{density_estimates}), assuming $R_p = 2.6 R_{\oplus}$ \citep{benneke2019}. This corresponds to masses of $M_p= 6.1,8.4,10.7,13.0,$ and $15.3 M_{\oplus}$.

\begin{table}[!t]
\centering
\caption{Uniform priors used by the free chemistry retrievals.}
\label{tab:free-chem-priors}
\begin{tabular}{lcc} \hline \hline
Fitting parameter&& Prior bounds\\
\hline
Planetary Radius ($R_p$)                         && $[2.0,3.4]$ $R_{\oplus}$ \\
Planetary Mass ($M_p$)&& $[3,17]$ $M_{\oplus}$\\ 
Atmospheric Temperature ($T$)                    && $[50,600]$ K\\ 
Active Molecule Volume Mixing Ratio ($\log \chi_i$)&& $[-12,-0.5]$\\
 Haze Number Density Scalar ($\log X_{haze}$)& &$[-5,3]$\\
 Pressure at Top of Haze Layer ($\log P_{deck}$ / bar)& &$[-9,-1]$\\
 Pressure at Base of Haze Layer ($\log P_{base}$ / bar)& &$[-9,1]$\\
 NIRSpec Data Offsets ($\delta_{NRS1,NRS2}$)& &$[-100,100]$ ppm\\
MIRI LRS Data Offsets ($\delta_{MIRI}$)&& $[-150,150]$ ppm\\
\hline \hline   
\end{tabular}
\end{table}

\section{Results}
\label{sec:results}

\subsection{Planetary Density Estimates}
\label{sec: results planetary density estimates}

Planetary properties are all measured as a function of the stellar quantities. Thus, it is important to check the robustness of our results against the uncertainties due to the methods used to derive the stellar parameters. For K2-18, \cite{j_maldonado_2020_5654735} computed the changes in the planetary density of planet b for various estimates of the stellar mass and radius. 

Specifically, the values obtained using the \citep[][hereafter MA15]{2015A&A...577A.132M} methodology are compared with those given by \cite{Benneke17} and \cite{2019A&A...621A..49C}.

The authors discuss that the main source of variations between different stellar parameters estimates arise from the use of different mass-luminosity relationships. In particular, they show that the masses and radius provided in \cite{Benneke17} are significantly lower than the rest of estimates. Lower stellar masses and radius translate into a higher overall density for the planet K2-18 b, and places it in a different position in the planetary radius versus planetary mass diagram (i.e. a different planetary composition) with respect to the other datasets of stellar parameters \citep[see][Fig.~2]{j_maldonado_2020_5654735}. Table~\ref{density_estimates} shows the different values of the stellar masses and radii and how they translate into different densities for the planet K2-18 b.

\begin{table}[!t]
\centering
\caption{Influence of the stellar parameters on the density estimation of planet K2-18 b.}
\label{density_estimates}
\begin{tabular}{lccc}
\hline\hline\noalign{\smallskip}
 Reference$^{\dag}$                  &  M$_{\star}$ (M$_{\odot})$  &  R$_{\star}$ (R$_{\odot})$ & $\rho_{\rm P}$ (gcm$^{\rm -3}$) \\
\hline
MA15  & 0.46  $\pm$   0.07  & 0.45  $\pm$   0.06  & 3.34 $\pm$ 1.44 \\
BE17  & 0.36  $\pm$   0.05  & 0.41  $\pm$   0.04  & 3.72 $\pm$ 1.19 \\
CL19  & 0.50  $\pm$   0.004 & 0.47  $\pm$   0.01  & 3.10 $\pm$ 0.53 \\
\hline
\end{tabular}
\tablecomments{$^{\dag}$ Reference for M$_{\star}$ and R$_{\star}$.
MA15: This work using \cite{2015A&A...577A.132M} methods; BE17: \cite{Benneke17}; CL19: \cite{2019A&A...621A..49C}}
\end{table}

\subsection{Atmospheric Retrieval Results}
\label{sec: retrieval results}
We perform atmospheric retrievals on the full JWST transmission spectrum using the baseline and reduced model setups described in Section \ref{sec:atmospheric retrieval}. For models including the full set of molecules, we find that CH$_4$ and CO$_2$ consistently dominate the opacity, with non-negligible contributions from C$_2$H$_4$ in the MIRI wavelengths. H$_2$O, CO, NH$_3$, H$_2$S, OCS, and C$_2$H$_2$ are poorly constrained and contribute negligibly (Figure \ref{fig:all mols haze-free cornerplot}). Including these additional unconstrained parameters increases the prior volume and dilutes the average likelihood, and can result in wider uncertainties for the well-constrained parameters. Comparing the Bayesian evidence, we find a difference of $\ln(B) = 1.34$ between the full and reduced models (see Table \ref{tab:lnz results}), indicating that the additional parameters do not significantly improve the fit. The evidence therefore favours the reduced model, effectively marginalising over the contribution of these species. Although less comprehensive, this approach improves the statistical robustness and enables a more focused assessment of molecules with detectable spectral signatures. We therefore adopt the reduced chemical set of CH$_4$, CO$_2$, and C$_2$H$_4$ for subsequent retrievals. 

\subsubsection{Effects of Hazes on Retrieved Atmospheric Properties}
\label{sec: retrievals of hazy atmospheres}
\begin{figure}[!b]
    \centering
    \includegraphics[width=1\linewidth]{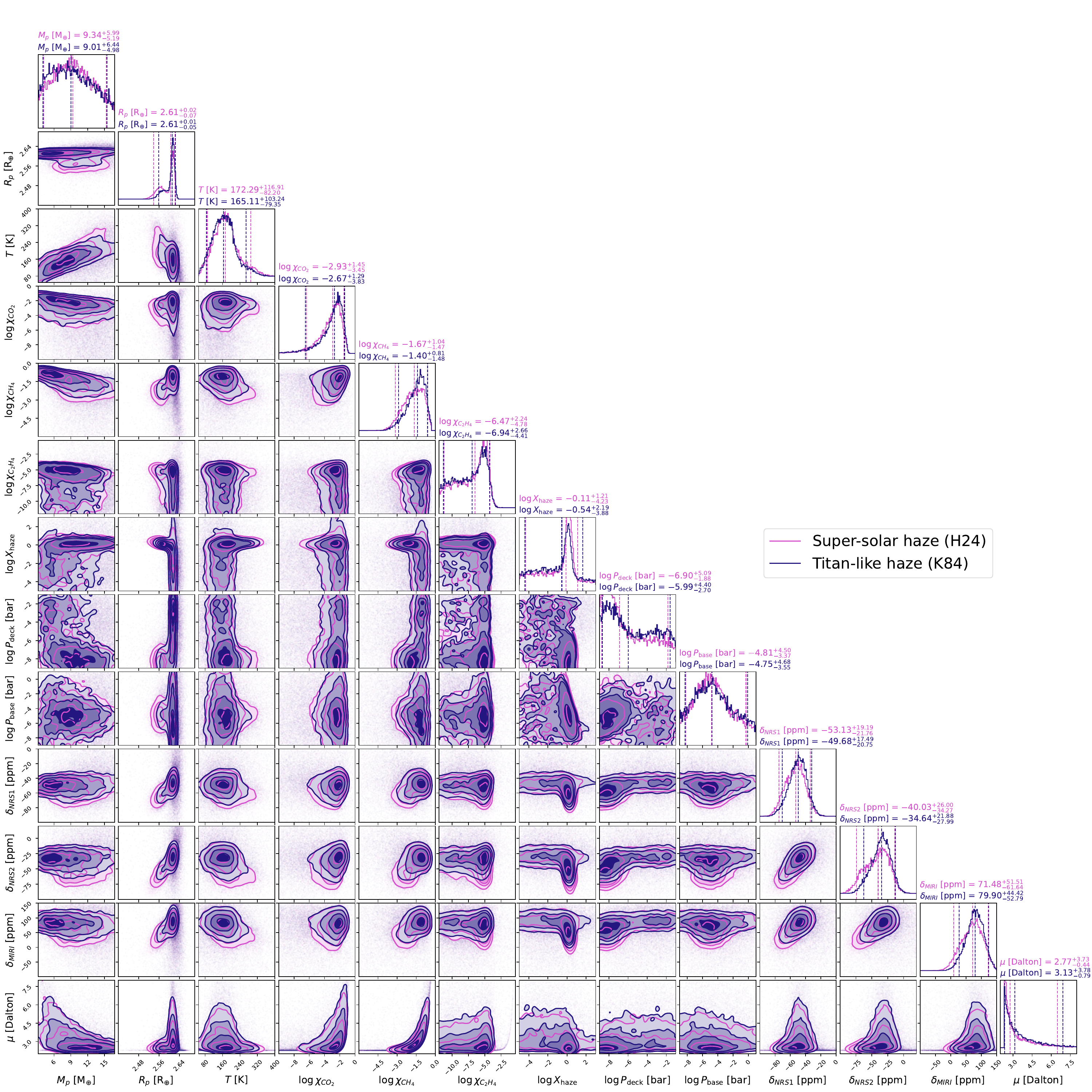}
    \caption{Comparison of posterior probability distributions for the reduced species (CH$_4$, CO$_2$, and C$_2$H$_4$) retrievals with Titan-like haze (\citetalias{khare1984}; purple) and super-solar haze (\citetalias{He24}; pink). The dashed lines represents the median values shown above each column along with $\pm2\sigma$ confidence intervals.}
    \label{fig:baseline cornerplots}
\end{figure}

\begin{figure}
    \centering
    \includegraphics[width=1\linewidth]{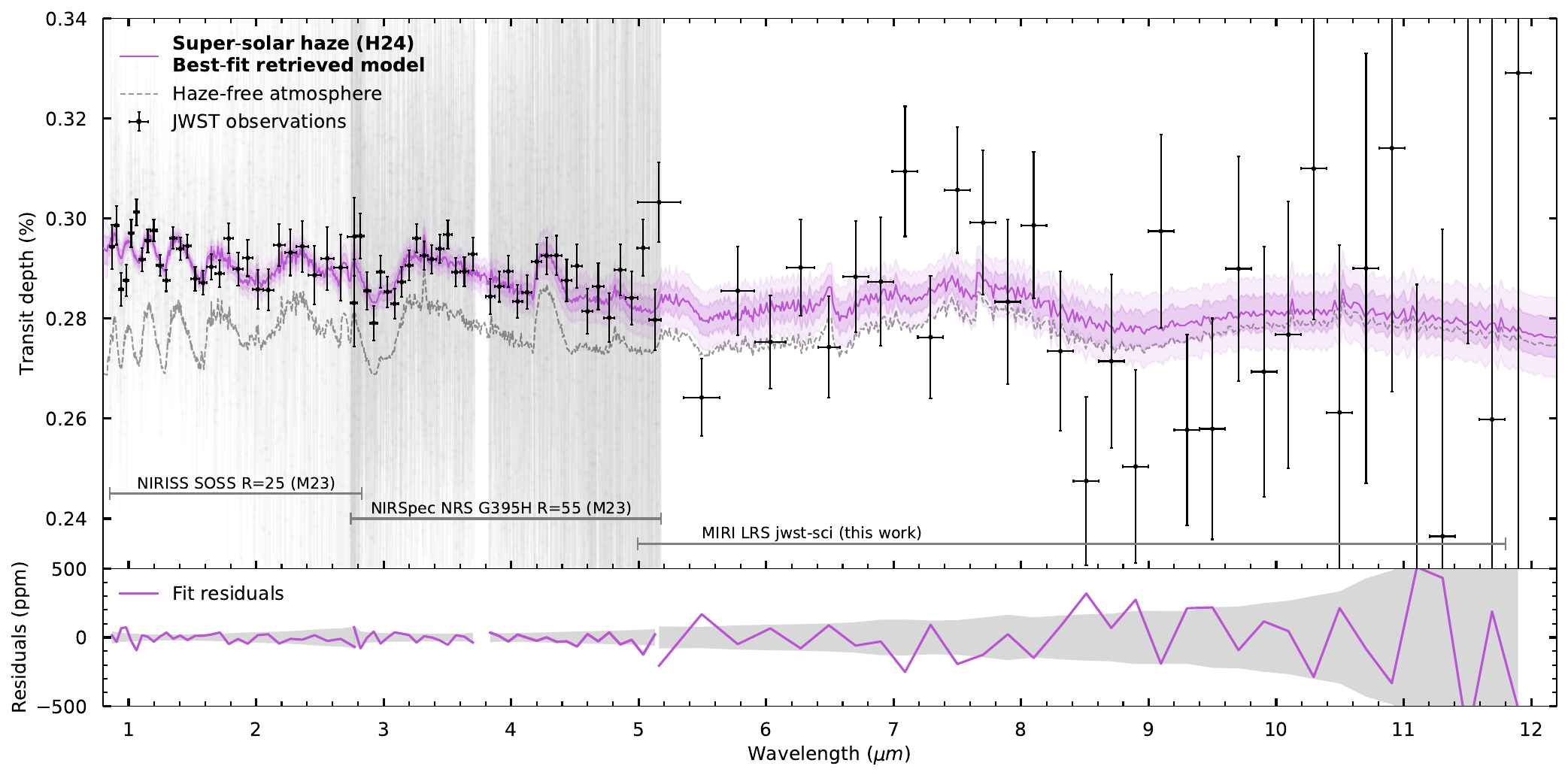}
    \caption{The best-fit (MAP) model of the super-solar haze (\citetalias{He24}) retrieval performed on the combined JWST NIRISS SOSS (\citetalias{madhusudhan2023}), NIRSpec G395H  (\citetalias{madhusudhan2023}) and MIRI LRS \texttt{JWST-sci} (Section \ref{sec:miri data reduction}) transmission spectrum. The MAP retrieved offsets have been applied to the NIRSpec NRS1, NRS2, and MIRI LRS datasets. and  \textit{Top}: The best-fit model, binned down to R=500 for clarity, is plotted in purple and the shaded regions represent the $1\sigma$ and $2\sigma$ model boundaries. The grey dashed line represents the best-fit model with a haze-free atmosphere. The NIRISS and NIRSpec observations, plotted as black error bars, are binned down to R=25 and R=55, respectively, for visual clarity. The native resolutions of the NIRISS and NIRSpec observations used in the retrievals are under-plotted in grey. \textit{Bottom}: The fit residuals (model - data) are plotted in purple and the grey region represents the data error envelope.}
    \label{fig:Best-fit H24 Spectrum}
\end{figure}

Figure \ref{fig:baseline cornerplots} shows the posterior distributions for the reduced retrieval setup (CH$_4$, CO$_2$, and C$_2$H$_4$) under the super-solar and Titan-like haze scenarios. Table \ref{tab:retrieval results (reduced setup)} summarises the median values with $\pm2\sigma$ model uncertainties alongside the Maximum A Posteriori (MAP) results. The MAP values trace regions of highest posterior probability density in parameter space, but we find they generally do not coincide with the median values due to the non-Gaussian nature of the posteriors. We therefore report the first ten MAP solutions for each retrieval following \citet{ma2025}, with the MAP$_1$ solution corresponding to the global maximum likelihood. These MAP$_1$ solutions are used to generate the best-fitting model spectra shown in Figures \ref{fig:Best-fit H24 Spectrum}, \ref{fig:Best-fit K84 Spectrum}, and \ref{fig:Best-fit haze-free Spectrum}. We note a significant spread among the MAP solutions, which highlights degeneracies between planetary mass, temperature, molecular abundances, and haze properties.

We find that the retrieved planetary mass, $M_p$, is poorly constrained across all model scenarios. Both haze-free and hazy retrievals exhibit asymmetric uncertainties, with the upper $2\sigma$ bounds broader than the lower limits. In the clear-atmosphere case, the upper mass uncertainty reaches as high as 63\%, while the Titan-like haze scenario yields an even larger upper uncertainty of 71\%, indicating strong degeneracy in the mass determination. This behaviour is reflected in the MAP solutions, where $M_p$ shows a substantial spread among the first ten solutions. For the haze-free retrieval, the difference between the largest and smallest MAP values is $\Delta M_p =2M_\oplus$ ($7.32$ and $5.32M_\oplus$). This spread increases to $\Delta M_p =3.2 M_\oplus$ for the super-solar haze case, and $\Delta M_p =5.1M_\oplus$ for the Titan-like haze scenario, which shows the largest variation. We further explore this behaviour using retrievals with fixed $M_p$ values in Section \ref{sec: retrievals with fixed Mp}.

For the MAP$_1$ solutions, the retrieved masses are $5.76 M_{\oplus}$ for the haze-free case, and $8.20 M_{\oplus}$ and $8.74 M_{\oplus}$ for the super-solar and Titan-like haze scenarios, respectively. In the presence of hazes, the planetary radius, $R_p$, deviates slightly from the clear-atmosphere value of $1.61 R_{\oplus}$, with secondary peaks manifesting in the planetary radius posterior distributions associated with the degenerate hazy and non-hazy states (Figure \ref{fig:baseline cornerplots}). As a result, the retrieval exhibits a bias toward smaller radii to reproduce the observed spectrum, resulting in MAP$_1$ values of $R_p=2.52R_{\oplus}$ and $2.57R_{\oplus}$ for the super-solar and Titan-like haze cases, respectively. These shifts correspond to modest variations of approximately 2–3\%. 

The atmospheric temperature is more tightly constrained in the haze-free scenario, with $T=164^{+64}_{-53}$ K and a MAP$_1$ value of $145$ K. In contrast, the hazy retrievals produce broader posterior distributions, converging toward warmer MAP$_1$ temperatures of $212$ K and $179$ K for the super-solar and Titan-like haze cases, respectively.

Molecular abundances show varying degrees of constraint. C$_2$H$_4$ contributes only significantly within the MIRI wavelength range and we only recover the upper $2\sigma$ limits, which is evident from their weakly peaked marginalised posteriors. Across all hazy and haze-free retrievals, we derive a consistent upper limit of $\lesssim -4.3$ dex. In contrast, CH$_4$ and CO$_2$ abundances are well constrained, with marginalised posteriors that are broadly consistent between haze-free and hazy retrievals. However, the MAP values for the hazy scenarios indicate a bias toward lower CH$_4$ and CO$_2$ abundances, resulting in a reduced mean molecular weight. This reflects a well-known degeneracy between mean molecular weight and haze opacity. For the haze-free model, our best-fit solution (MAP$_1$) yields $\log \chi_{CH_4} = -0.92$, $\log \chi_{CO_2} = -1.73$, and $\mu = 4.72$ Da.
For the super-solar haze case, the best-fit values are $\log \chi_{CH_4} = -2.07$, $\log \chi_{CO_2} = -2.44$, and $\mu = 2.58$ Da. For the Titan-like haze case, the best-fit values are $\log \chi_{CH_4} = -1.68$, $\log \chi_{CO_2} = -2.05$, and $\mu = 2.97$ Da.

The hazy-atmosphere retrievals show that both hazy scenarios allow aerosols to be present in the upper atmosphere at levels that significantly alter the transmission spectrum. The super-solar haze case yields a MAP$_1$ haze layer extending from $\log P_{base}=-6.44$ to $\log P_{deck} = -7.69$ bar, with a number density scaling of $\log X_{haze}= 0.74$, corresponding to approximately 5.5 times the initial reference density (Table \ref{tab:haze particle radii and densities}). The Titan-like haze scenario produces a thicker vertical extent, spanning $\log P_{base}=-5.91$ to $\log P_{deck} = -8.66$ bar, but with a lower scaling of $\log X_{haze}= 0.121$, or roughly 1.3 times the initial density. 

\begin{figure}
    \centering
    \includegraphics[width=0.9\linewidth]{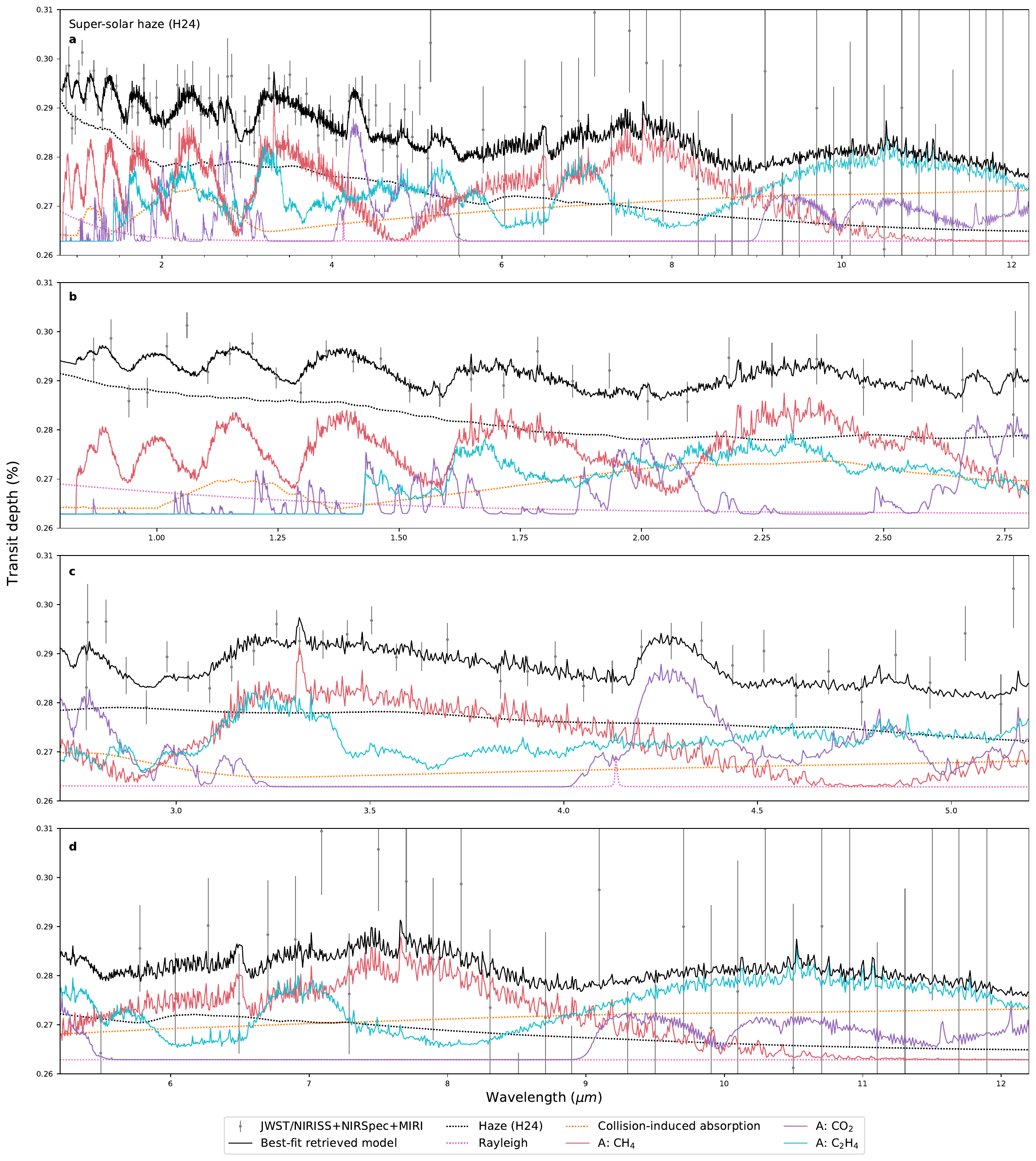}
    \caption{Opacity contributions of the best-fit super-solar haze (\citetalias{He24}) model plotted as a solid black line (binned to R = 1000). Each curve represents a contribution from A: individual gas-phase absorption; haze particle absorption and scattering; Rayleigh scattering; and collision-induced absorption. The NIRISS (R=25), NIRSpec (R=55) and MIRI data points are underplotted in grey. \textit{Panel \textbf{a}}: the broad wavelength range (0.8-12 $\mu$m) of observations. \textit{Panel \textbf{b}}: the NIRISS SOSS observations. \textit{Panel \textbf{c}}: the NIRSpec G395H observations. \textit{Panel \textbf{d}}: the MIRI LRS observations.}
    \label{fig:Best-fit H24 contributions}
\end{figure}

Overall, the retrieved atmospheric properties across all scenarios are consistent with a sub-Neptune planet possessing an H$_2$-dominated atmosphere. All models reproduce the main spectral features observed in the NIRISS and NIRSpec data. The best-fit (MAP$_1$) transmission spectra for the super-solar and Titan-like haze cases (Figures \ref{fig:Best-fit H24 Spectrum} and\ref{fig:Best-fit K84 Spectrum}) match the observed data within $2\sigma$ model uncertainties. Figures  \ref{fig:Best-fit H24 contributions} and \ref{fig:Best-fit K84 contributions} show the spectral contributions from molecular absorption, haze, Rayleigh scattering, and collision-induced absorption. CH$_4$ and CO$_2$ dominate the transmission spectrum, with C$_2$H$_4$ strongly contributing in the MIRI region. The haze-free best-fit spectrum (Fig \ref{fig:Best-fit haze-free Spectrum}) appears to be flatter in the MIRI wavelength range, primarily due to a higher mean molecular weight ($4.72$ Dalton) and a lower atmospheric temperature ($145$ K), both of which reduce the atmospheric scale height. The retrieval compensates for this by favouring a lower planetary mass ($5.76 M_{\oplus}$) to prevent the scale height from becoming too small. However, in hazy retrievals, the scale height is less suppressed. In these cases, a continuum of haze particles spanning multiple sizes is required to achieve an optimal fit across both the NIR and MIRI wavelengths. The inclusion of this haze continuum introduces strong Rayleigh scattering at short wavelengths, which reduces feature amplitudes in the NIR while preserving stronger features in the MIRI range, leading to the most significant improvement in the fit at longer wavelengths. Comparing the Bayesian evidences (Table \ref{tab:lnz results}), both haze scenarios are statistically preferred over the haze-free case, with Bayes factors of $\ln(B) = 1.84$ and $1.73$ for the super-solar and Titan-like haze models, respectively.

\subsubsection{Sensitivity of Atmospheric Retrievals to Planetary Mass}
\label{sec: retrievals with fixed Mp}

The results presented in Section \ref{sec: retrievals of hazy atmospheres} and Table \ref{tab:retrieval results (reduced setup)} indicate that planetary mass is the least well-constrained parameter in both haze-free and hazy retrievals. This arises because $M_p$ is inversely proportional to the atmospheric scale height and is therefore strongly coupled to temperature and mean molecular weight, such that an increase in mass can mimic the effect of a lower temperature or a heavier atmosphere. Combined with degeneracies between haze opacity and molecular abundances, this makes it difficult to isolate the dominant drivers of variation in the MAP solutions.

\begin{figure}
    \centering
    \includegraphics[width=1\linewidth]{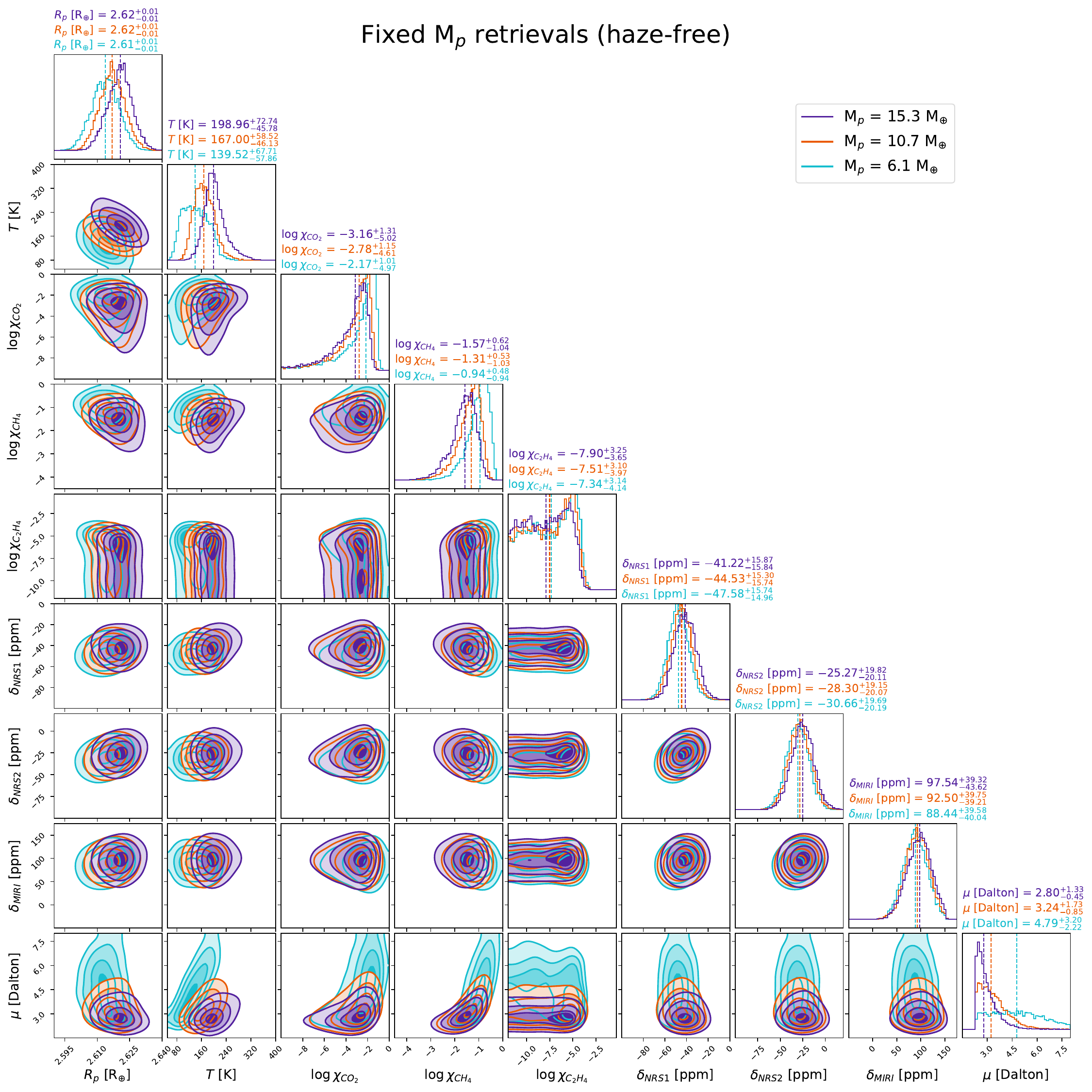}
    \caption{Comparison of posterior probability distributions for haze-free retrievals performed with fixed $M_p =6.1, 10.7,$ and $15.3M_{\oplus}$. The dashed lines represent the median, corresponding to the values shown above each column along with $\pm2\sigma$ confidence intervals.}
    \label{fig:hazefree_fixed_mp_cornerplots}
\end{figure}

To address this, we perform a series of retrievals in which $M_p$ is fixed to known values. Specifically, for each atmospheric scenario, we explore masses spanning the $\pm0.5\sigma$ and $\pm1\sigma$ bounds derived from density estimates  (MA15; Table \ref{density_estimates}), assuming $R_p = 2.6 R_{\oplus}$ \citep{benneke2019}, corresponding to $M_p= 6.1,8.4,10.7,13.0,$ and $15.3 M_{\oplus}$. The resulting posterior distributions are shown in Figures \ref{fig:hazefree_fixed_mp_cornerplots}, \ref{fig:h24_fixed_mp_cornerplots}, and \ref{fig:k84_fixed_mp_cornerplots}, with a subset of cases ($M_p= 6.1, 10.7,$ and $15.3 M_{\oplus}$) plotted for clarity. Tables \ref{tab:retrieval results hazefree fixed mp}, \ref{tab:retrieval results h24 fixed mp}, and \ref{tab:retrieval results k84 fixed mp} summarises the median values with $\pm2\sigma$ model uncertainties along with the first five MAP solutions for each case for the haze-free, super-solar haze, and Titan-like haze cases, respectively.

We find that fixing the planetary mass introduces systematic shifts in the retrieved atmospheric properties. Across all scenarios, higher $M_p$ values drive the retrieval toward warmer temperatures and lower mean molecular weights to maintain a consistent scale height. For the haze-free case, the median temperature varies by $\sim60$ K between the lowest and highest fixed-$M_p$ models, while the effect is more pronounced in the presence of hazes, reaching $\sim92$ K for the Titan-like haze and up to $\sim166$ K for the super-solar haze scenario. In addition, the temperature posteriors broaden with increasing mass, and in the super-solar haze case begin to show bimodal structure. The molecular abundances also respond differently depending on the atmospheric model. In the haze-free retrievals, CO$_2$ displays the strongest sensitivity across retrievals with different $M_p$, varying by $\sim1$ dex, while CH$_4$ changes more modestly ($\sim0.6$ dex), leading to differences in the mean molecular weight of up to 2 Da. In contrast, for hazy atmospheres, CH$_4$ becomes more sensitive to the assumed mass, varying by $0.92$ and $0.82$ dex for the super-solar and Titan-like haze cases, respectively.

Analysis of the first five MAP solutions for each fixed-$M_p$ retrieval (Tables \ref{tab:retrieval results hazefree fixed mp}, \ref{tab:retrieval results h24 fixed mp}, and \ref{tab:retrieval results k84 fixed mp}) shows that, although the MAP temperatures exhibit some residual spread, they are more tightly clustered than in the free-$M_p$ case. In the haze-free retrievals, the CO$_2$ abundance remains unstable, varying by up to $\sim1$ dex across the MAP solutions for a given fixed mass. Given the relatively high molecular weight of CO$_2$ these variations lead to noticeable differences in the derived mean molecular weight. In contrast, CH$_4$ shows smaller variations (typically $<0.4$ dex) and stays close to its median value, suggesting it is more reliably constrained when $M_p$ is fixed. For the super-solar and Titan-like haze cases, similar variations are seen between temperature, CO$_2$, and CH$_4$, with the latter showing larger spreads of up to $0.8$ dex and $1$ dex, respectively. The haze parameters, particularly the number density scaling and base pressure, also vary across MAP solutions, pointing to degeneracies in the vertical extent and optical thickness of the haze layer. Notably, this behaviour is also present in the free-$M_p$ retrievals (Table \ref{tab:retrieval results (reduced setup)}), suggesting that haze-related degeneracies are largely decoupled from the influence of planetary mass uncertainty on the other atmospheric parameters.

\newpage
\section{Discussion} 
\label{sec:discussion}

\subsection{Comparison with Previous Literature}
\label{comparison with literature}

Here, we compare our results with those reported in the literature. We note that differences in data reduction, retrieval frameworks, and datasets can affect direct comparisons. For example, \citet{schmidt2025}'s analysis uses a different data reduction of the NIR datasets, and \citet{hu2025} includes more recent observations from NIRSpec G395H and G235H (GO Program 2372). For consistency, we adopt the results from our fixed-mass retrievals at $M_p=8.4M_{\oplus}$ (Tables \ref{tab:retrieval results hazefree fixed mp}, \ref{tab:retrieval results h24 fixed mp}, and \ref{tab:retrieval results k84 fixed mp}), close to the value of $8.63M_\oplus$ \citep{benneke2019}, which is commonly assumed across the studies discussed here.

For the haze-free case, our retrieved CH$_4$ abundance ($\log\chi_{\mathrm{CH_4}} = -1.12^{+0.51}_{-1.03}$, MAP$_1=-0.98$) is consistent with combined NIR+MIRI studies such as \citet{luque2025} ($-0.85^{+0.20}_{-0.32}$; \texttt{PLATON}) and \citet{jaziri+drant2025} ($-1.20^{+0.41}_{-0.62}$) which use the same NIR datasets but adopt a different reduction of the MIRI LRS data, with \citet{jaziri+drant2025} additionally fixing instrumental offsets to $-41$ ppm (NIRSpec) and $+160$ ppm (MIRI). Our results are also consistent with NIR-focused studies, including \citet{schmidt2025} ($-1.15^{+0.40}_{-0.52}$; \texttt{POSEIDON}) and \citet{hu2025} ($-1.06^{+0.24}_{-0.37}$; \texttt{ExoTR}), and agree within $2\sigma$ of \citetalias{madhusudhan2023} ($-1.89^{+0.63}_{-0.71}$; two offset case). Differences in CH$_4$ abundance across these works using the same dataset have been attributed to modelling choices, particularly the use of different opacity databases (e.g. \citetalias{madhusudhan2023} adopt the \citet{huang2013,huang2017} line list, while others use ExoMol; \citealt{yurchenko2024}). Our CO$_2$ constraints ($\log\chi_{\mathrm{CO_2}} = -2.67^{+1.11}_{-4.67}$, MAP$_1$ = $-1.68$) are broadly consistent with \citetalias{madhusudhan2023}, the upper limits reported by \citet{schmidt2025}, and the values from \citet{hu2025} and \citet{luque2025}, while our upper limits on C$_2$H$_4$ are in agreement with \citet{jaziri+drant2025}.

For the hazy retrievals, both super-solar and Titan-like cases yield CH$_4$ and CO$_2$ abundances that are systematically lower by $\sim 0.5$--$1$ dex compared to haze-free results in the literature, leading to reduced mean molecular weights. This behaviour is consistent with the well-known degeneracy between haze opacity and atmospheric composition, where a haze layer suppresses spectral features and can be offset by a lighter atmosphere. A similar trend is reported by \citet{jaziri+drant2025}, who include haze refractive indices derived from a  $95\% $ Ar $+$ $5\%$ CH$_4$ mixture and obtain $\log\chi_{\mathrm{CH_4}} = -3.53^{+0.58}_{-0.53}$ with $\mu=2.32^{+0.04}_{-0.01}$ Da, slightly lower than our haze cases ($\mu=2.61^{+2.38}_{-0.30}$ and $2.99^{+2.42}_{-0.66}$ Da), likely due to differences in haze opacities, datasets, and retrieval setups.

For temperature, our haze-free retrieval yields $150^{+66}_{-46}$ K (MAP$_1 =200$ K), with hazy cases producing comparable values at $M_p = 8.4M_{\oplus}$, in agreement with \citet{schmidt2025} and \citet{hu2025}, and within $2\sigma$ of \citetalias{madhusudhan2023} ($235^{+78}_{-56}$ K). Warmer temperatures in hazy atmospheres, as reported by \citet{jaziri+drant2025}, are only reproduced in our retrievals for $M_p > 8.4M_{\oplus}$ or when $M_p$ is treated as a free parameter. Moreover, we assumed an isothermal temperature profile, whereas \citet{jaziri+drant2025} used a 4-point temperature profile that allows for a thermal inversion. Finally, the MIRI-only analyses (e.g. \citetalias{madhusudhan2025}) suggest much higher temperatures (e.g. $\sim420$ K at 1 mbar) than what is seen in our retrievals and in the literature. These are likely driven by larger feature amplitudes in the MIRI data. Given the known discrepancies of the MIRI-only studies, we limit our comparison to studies using NIR or joint NIR+MIRI datasets.

\subsection{Instrumental Offsets and Systematics}
\label{instrumental offsets and systematics}

Our retrieval framework includes wavelength-dependent offsets between JWST instruments, which are known to influence inferred atmospheric properties. In particular, NIRSpec systematics are well documented in transit spectroscopy \citep[e.g.,][]{moran2023, may2023, gressier2024, alderson2025}, and previous studies have shown that at least one free offset is required to achieve consistent fits to NIR data \citep{madhusudhan2023, schmidt2025}. For example, \citetalias{madhusudhan2023} reported a NIRSpec offset of $-$41 ppm relative to NIRISS. More recent analyses \citep{hu2025} further identify visit-to-visit inconsistencies in the NIRSpec G235H and G395H observations, with detector-level offsets (NRS1 vs. NRS2) reaching up to $\sim$90 ppm. Offsets have also been suggested for MIRI, with reported values of $\sim$ 110--165 ppm \citep{luque2025}, although their origin is less well understood given that independent reductions yield consistent mean fluxes and MIRI is less sensitive to stellar heterogeneity. This raises the possibility that time-dependent or reduction-dependent systematics may play a role \citep{luque2025,stevenson2025}.

In our retrievals, allowing for instrumental offsets is necessary to reproduce the observed spectra. When offsets are not included, the inferred molecular abundances --- particularly CO$_2$, which has strong absorption between 4--4.5 $\mu$m in the NRS2 band --- are especially biased \citep[see e.g.][]{madhusudhan2023, schmidt2025}.
For the haze-free retrievals, we consistently retrieve offsets of $\sim-$45 ppm (NRS1), $\sim-$30 ppm (NRS2), and $\sim+$90 ppm (MIRI). In contrast, the hazy retrievals require larger negative offsets for NIRSpec (NRS1: $\sim-$75 ppm and $-$65 ppm; NRS2: $\sim-$70 ppm and $\sim-$55 ppm for the super-solar and Titan-like cases, respectively), while the MIRI offsets are reduced to  $\sim+$20--40 ppm and  $\sim+$30--60 ppm. These differences reflect the degeneracy between haze opacity and mean molecular weight: in hazy models, the abundance of CO$_2$ and therefore its NIRSpec features are suppressed, requiring larger offsets to match the data. Conversely, haze-free models produce flatter spectra in the MIRI range, leading to larger positive offsets to reconcile the model with the observations, whereas hazy models allow the spectrum to display stronger features and therefore require smaller corrections.

Despite this, the MIRI data remain sensitive to reduction choices and residual systematics. As discussed in Sections \ref{sec:pylightcurve} and \ref{appendix: miri: no. of integrations masked}, individual MIRI data points can vary depending on the treatment of time-correlated noise and the number of discarded initial integrations. Other studies have shown that spectral binning can also affect the inferred atmospheric properties \citep{stevenson2025}, and the overall constraining power of MIRI remains limited \citep{welbanks2025, luque2025}. Additional MIRI observations will therefore be important to better constrain the relative amplitudes and improve consistency with the NIR datasets.

\subsection{Hazes on K2-18b?}
\label{plausibility of haze}

Our retrievals show that both super-solar and Titan-like haze scenarios can reproduce the combined JWST observations. While these hazy models are statistically preferred over the haze-free case ($\ln(B) = 1.84$ and $1.73$; Table \ref{tab:lnz results}), the evidence is weak-to-moderate based on standard Bayesian criteria \citep{sellke2001, benneke_and_seager2013}. We emphasise that these metrics are typically used for molecular detections, whereas our goal is to test whether hazes can provide a consistent explanation for the differing feature amplitudes between the NIR and MIRI datasets. As such, the Bayes factors indicate a preference for hazes but are not sufficient to claim a detection or rule out alternative interpretations.

From a physical standpoint, hazes on K2-18b are plausible. Several studies suggest that the planet’s atmosphere is in disequilibrium \citep[e.g.][]{wogan2024, jaziri2025}, and methane --- which is expected to dominate the carbon budget at these temperatures \citep{moses2013} --- can drive hydrocarbon production through photochemistry in an H$_2$-rich atmosphere. Recent work by \citet{lavvas2026} shows that CH$_4$-driven photochemistry can produce significant haze abundances that alter the transmission spectrum and influence the thermal structure. They also find that haze cooling may enable the condensation of NH$_4$SH, offering a possible explanation for the lack of observed NH$_3$. In addition, the relatively active host star likely enhances UV and EUV fluxes \citep{guinan2019, Santos20}, promoting the production of hydrocarbons such as C$_2$H$_2$, C$_2$H$_4$, C$_2$H$_6$ via collisions with photoelectrons \citep{lavvas2026}. These species are known precursors to haze formation and are observed in Solar System atmospheres such as Titan and Neptune \citep{waite2005, meadows2008}. In our retrievals, C$_2$H$_2$ is not well constrained, and C$_2$H$_6$ is not included due to lack of available opacity data for H$_2$-dominated atmospheres. C$_2$H$_4$ shows strong features in the MIRI range but is only weakly detected due to its broad spectral signature and the limited signal-to-noise. Constraining these species will be important for understanding haze formation pathways. Recent studies \citep{stevenson2025, jaziri2025} also support a role for C$_2$H$_4$ in explaining the MIRI features.

We do not find a strong preference between the super-solar and Titan-like haze scenarios, as they share broadly similar optical properties, with the main difference being enhanced absorption at $\sim$6–8 $\mu$m in the Titan-like case due to enhanced nitrogen-bearing functional groups. Other studies using laboratory-based haze optical properties \citep{drant2026} find that CH$_4$-rich, N$_2$-poor hazes can reproduce the combined observations, and explains the MIRI feature at $\sim$7 $\mu$m due to C-H bending absorption of the haze \citep{jaziri+drant2025}. Together, these results highlight that haze composition and optical properties play a key role in interpreting the spectrum of K2-18b, and likely other temperate sub-Neptunes. 
Therefore, we advocate for more laboratory work to better constrain constrain haze properties and reduce current degeneracies in atmospheric retrievals.

Alternative mechanisms that could mimic haze-like signatures must also be considered. Unocculted starspots, for instance, can introduce a blueward slope in the NIRISS wavelength range \citep{moran2023}, but the combined 0.85–12 $\mu$m dataset demands a slope that extends to longer wavelengths, which is more naturally explained by the presence of haze. Additionally, both \citet{hu2025} and \citetalias{madhusudhan2023} found that inhomogeneous clouds or hazes are favoured over stellar heterogeneity in their models. Another possibility is a circumplanetary dust ring, which could produce a broad continuum with composition-dependent features \citep{ohno2022a,ohno2022b}. Since gas giants in the Solar System host both hazes and tenuous rings, a combined scenario for K2-18b warrants future investigation.

\subsection{Impact of Planetary Mass Uncertainties on Atmospheric Retrievals}
\label{sec:planetary mass uncertainties}

In Section \ref{sec: retrievals of hazy atmospheres}, we present retrievals where the planetary mass, $M_p$, is treated as a free parameter with uniform priors. This choice is motivated by the large uncertainties in the bulk density estimates (Section \ref{sec: results planetary density estimates}), which arise from differences in stellar parameters reported in the literature (e.g. \citealt{Benneke17, cloutier2019}) and the use of different mass-luminosity relations. These propagate directly into a wide range of inferred planetary masses. Fixing $M_p$ a priori can therefore bias the atmospheric scale height and, in turn, the retrieved temperatures and compositions. Consistent with this, we find that $M_p$ cannot be robustly constrained from the transmission spectra alone and remains the least well-constrained parameter in both haze-free and hazy retrievals, with uncertainties reaching up to $\sim 71\%$. Our density-based estimate of $M_p = 10.7 \pm 4.6M_{\oplus}$ (assuming $R_p = 2.6R_{\oplus}$; \citealt{benneke2019}) spans a similar range to the freely retrieved values. Notably, many of the haze-free MAP solutions fall below this range (Table \ref{tab:retrieval results (reduced setup)}), with only one solution (MAP$_6$; $M_p=7.32 M_{\oplus}$) consistent within $1\sigma$ of the commonly adopted value of $8.63 \pm 1.35 M_{\oplus}$ \citep{benneke2019}, while hazy retrievals tend to favour higher masses, with their best-fit solutions consistent with this literature value. 

To explore the impact of this uncertainty, we performed retrievals with fixed masses spanning $6.1$ to $15.3$ $M_{\oplus}$, corresponding to the upper and lower bounds of our density-based mass estimate (Section \ref{sec: retrievals with fixed Mp}). Fixing $M_p$ introduces systematic shifts in the inferred atmospheric properties, with higher masses generally leading to warmer temperatures and lower mean molecular weights to maintain a similar scale height. Despite this, the Bayesian evidence does not strongly favour any particular mass assumption in the hazy cases (Table \ref{tab:lnz results fixed mp}). For the haze-free scenario, lower-mass solutions ($M_p = 6.1M_{\oplus}$) provide a marginally better statistical fit, but yield unrealistically cool temperatures ($\sim 140$ $\mathrm{K}$), well below the expected equilibrium temperature ($\sim 281$ $\mathrm{K}$; \citealt{howard2025}). More generally, haze-free models tend to produce systematically cooler ($<205$ $\mathrm{K}$) atmospheres for $M_p \leq 13.0M_{\oplus}$, although this may partly reflect our assumption of an isothermal temperature profile. In contrast, recent work by \citet{lavvas2026} suggests that the atmospheric region probed by transmission spectroscopy should have temperatures of $\sim 250$--$300$ $\mathrm{K}$. Our hazy retrievals are broadly consistent with this expectation only for higher mass assumptions, with the super-solar haze case satisfying this range for $M_p \geq 10.7,M_{\oplus}$ and the Titan-like haze case for $M_p \geq 13.0,M_{\oplus}$.

These results highlight the well-known degeneracy between planetary mass and atmospheric scale height \citep{changeat2022}. Moreover, in their sensitivity study using simulated ARIEL spectra over $0.5$--$8\mu\mathrm{m}$, \citet{changeat2022} showed that planetary mass can be accurately retrieved for clear, cloud-free atmospheres. However, when high-altitude clouds or hazes are included, the uncertainty in the retrieved mass increases substantially --- reaching up to $\sim 60\%$ in extreme cases --- due to the suppression of spectral features and the resulting degeneracy with mean molecular weight and temperature. This behaviour is consistent with our findings, where the inclusion of hazes both broadens the allowed $M_p$ range and amplifies degeneracies with other atmospheric parameters. 

Our results further highlight the strong degeneracy between planetary mass and atmospheric scale height in transmission spectroscopy retrievals, particularly in the presence of hazes. We therefore stress the need for improved independent constraints on $M_p$. Current mass estimates for K2-18b still carry substantial uncertainties, driven in part by the low signal-to-noise of radial velocity measurements of its relatively faint host star, as well as differences in stellar parameter estimates arising from the use of different mass-luminosity relations. These factors propagate directly into the derived planetary mass and density, complicating comparisons across studies. We therefore encourage future work to carefully account for these uncertainties when interpreting atmospheric retrievals, and to prioritise improved mass measurements, as tighter constraints on $M_p$ will be key to breaking degeneracies and obtaining more reliable atmospheric properties.

\subsection{Towards Self-consistency}
\label{towards self consistency}

A number of simplified assumptions were adopted in this study, and we acknowledge the limitations of our models while outlining directions for improvement. Key questions raised by our results --- such as how hazes influence the abundances of CH$_4$ and CO$_2$, and what conditions are required to produce haze in quantities consistent with the observed transmission spectrum --- ultimately require physically self-consistent modelling. Our retrieval framework uses a free-chemistry approach, which does not enforce chemical or radiative self-consistency. While this provides flexibility and computational efficiency, modelling hazy atmospheres likely requires stronger physical constraints. That said, we adopt several physically motivated choices: the initial set of molecular opacities (H$_2$O, CH$_4$, CO$_2$, CO, NH$_3$, OCS, H$_2$S, C$_2$H$_2$, and C$_2$H$_4$) is guided by the self-consistent disequilibrium models of \citet{lavvas2026}. Accounting for disequilibrium chemistry is important, as it can significantly alter the abundances of key species such as NH$_3$, OCS, and CO$_2$ relative to equilibrium predictions. However, further work is needed to explain features such as the apparent depletion of H$_2$O and the absence of NH$_3$ in the observed spectra. More broadly, current disequilibrium models may rely on incomplete chemical networks, which can bias outcomes. Therefore, studies focusing on model inter-comparisons would be particularly valuable for the wider community.

In addition, our assumption of an isothermal temperature profile may not adequately represent hazy atmospheres. Hazes can heat the upper atmosphere through enhanced UV absorption, resulting in vertical energy redistribution, and potentially producing thermal inversions \citep{arney2016}. A warmer upper atmosphere would increase the local scale height, affecting the interpretation of spectral features. Temperature profiles that allow for such structure (e.g. \citealt{jaziri+drant2025}) are therefore likely more appropriate when modelling high-altitude hazes. Similarly, our treatment of haze is simplified and parametric. We adopt haze opacities, particle sizes, and number densities broadly based on \citet{lavvas2026}, but represent the haze using only five discrete particle sizes with fixed relative abundances, uniform with altitude, and allow only the total abundance and pressure boundaries to vary. This is only one possible representation, and alternative particle size distributions and number densities could produce similar opacity profiles. In reality, haze properties depend on vertical transport and microphysics, with particle sizes and abundances varying with pressure–temperature structure, eddy diffusion, haze production rates, and stellar UV irradiation \citep[e.g.][]{ohno2020,kawashima2019}. Multiple haze populations with different compositions may also coexist. Fully self-consistent retrievals that couple these processes with radiative transfer and chemistry remain computationally expensive, but recent advances in machine learning for accelerating forward models \citep[e.g.][]{Paraskevaidou2025,Vojtekova2025} offer a promising path towards more physically-motivated atmospheric characterisation in future studies.

\section{Conclusions}
\label{sec:conclusions}

We have presented a new analysis of K2-18b’s JWST transmission spectrum, combining an independent reduction of the MIRI LRS observations with previously published NIRISS and NIRSpec data \citepalias{madhusudhan2023}. Using free-chemistry Bayesian retrievals, we investigated the role of hydrocarbon hazes in shaping the transmission spectrum across $0.85$--$12 \mu$m. We also assessed the impact of stellar parameter uncertainties on the derived bulk properties of the planet. Using the stellar parameters derived in this work, we obtain a planetary density of $\rho_P = 3.34 \pm 1.44$ g cm$^{-3}$. This value is consistent with previous estimates, although differences in adopted stellar mass--radius relations lead to variations in the inferred planetary density and therefore the interpretation of K2-18b’s bulk composition. 

Our retrievals show that both super-solar and Titan-like haze scenarios can reproduce the observed spectral features and provide a consistent explanation for the differing feature amplitudes between the near- and mid-infrared datasets. In particular, the inclusion of a haze continuum suppresses spectral features in the NIR while preserving stronger absorption signatures in the MIRI wavelength range, improving the overall agreement with the combined JWST spectrum. While the statistical preference over haze-free models is modest, the physical consistency of hazy solutions, combined with their ability to match the full spectral range, supports their interpretation as a viable atmospheric scenario. 

Across all retrievals, K2-18b is consistent with an H$_2$-dominated sub-Neptune atmosphere, with CH$_4$ and CO$_2$ as the dominant absorbers, consistent with previous literature findings. However, the retrieved abundances are strongly coupled to haze opacity, with hazy models favouring systematically lower molecular abundances, reflecting the well-known degeneracy between haze/cloud opacity and mean molecular weight. Our best-fitting solutions yield $\log \chi_{CH_4} = -2.07$, $\log \chi_{CO_2} = -2.44$, and $\mu = 2.58$ Da for the super-solar haze case, and $\log \chi_{CH_4} = -1.68$, $\log \chi_{CO_2} = -2.05$, and $\mu = 2.97$ Da for the Titan-like haze case. Across all hazy and haze-free retrievals, we derive a consistent upper limit of $\log \chi_{C_2H_4} < -4.3$. In agreement with previous studies, H$_2$O, NH$_3$, CO, OCS, and H$_2$S remain unconstrained.

Our analysis further highlights the strong degeneracy between planetary mass and atmospheric scale height in transmission spectroscopy retrievals. Although this effect has been explored in previous sensitivity studies (e.g. \citealt{changeat2022}), it remains less frequently discussed than other retrieval degeneracies, despite its potentially significant impact on atmospheric interpretations. We demonstrate that this is particularly important for K2-18b, where both the independently measured planetary mass and the retrieved masses carry substantial uncertainties. We find that the planetary mass, $M_p$, is poorly constrained by the data, with uncertainties reaching up to $\sim 71\%$. We identify strong degeneracies between $M_p$, atmospheric temperature, mean molecular weight, and haze opacity, which permit a wide range of atmospheric solutions with similar likelihoods. Retrievals with fixed $M_p$ demonstrate that assumed masses spanning the uncertainty range of our density-based estimate ($M_p = 10.7 \pm 4.6 M_{\oplus}$) can significantly bias the inferred atmospheric properties. In particular, higher assumed masses favour warmer and lower mean molecular weight atmospheres in order to maintain a consistent atmospheric scale height. In the most extreme case, the median retrieved temperature varies by up to $\sim 166$ K, while CO$_2$ and CH$_4$ abundances can differ by up to $\sim 1$ dex between retrievals. 

From a physical perspective, hazes are supported by disequilibrium photochemistry models \citep[e.g.][]{lavvas2026}, where CH$_4$-driven hydrocarbon chemistry can produce high-altitude hazes which can significantly alter the transmission spectrum. However, the current observations do not allow us to distinguish between different haze compositions or definitively confirm their presence, and multiple physically distinct atmospheric scenarios remain viable. More broadly, this study highlights the challenges of interpreting transmission spectra in the presence of strong parameter degeneracies and limited signal-to-noise. Progress will require improved stellar characterisation to obtain more precise independent mass measurements, a better understanding of instrumental systematics, and continued efforts in self-consistent modelling studies to address the apparent lack of H$_2$O and NH$_3$ in the observed spectra. Inter-comparison studies between such models will also be valuable for assessing systematic differences in atmospheric interpretations. Future JWST observations, particularly at mid-infrared wavelengths, together with laboratory measurements of the optical properties of a wider range of haze compositions, will be essential for advancing our understanding of K2-18b and other similar sub-Neptune atmospheres.

\section*{Acknowledgements}
The authors would like to thank Ahmed Al-Refaie, Kai H. Yip, Sergey Yurchenko, Jonathan Tennyson, and Charles Bowesman for helpful discussions. R. L. acknowledges support from NERC through the London NERC Doctoral Training Partnership (grant NE/ S007229/1). G. T. acknowledges support from UK Space Agency (grant ST/X002616/1). J. M. acknowledges support from the Italian Ministero dell'Università e della Ricerca and the European Union - Next Generation EU through project PRIN 2022 PM4JLH ``Know your little neighbours: characterising low-mass stars and planets in the Solar neighbourhood''.
\vspace{0.7mm}
\newline

\noindent \textit{Software:} \texttt{Numpy} \citep{harris2020}, \texttt{scipy} \citep{virtanen2020}, \texttt{TauREx 3.1} \citep{al-refaie2021}, \texttt{YunMa} \citep{ma2023}, \texttt{jwst} \citep{bushouse2020}, \texttt{PyLightcurve} \citep{tsiaras2016-pylightcurve}, \texttt{PyMultinest} \citep{buchner2014}, \texttt{astropy}  \citep{astropy:2013,astropy:2018,astropy:2022}, \texttt{numba} \citep{lam2015numba}.
\vspace{0.7mm}
\newline

\noindent \textit{Facilities:} JWST (NIRISS, NIRSpec, MIRI)
\vspace{0.7mm}
\newline

\noindent \textit{Data:} This work is based on observations made with the NASA/ESA/CSA James Webb Space Telescope. The data were obtained from the Mikulski Archive for Space Telescopes at the Space Telescope Science Institute, which is operated by the Association of Universities for Research in Astronomy, Inc., under NASA contract NAS 5-03127 for JWST. The MIRI LRS observations presented in this work are associated with Cycle 1 GO Program 2722. We are thankful to those who operate this archive, the public nature of which increases scientific productivity and accessibility. Additionally, we express gratitude to the Principal Investigators (PIs) of the observations used in study, as well as their respective teams, for requesting the observations that have and will continue to expand the characterisation of exoplanetary atmospheres. The observations analysed in this study can be accessed via\dataset[10.17909/y7qm-z707]{https://doi.org/10.17909/y7qm-z707}.
\vspace{0.7mm}
\newline

\bibliographystyle{aasjournal}
\bibliography{refs_K2-18}

\appendix

\section{Sensitivity Tests on the MIRI data}
\label{appendix: data reduction}

\subsection{Removing Cosmic Rays with Jump Detection}
\label{appendix: including jump step}
Here, we perform an additional data reduction method that replaces the \texttt{OutlierDetectionStep} with the \texttt{jump} step in Stage 1 of the JWST Science Calibration Pipeline, described in Section \ref{sec:jwst-pipeline}. The Stage 1 steps are performed in the following order: data quality initialization (\texttt{dq\_init}), electromagnetic interference correction (\texttt{emicorr}), saturation flagging (\texttt{saturation}), last frame flagging (\texttt{lastframe}), reset correction (\texttt{reset}), linearity correction (\texttt{linearity}), dark current subtraction (\texttt{dark\_current}), a custom jump detection step, and ramp fitting (\texttt{ramp\_fitting}). 

We use the jump detection step included in the \texttt{transitspectroscopy} package \citep{espinoza_nestor2022}, which has the same functionality as the \texttt{jump} step in the JWST \texttt{Detector1Pipeline}, but adapted specifically for Time-Series Observations (TSOs). We set a sigma threshold of 10-$\sigma$ for outlier detection and a window of 150 integrations for the median filter (calculated from window/nints $\sim$ 1/30, which works well for transits). The main advantage of using the jump detection step is that it takes advantage of JWST's up-the-ramp sampling by detecting differences ('jumps') between adjacent groups in an integration to find outliers --- typically pixels hit by cosmic rays --- in each frame. This is done before ramp fitting is applied, where the ramps are converted into slopes for each integration in the exposure. 

In Stage 2, we assign the world coordinate system (\texttt{assign\_wcs}), set the source type to "POINT" (\texttt{srctype}), apply the flat fielding correction to the data products, and lastly, the \texttt{PixelReplaceStep} to interpolate the flux values of the "NaN" pixels using adjacent profile approximation. The Stage 3 and light curve fitting methods follow the same procedure outlined in Section \ref{sec:jwst-pipeline} and \ref{sec:pylightcurve}.

\begin{figure}[!hb]
    \centering
    \includegraphics[width=1\linewidth]{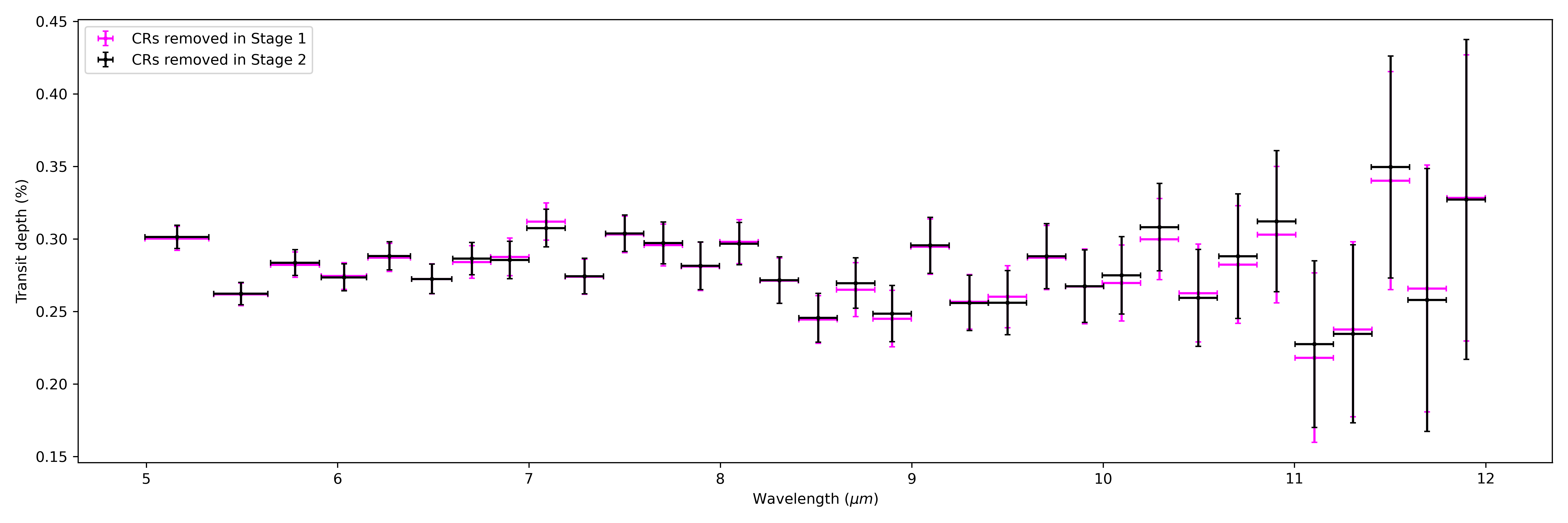}
    \caption{Comparison between the MIRI LRS spectra extracted from two data-reduction methods that involve either: the \texttt{OutlierDetectionStep} in Stage 2 (black); or the jump detection step in Stage 1 (magenta), to remove pixels affected by cosmic rays.}
    \label{fig:jump step vs outlier detection}
\end{figure}

Figure \ref{fig:jump step vs outlier detection} shows the resulting transmission spectrum from this data reduction, compared with the reduction described in the main text which uses the \texttt{OutlierDetectionStep} at Stage 2 instead of the \texttt{jump} step at Stage 1. The two spectra are in close agreement, with small variations of $\sim$15ppm on average for data points $<10$ $\mu$m. We conclude that, for the two methods explored, the resulting MIRI spectra is not sensitive to whether cosmic rays are removed in Stage 1 or Stage 2. We therefore adopt the spectra extracted using the \texttt{OutlierDetectionStep} as our nominal dataset.

\subsection{Number of Integrations Masked At The Start}
\label{appendix: miri: no. of integrations masked}
We investigate the sensitivity of the extracted transit depths to the number of integrations masked at the beginning during the light curve fitting stage in Section \ref{sec:pylightcurve}. Time-series observations taken with MIRI LRS are affected by the settling of the detector, which results in a prominent exponential-like ramp effect seen in the light curves at the beginning of the observation \citep{bouwman2023,greene2023}. The strength and duration of these ramps are not well understood, but observations have shown that they can vary as a function of visit \citep{august2025}, stellar brightness \citep{connors2025}, and the previous filter used by MIRI \citep{fortune2025}.

\begin{figure}[!tb]
    \centering
    \includegraphics[width=1\linewidth]{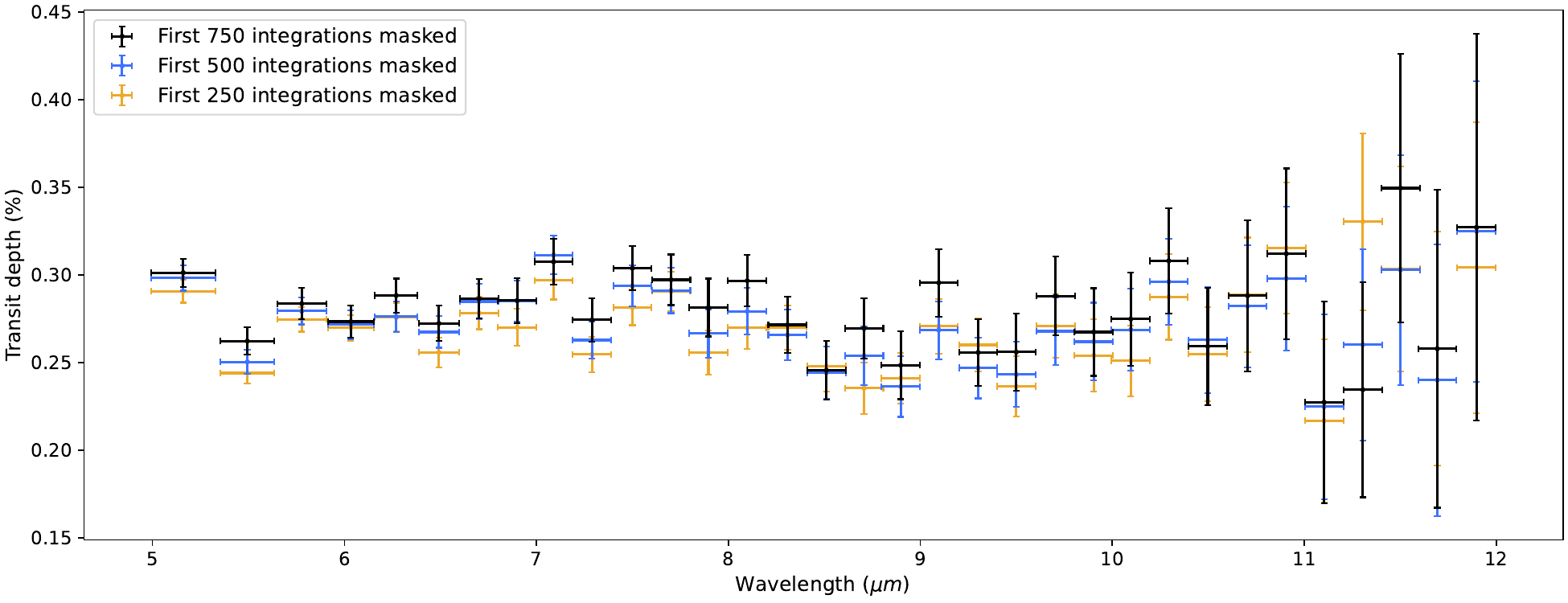}
    \caption{Comparison of MIRI LRS spectra from light curves extracted with either 250 (red), 500 (green), or 750 (black) integrations masked at the beginning. 250 integrations corresponds to 17 minutes.}
    \label{fig:no. of integrations masked}
\end{figure}

Using a quadratic detrending function, we obtain spectra (Figure \ref{fig:no. of integrations masked}) extracted from light curves with either 250, 500, or 750 integrations masked, where 750 integrations removes the ramp entirely. We find that the number of initial masked integrations can affect the derived transit depth, which appears to be more sensitive at certain wavelengths. Variations averaging $\sim$92 ppm and $\sim$143 ppm are seen when comparing the 750-integrations-masked spectrum to the 500- and 250-integrations-masked spectra, respectively.

We choose the model with 750 initial integrations masked as our nominal spectrum, as this does not call for fitting functions more complex than a quadratic, which otherwise would inadequately fit the lightcurves with 250 and 500 initial masked integrations as remnants of an exponential ramp are still evident in their light curves.

\subsection{Spectral Light Curve Binning}
\label{appendix: miri: spectral binning}

\begin{figure}[!hb]
    \centering
    \includegraphics[width=1\linewidth]{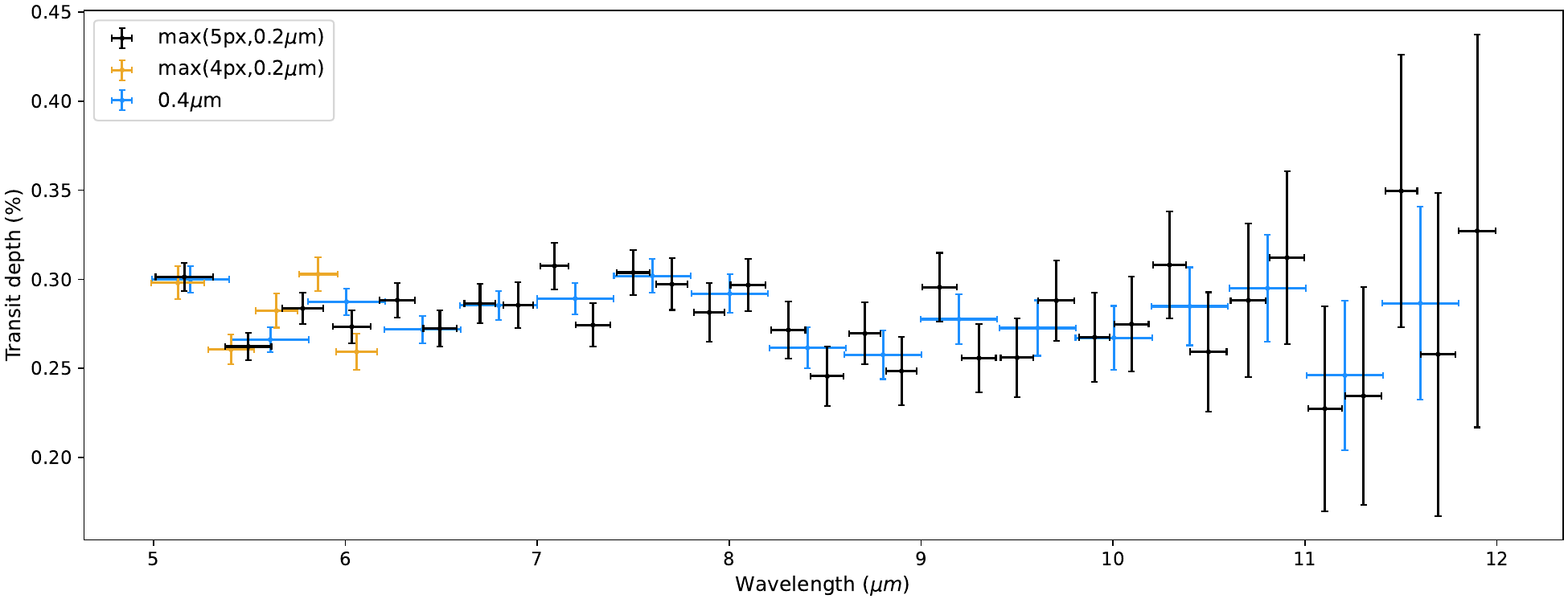}
    \caption{Comparison of MIRI LRS spectra from light curves extracted using the following binning widths: maximum of 0.2 $\mu$m or 4 pixels (magenta); maximum of 0.2 $\mu$m or 5 pixels (black); and 0.4 $\mu$m (blue).}
    \label{fig:light curve binning widths}
\end{figure}

Here, we explore spectral binning with different widths following the procedure outlined in \citetalias{madhusudhan2025} for the widths: maximum of 0.2 $\mu$m or 4 pixels, maximum of 0.2 $\mu$m or 5 pixels, and 0.4 $\mu$m. Figure \ref{fig:light curve binning widths} shows that the resulting spectra are consistent between different bin widths, including the region below 5.6 $\mu$m for which \citetalias{madhusudhan2025} discarded in their analyses due to inconsistencies with their bin widths. We do not discard these data points in the process, as we find that the transit depths are consistent in the region where the NIRSpec observations briefly overlap with MIRI, at around $\sim$5.16 $\mu$m (see Figure \ref{fig:nir+miri spectra comparison}).  

\begin{sidewaystable}
\section{Atmospheric Retrieval Results}
\label{appendix: atmospheric retrieval}

    \centering
    
    \caption{Summary of the Maximum A Posteriori (MAP) values and median values with $\pm2\sigma$ uncertainties, or 95th percentile upper limits, from the atmospheric retrievals (reduced molecule setup)}
    \label{tab:retrieval results (reduced setup)}
    \setlength{\tabcolsep}{3pt}
    \resizebox{1\textwidth}{!}{%
    \hspace*{-3.8cm}
    \begin{tabular}{l|cccccccccccc|cc} \hline \hline
          &  $M_p$&  $R_p$&  $T$&  $\log \chi_{\text{CO}_2}$&  $\log \chi_{\text{CH}_\text{4}}$& $\log \chi_{\text{C}_2\text{H}_4}$&$\log \text{X}_\text{haze}$ & $\log P_\text{deck}$& $\log P_\text{base}$& $\delta_\text{NRS1}$& $\delta_\text{NRS2}$&$\delta_\text{MIRI}$ &$\mu$   &Weight\\
          Model Solutions&  $[M_{\oplus}]$&  $[R_{\oplus}]$&  $[\text{K}]$&  &  & & & $[\text{bar}]$& $[\text{bar}]$& $[\text{ppm}]$& $[\text{ppm}]$&$[\text{ppm}]$ & $[\text{Dalton}]$  &\\
          \hline
 \multicolumn{14}{c}{Haze-free} &\\
         \hline
          \multirow{2}{*}{Median $\pm2\sigma$}&  \multirow{2}{*}{$9.53^{+6.00}_{-4.14}$}&  \multirow{2}{*}{$2.62^{+0.01}_{-0.01}$}&  \multirow{2}{*}{$164.41^{+63.54}_{-53.40}$}&  \multirow{2}{*}{$-2.63^{+1.19}_{-4.98}$}&  \multirow{2}{*}{$-1.19^{+0.57}_{-1.09}$}& \multirow{2}{*}{$<-4.33$}&\multirow{2}{*}{...}& \multirow{2}{*}{...}& \multirow{2}{*}{...}& \multirow{2}{*}{$-45.04^{+15.49}_{-15.08}$}& \multirow{2}{*}{$-28.92^{+19.64}_{-19.71}$}&\multirow{2}{*}{$91.63^{+41.04}_{-39.25}$}& \multirow{2}{*}{$3.52^{+2.95}_{-1.11}$} &...\\
 & & & & & & & & & & & & &  &\\
 MAP$_1$& $ 5.76$& $ 2.61$& $ 144.87$& $-1.73$& $-0.92$& $-4.60$& ...& ...& ...& $-53.54$& $ -30.53$& $85.89$&$4.72$ &$4.45\times 10^{-4}$\\
 MAP$_2$& $5.61$& $2.61$& $122.26$& $-2.05$& $-0.90$& $-4.93$& ...& ...& ...& $-45.62$& $-28.55$& $106.00$& $4.39$&$3.73\times 10^{-4}$\\
 MAP$_3$& $6.25$& $2.61$& $142.74$& $-2.30$& $-0.85$& $-4.59$& ...& ...& ...& $-52.12$& $-32.13$& $73.64$& $4.46$&$3.68\times 10^{-4}$\\
 MAP$_4$& $5.32$& $ 2.61$& $164.73$& $-2.47$& $-0.54$& $-4.10$& ...& ...& ...& $-45.67$& $-29.08$& $86.82$& $6.38$&$3.37\times 10^{-4}$\\
 MAP$_5$& $5.75$& $2.61$& $121.85$& $-2.51$& $-0.89$& $-4.94$& ...& ...& ...& $-49.91$& $-38.28$& $87.99$& $4.22$&$3.29\times 10^{-4}$\\
 MAP$_6$& $7.32$& $2.61$& $151.59$& $-2.13$& $-0.94$& $-5.16$& ...& ...& ...& $-51.78$& $-30.10$& $80.96$& $4.18$&$3.23\times 10^{-4}$\\
 MAP$_7$& $6.01$& $ 2.61$& $176.43$& $-2.01$& $-0.69$& $-4.38$& ...& ...& ...& $-43.63$& $-25.32$& $104.50$& $5.55$&$3.21\times 10^{-4}$\\
 MAP$_8$& $7.06$& $2.61$& $150.79$& $-2.22$& $-1.01$& $-4.50$& ...& ...& ...& $-52.14$& $-31.63$& $91.50$& $3.90$&$3.14\times 10^{-4}$\\
 MAP$_9$& $6.29$& $2.61$& $143.89$& $-2.03$& $-0.88$& $-4.70$& ...& ...& ...& $-54.26$& $-28.48$& $107.69$& $4.51$&$3.01\times 10^{-4}$\\
 MAP$_{10}$& $5.33$& $2.61$& $156.99$& $-1.53$& $-0.81$& $-6.10$& ...& ...& ...& $-51.60$& $-36.14$& $83.03$& $5.65$&$2.86\times 10^{-4}$\\
  \hline
 \multicolumn{14}{c}{With super-solar haze} &\\
  \hline
 \multirow{2}{*}{Median $\pm2\sigma$}& \multirow{2}{*}{$9.34^{+5.99}_{-5.19}$}& \multirow{2}{*}{$2.61^{+0.02}_{-0.07}$}& \multirow{2}{*}{$172.29^{+116.91}_{-82.20}$}& \multirow{2}{*}{$-2.93^{+1.45}_{-3.45}$}& \multirow{2}{*}{$-1.67^{+1.04}_{-1.47}$}& \multirow{2}{*}{$<-4.23$}& \multirow{2}{*}{$<1.32$}& \multirow{2}{*}{$-6.90^{+5.09}_{-1.88}$}& \multirow{2}{*}{$-4.81^{+4.50}_{-3.37}$}& \multirow{2}{*}{$-53.13^{+19.19}_{-21.76}$}& \multirow{2}{*}{$-40.03^{+26.00}_{-34.27}$}& \multirow{2}{*}{$71.48^{+51.51}_{-61.64}$}&\multirow{2}{*}{$2.77^{+3.73}_{-0.44}$} &...\\
 & & & & & & & & & & & & & &\\
          MAP$_1$&  $8.20$&  $2.52$&  $212.20$&  $-2.44$&  $-2.07$& $-4.74$&$0.74$& $-7.69$& $-6.44$& $-71.24$& $-73.32$&$ 19.07$& $ 2.58$&$4.64\times 10^{-4}$\\
 MAP$_2$& $5.04$& $2.53$& $149.43$& $-2.28$& $-1.22$& $-4.17$& $ 0.57$& $-7.56$& $-6.19$& $-72.55$& $-66.99$& $15.25$&  $3.36$&$4.16\times 10^{-4}$\\
 MAP$_3$& $7.39$& $2.54$& $ 172.72$& $-2.37$& $-1.83$& $-4.5$&$0.42$& $ -8.82$& $-7.02$& $-75.35$& $-66.68$&$10.7$&  $ 2.69$&$3.34\times 10^{-4}$\\
 MAP$_4$& $7.39$& $2.54$& $205.17
 $& $-1.72$& $-1.93$& $-4.29$& $0.60$& $-8.58$& $-7.40$& $-62.77$& $-51.04$& $31.28$& $3.27$&$3.31\times 10^{-4}$\\
 MAP$_5$& $5.15$& $2.53$& $145.20$& $-2.74$& $-1.55$& $-4.34$& $0.66$& $-5.94$& $-4.79$& $-66.11$& $-62.38$& $37.79$& $2.77$&$3.19\times 10^{-4}$\\
 MAP$_6$& $6.04$& $2.55$& $141.51$& $-2.11$& $-1.78$& $-4.86$& $0.70$& $ -7.79$& $-6.73$& $-78.88$& $-65.76$& $3.14$& $2.86$&$3.13\times 10^{-4}
$\\
 MAP$_7$& $6.97$& $2.54$& $192.49$& $-1.79$& $-1.76$& $-5.15$& $0.89$& $-8.05$& $-7.29$& $-67.63$& $-69.18$& $ 12.02$& $3.22$&$2.89\times 10^{-4}$\\
 MAP$_8$& $  6.89$& $2.54$& $161.33$& $-3.15$& $-2.21$& $-5.07$& $0.12$& $-8.12$& $-4.33$& $-70.94
 $& $-70.49$& $4.52$& $2.42$&$2.83\times 10^{-4}
$\\
 MAP$_9$& $8.19$& $2.51$& $205.10$& $-3.12$& $-2.54$& $-4.79$& $0.56$& $-7.50$& $-5.52$& $-73.75$& $-79.13$& $-0.80$& $2.38$&$2.73\times 10^{-4}$\\
 MAP$_{10}$& $7.76$& $2.53$& $230.38$& $-1.93$& $-1.58$& $-4.29$& $0.07$& $-8.43$& $-3.97$& $-73.27$& $-62.43$& $15.08
 $& $3.16$&$2.67\times 10^{-4}$\\
 \hline
 \multicolumn{14}{c}{With Titan-like haze} &\\
 \hline
 \multirow{2}{*}{Median $\pm2\sigma$}& \multirow{2}{*}{$9.01^{+6.44}_{-4.98}$}& \multirow{2}{*}{$2.61^{+0.01}_{-0.05}$}& \multirow{2}{*}{$165.11^{+103.24}_{-79.35}$}& \multirow{2}{*}{$-2.67^{+1.29}_{-3.83}$}& \multirow{2}{*}{$-1.40^{+0.81}_{-1.48}$}& \multirow{2}{*}{$<-4.28$}& \multirow{2}{*}{$<1.65$}& \multirow{2}{*}{$-5.99^{+4.40}_{-2.70}$}& \multirow{2}{*}{$-4.75^{+4.68}_{-3.55}$}& \multirow{2}{*}{$-49.68^{+17.49}_{-20.75}$}& \multirow{2}{*}{$-34.64^{+21.88}_{-27.99}$}& \multirow{2}{*}{$79.90^{+44.42}_{-52.79}$}& \multirow{2}{*}{$3.13^{+3.78}_{-0.79}$} &...\\
 & & & & & & & & & & & & & &\\
 MAP$_1$& $ 8.74$& $ 2.57$& $ 179.36$& $ -2.05$& $ -1.68$& $ -4.35$& $ 0.12$& $ -8.66$& $ -5.91$& $ -63.81$& $ -56.38$& $ 61.62$& $ 2.97$&$4.69\times 10^{-4}$\\
 MAP$_2$& $8.26$& $2.55$& $183.09$& $-2.23$& $-2.33$& $-5.03$& $0.22$& $-8.36$& $-5.57$& $-68.99$& $-61.40$& $14.69$&  $2.61$&$3.37\times 10^{-4}$\\
 MAP$_3$& $9.34$& $2.57$& $186.11$& $-2.53$& $-1.48$& $-4.46$& $0.79$& $-7.43$& $-6.90$& $-59.24$& $-50.52 $& $41.00$&  $2.89$&$3.31\times 10^{-4}$\\
 MAP$_4$& $10.23$& $2.58$& $216.73$& $-1.79$& $-1.51$& $-4.67$& $0.15$& $-8.92$& $-6.54$& $-55.56$& $-45.67$& $ 68.68$&  $3.41$&$2.85\times 10^{-4}
$\\
 MAP$_5$& $7.98$& $2.55$& $179.88$& $-2.21$& $-2.03$& $-5.80$& $0.14$& $-8.54$& $-5.19$& $-74.47$& $-69.17$& $31.66$& $ 2.69$&$2.55\times 10^{-4}
$\\
 MAP$_6$& $7.75$& $2.56$& $167.79$& $-2.31$& $-1.67$& $-4.92$& $0.98$& $-7.21$& $-6.77$& $-71.29$& $-58.67$& $70.02$& $2.81$&$2.49\times 10^{-4}
$\\
 MAP$_7$& $7.90$& $2.57$& $153.07$& $-3.52$& $-1.89$& $-4.39$& $0.15$& $-7.35$& $-4.77$& $-63.8$& $-50.39$& $55.32$& $2.50$&$2.42\times 10^{-4}$\\
 MAP$_8$& $5.17$& $2.57$& $170.94$& $-1.34$& $-1.68$& $-5.98$& $0.37$& $-7.34$& $-6.19$& $-65.44$& $-52.77$& $57.80$& $4.49$&$2.37\times 10^{-4}$\\
 MAP$_9$& $6.29$& $2.56$& $141.25$& $-2.61$& $-1.71$& $-4.49$& $0.26$& $-6.90$& $-4.50$& $-62.71$& $-56.71$& $41.67$& $2.68$&$2.36\times 10^{-4}$\\
  MAP$_{10}$& $6.95$& $2.56$& $141.12$& $-3.37$& $-2.08$& $-5.00$& $0.06$& $-7.24 $& $-4.06$& $-69.4$& $-52.76$& $63.59$&  $2.44$&$2.35\times 10^{-4}$\\
\hline \hline
\end{tabular}
    }
\end{sidewaystable}
\clearpage

\begin{table}
    \centering
    \caption{Log-evidences ($\ln(Z)$) for each retrieval presented in Section \ref{sec: retrievals of hazy atmospheres}}
    \label{tab:lnz results}
\begin{tabular}{lcc}
    \hline \hline
         Model&  $\ln(Z)$&  $\ln(B)$\\
         \hline 
         All molecules&  28947.22&  Reference\\
         CH$_4$+CO$_2$+C$_2$H$_4$&  28948.56&  1.34\\
         \hline
         CH$_4$+CO$_2$+C$_2$H$_4$&  28948.56&  Reference\\
 CH$_4$+CO$_2$+C$_2$H$_4$ + super-solar haze& 28950.40&1.84\\
 CH$_4$+CO$_2$+C$_2$H$_4$ + Titan-like haze& 28950.29&1.73\\
 CH$_4$+CO$_2$& 28948.54&-0.02\\
 \hline \hline
\end{tabular}
\end{table}

\begin{table}[h]
    \centering
    \caption{Log-evidences ($\ln(Z)$) for each fixed-$M_p$ retrieval presented in Section \ref{sec: retrievals with fixed Mp}, compared to the reference free-$M_p$ retrievals }
    \label{tab:lnz results fixed mp}
\begin{tabular}{lcc}
    \hline \hline
         Model&  $\ln(Z)$&  $\ln(B)$\\
         \hline
         CH$_4$+CO$_2$+C$_2$H$_4$&  28948.56&  Reference\\
 Fixed $M_p = 6.1 M_{\oplus}$& 28949.67&1.11\\
 Fixed $M_p = 8.4 M_{\oplus}$& 28949.18&0.62\\
 Fixed $M_p = 10.7 M_{\oplus}$& 28948.90&0.34\\
 Fixed $M_p = 13.0 M_{\oplus}$& 28948.62&0.05\\
 Fixed $M_p = 15.3 M_{\oplus}$& 28948.27&-0.29\\
 \hline
 CH$_4$+CO$_2$+C$_2$H$_4$ + super-solar haze& 28950.40&Reference\\
 Fixed $M_p = 6.1 M_{\oplus}$& 28950.49
&0.09
\\
 Fixed $M_p = 8.4 M_{\oplus}$& 28950.69
&0.29
\\
 Fixed $M_p = 10.7 M_{\oplus}$& 28950.33
&-0.07
\\
 Fixed $M_p = 13.0 M_{\oplus}$& 28950.29
&-0.11
\\
 Fixed $M_p = 15.3 M_{\oplus}$& 28949.85&-0.55\\
 \hline
 CH$_4$+CO$_2$+C$_2$H$_4$ + Titan-like haze& 28950.29&Reference\\
 Fixed $M_p = 6.1 M_{\oplus}$& 28949.99&-0.29\\
 Fixed $M_p = 8.4 M_{\oplus}$& 28950.17&-0.11\\
 Fixed $M_p = 10.7 M_{\oplus}$& 28949.90&-0.38\\
 Fixed $M_p = 13.0 M_{\oplus}$& 28949.58&-0.70\\
 Fixed $M_p = 15.3 M_{\oplus}$& 28949.46&-0.83\\
 \hline \hline
\end{tabular}
\end{table}
\clearpage

\begin{figure}
    \centering
    \includegraphics[width=1\linewidth]{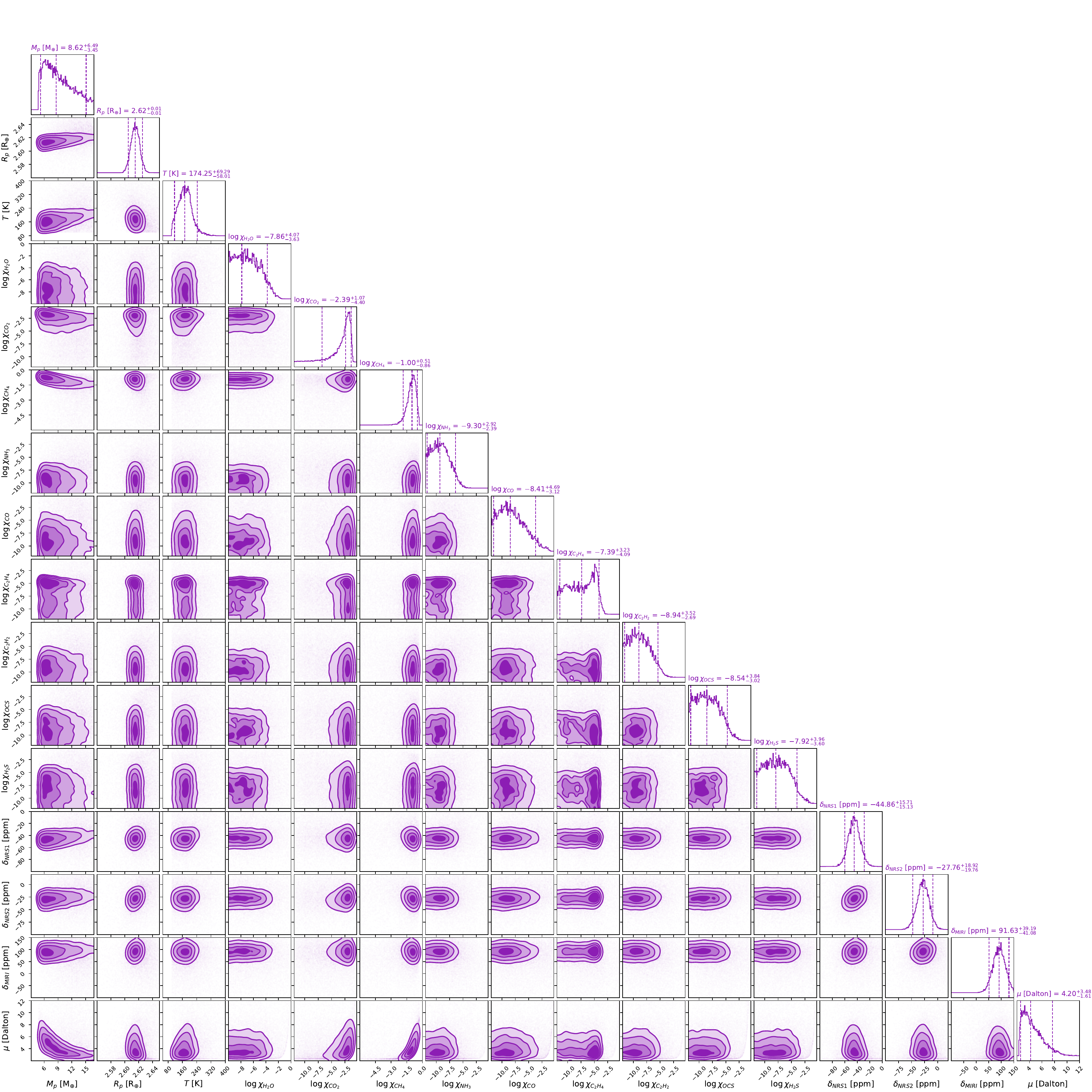}
    \caption{Posterior probability distribution for the haze-free retrieval, with all molecules included. The dashed lines represent the median, corresponding to the values shown above each column along with $\pm1\sigma$ errors.}
    \label{fig:all mols haze-free cornerplot}
\end{figure}

\begin{figure}
    \centering
    \includegraphics[width=1\linewidth]{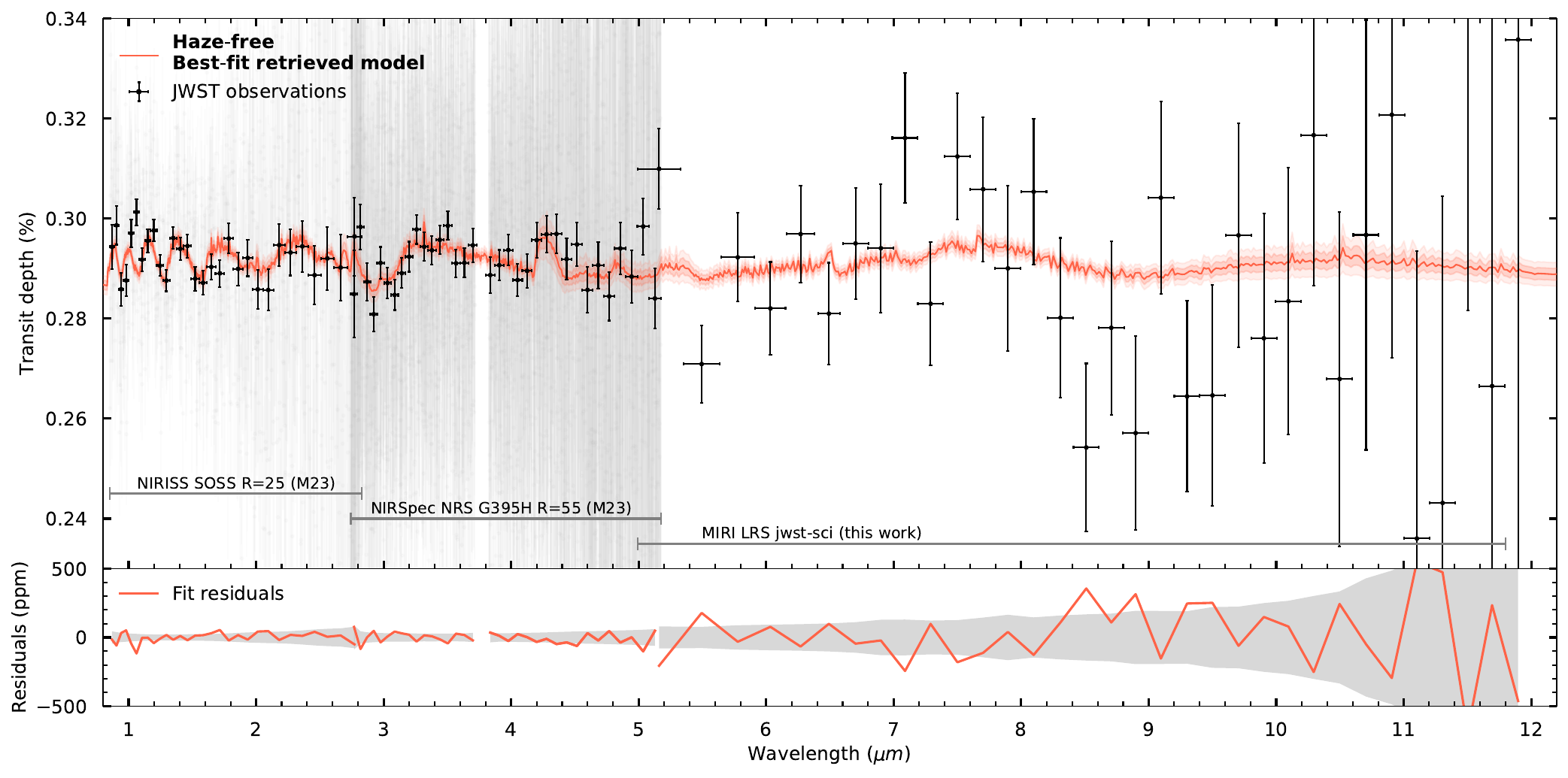}
    \caption{The best-fit (MAP) model of the haze-free retrieval performed on the combined JWST NIRISS SOSS (\citetalias{madhusudhan2023}), NIRSpec G395H (\citetalias{madhusudhan2023}) and MIRI LRS \texttt{JWST-sci} (Section \ref{sec:miri data reduction}) transmission spectrum. The MAP retrieved offsets have been applied to the NIRSpec NRS1, NRS2, and MIRI LRS datasets. All other elements are the same as Figure \ref{fig:Best-fit H24 Spectrum}.}
    \label{fig:Best-fit haze-free Spectrum}
\end{figure}

\begin{figure}
    \centering
    \includegraphics[width=1\linewidth]{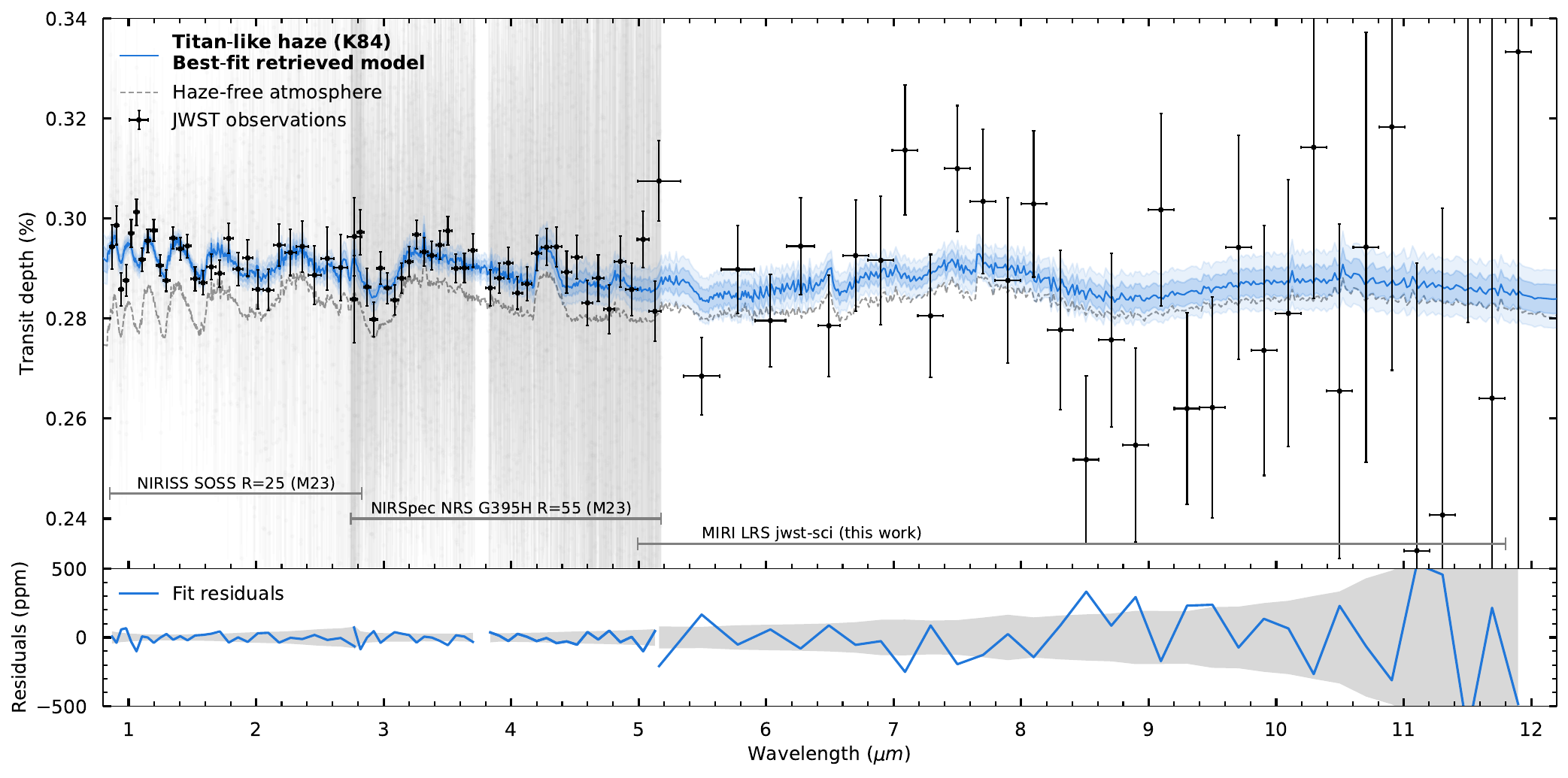}
    \caption{The best-fit (MAP) model of the Titan-like haze (\citetalias{khare1984}) retrieval performed on the combined JWST NIRISS SOSS (\citetalias{madhusudhan2023}), NIRSpec G395H  (\citetalias{madhusudhan2023}) and MIRI LRS \texttt{JWST-sci} (Section \ref{sec:miri data reduction}) transmission spectrum. The MAP retrieved offsets have been applied to the NIRSpec NRS1, NRS2, and MIRI LRS datasets. All other elements are the same as Figure \ref{fig:Best-fit H24 Spectrum}.}
    \label{fig:Best-fit K84 Spectrum}
\end{figure}

\begin{figure}
    \centering
    \includegraphics[width=0.9\linewidth]{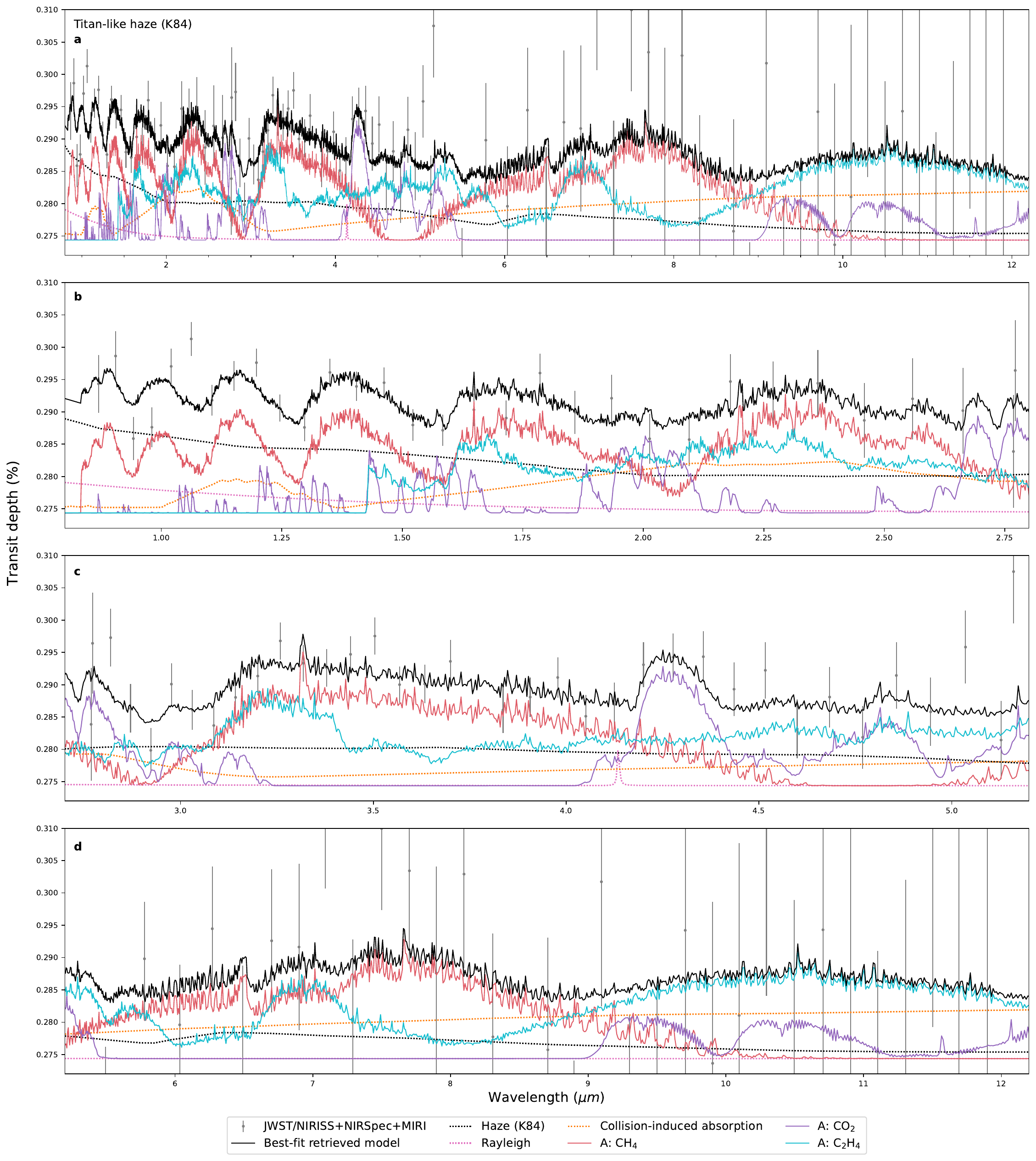}
    \caption{Opacity contributions of the best-fit model for the baseline Titan-like haze case shown in Figure \ref{fig:Best-fit K84 Spectrum} plotted as a solid black line (binned to R = 1000). All other elements are the same as Figure \ref{fig:Best-fit H24 contributions}}
    \label{fig:Best-fit K84 contributions}
\end{figure}

\begin{figure}
    \centering
    \includegraphics[width=1\linewidth]{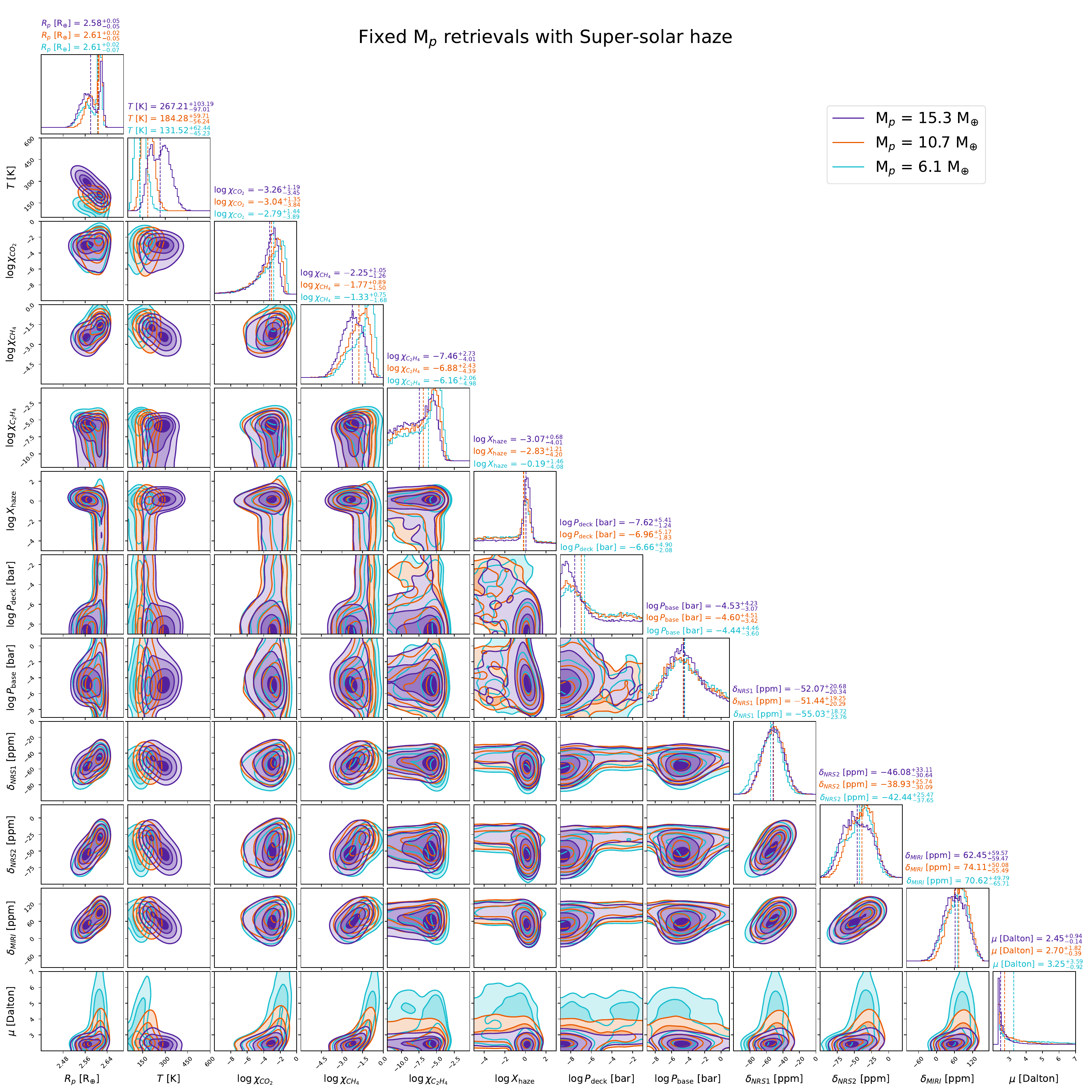}
    \caption{Comparison of posterior probability distributions for retrievals performed with super-solar haze and with fixed $M_p =6.1, 10.7,$ and $15.3M_{\oplus}$. The dashed lines represent the median, corresponding to the values shown above each column along with $\pm2\sigma$ confidence intervals.}
    \label{fig:h24_fixed_mp_cornerplots}
\end{figure}

\begin{figure}
    \centering
    \includegraphics[width=1\linewidth]{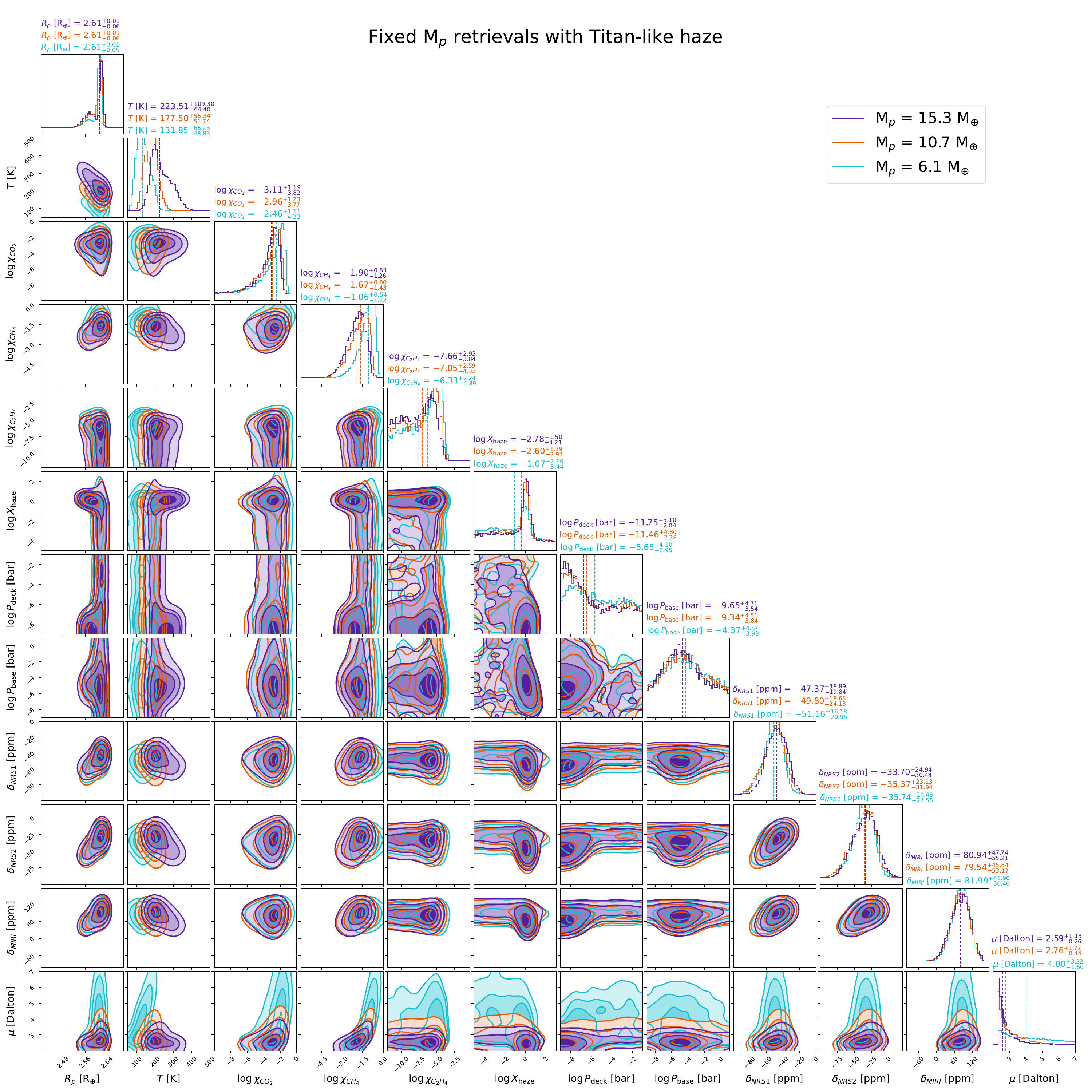}
    \caption{Comparison of posterior probability distributions for retrievals performed with Titan-like haze and with fixed $M_p =6.1, 10.7,$ and $15.3M_{\oplus}$. The dashed lines represent the median, corresponding to the values shown above each column along with $\pm2\sigma$ confidence intervals.}
    \label{fig:k84_fixed_mp_cornerplots}
\end{figure}

\begin{sidewaystable}
    \centering
    \caption{Summary of the Maximum A Posteriori (MAP) values and median values with $\pm2\sigma$ uncertainties, or 95th percentile upper limits, from the haze-free atmospheric retrievals with fixed planetary masses}
    \label{tab:retrieval results hazefree fixed mp}
    \setlength{\tabcolsep}{3pt}
    \resizebox{0.85\textwidth}{!}{%
    \hspace*{-2cm}
    \begin{tabular}{l|cccccccc|cc} \hline \hline
           &  $R_p$&  $T$&  $\log \chi_{\text{CO}_2}$&  $\log \chi_{\text{CH}_\text{4}}$& $\log \chi_{\text{C}_2\text{H}_4}$& $\delta_\text{NRS1}$& $\delta_\text{NRS2}$&$\delta_\text{MIRI}$ &$\mu$   &Weight\\
           Model Solutions&  $[R_{\oplus}]$&  $[\text{K}]$&  &  & & $[\text{ppm}]$& $[\text{ppm}]$&$[\text{ppm}]$ & $[\text{Dalton}]$  &\\
          \hline
  \multicolumn{11}{c}{$M_p = 6.1 M_{\oplus}$}\\
         \hline
\multirow{2}{*}{Median $\pm2\sigma$}
& \multirow{2}{*}{$2.61^{+0.01}_{-0.01}$}
& \multirow{2}{*}{$139.52^{+67.71}_{-57.86}$}
& \multirow{2}{*}{$-2.17^{+1.01}_{-4.97}$}
& \multirow{2}{*}{$-0.94^{+0.48}_{-0.94}$}
& \multirow{2}{*}{$<-4.20$}& \multirow{2}{*}{$-47.58^{+15.74}_{-14.96}$}
& \multirow{2}{*}{$-30.66^{+19.69}_{-20.19}$}
& \multirow{2}{*}{$88.44^{+39.58}_{-40.04}$}
& \multirow{2}{*}{$4.79^{+3.20}_{-2.22}$}
& ...\\
& & & & & & & & & &\\
MAP$_1$& 2.61 & 134.84 & -1.99 & -0.90 & -5.22 & -47.28 & -26.23 & 88.14 & 4.47 & 4.22$\times 10^{-4}$\\
MAP$_2$& 2.61 & 131.68 & -1.88 & -1.10 & -5.19 & -48.28 & -26.32 & 79.19 & 3.95 & 3.97$\times 10^{-4}$\\
MAP$_3$& 2.61 & 153.64 & -2.27 & -0.71 & -4.49 & -46.99 & -23.44 & 112.29 & 5.21 & 3.82$\times 10^{-4}$\\
MAP$_4$& 2.61 & 170.62 & -1.52 & -0.71 & -7.68 & -51.17 & -29.89 & 87.86 & 6.25 & 3.77$\times 10^{-4}$\\
MAP$_5$& 2.61 & 125.19 & -1.92 & -0.99 & -5.31 & -52.79 & -26.50 & 96.39 & 4.23 & 3.76$\times 10^{-4}$\\
  \hline
  \multicolumn{11}{c}{$M_p = 8.4 M_{\oplus}$}\\
  \hline
\multirow{2}{*}{Median $\pm2\sigma$}
& \multirow{2}{*}{$2.62^{+0.01}_{-0.01}$}
& \multirow{2}{*}{$149.50^{+65.78}_{-46.03}$}
& \multirow{2}{*}{$-2.67^{+1.11}_{-4.67}$}
& \multirow{2}{*}{$-1.12^{+0.51}_{-1.03}$}
& \multirow{2}{*}{$<-4.26$}& \multirow{2}{*}{$-46.27^{+15.58}_{-15.08}$}
& \multirow{2}{*}{$-29.89^{+19.30}_{-19.73}$}
& \multirow{2}{*}{$90.19^{+38.78}_{-41.53}$}
& \multirow{2}{*}{$3.69^{+2.38}_{-1.25}$}
&...\\
  & & & & & & & & & &\\
MAP$_1$& 2.61 & 200.37 & -1.68 & -0.98 & -5.07 & -49.66 & -24.33 & 84.34 & 4.60 & 4.36$\times 10^{-4}$\\
MAP$_2$& 2.61 & 187.87 & -1.72 & -1.07 & -4.90 & -44.04 & -27.32 & 74.78 & 4.27 & 4.03$\times 10^{-4}$\\
MAP$_3$& 2.61 & 174.88 & -2.52 & -0.96 & -4.48 & -45.49 & -22.24 & 92.57 & 3.94 & 3.96$\times 10^{-4}$\\
MAP$_4$& 2.61 & 161.56 & -2.32 & -0.94 & -5.09 & -50.56 & -34.70 & 87.05 & 4.08 & 3.85$\times 10^{-4}$\\
MAP$_5$& 2.61 & 172.15 & -1.76 & -0.99 & -5.31 & -45.88 & -33.08 & 89.00 & 4.44 & 3.50$\times 10^{-4}$\\
  \hline
  \multicolumn{11}{c}{$M_p = 10.7M_{\oplus}$}\\
  \hline
\multirow{2}{*}{Median $\pm2\sigma$}
& \multirow{2}{*}{$2.62^{+0.01}_{-0.01}$}
& \multirow{2}{*}{$167.00^{+58.52}_{-46.13}$}
& \multirow{2}{*}{$-2.78^{+1.15}_{-4.61}$}
& \multirow{2}{*}{$-1.31^{+0.53}_{-1.03}$}
& \multirow{2}{*}{$<-4.41$}& \multirow{2}{*}{$-44.53^{+15.30}_{-15.74}$}
& \multirow{2}{*}{$-28.30^{+19.15}_{-20.07}$}
& \multirow{2}{*}{$92.50^{+39.75}_{-39.21}$}
& \multirow{2}{*}{$3.24^{+1.73}_{-0.85}$}
&...\\
  & & & & & & & & & &\\
MAP$_1$& 2.61 & 187.40 & -2.29 & -1.27 & -4.71 & -40.10 & -26.80 & 92.02 & 3.25 & 4.35$\times 10^{-4}$\\
MAP$_2$& 2.62 & 183.94 & -2.05 & -1.08 & -5.22 & -44.49 & -27.29 & 77.58 & 3.82 & 4.00$\times 10^{-4}$\\
MAP$_3$& 2.61 & 172.47 & -2.41 & -1.44 & -5.22 & -41.08 & -26.88 & 96.97 & 2.98 & 3.92$\times 10^{-4}$\\
MAP$_4$& 2.61 & 174.40 & -2.75 & -1.21 & -5.17 & -44.67 & -26.16 & 80.43 & 3.24 & 3.83$\times 10^{-4}$\\
MAP$_5$& 2.61 & 195.35 & -2.01 & -1.13 & -5.10 & -52.82 & -25.95 & 89.07 & 3.74 & 3.45$\times 10^{-4}$\\
    \hline
 \multicolumn{11}{c}{$M_p = 13. 0M_{\oplus}$}\\
   \hline
\multirow{2}{*}{Median $\pm2\sigma$}
& \multirow{2}{*}{$2.62^{+0.01}_{-0.01}$}
& \multirow{2}{*}{$183.83^{+62.29}_{-46.64}$}
& \multirow{2}{*}{$-3.03^{+1.21}_{-5.00}$}
& \multirow{2}{*}{$-1.46^{+0.61}_{-1.06}$}
& \multirow{2}{*}{$<-4.47$}& \multirow{2}{*}{$-43.16^{+16.13}_{-15.87}$}
& \multirow{2}{*}{$-26.63^{+19.60}_{-20.20}$}
& \multirow{2}{*}{$-94.13^{+38.83}_{-41.04}$}
& \multirow{2}{*}{$2.95^{+1.53}_{-0.58}$}
&...\\
  & & & & & & & & & &\\
MAP$_1$& 2.62 & 190.34 & -2.48 & -1.21 & -5.13 & -46.02 & -26.66 & 102.00 & 3.30 & 4.40$\times 10^{-4}$\\
MAP$_2$& 2.61 & 202.47 & -2.62 & -1.34 & -5.26 & -45.41 & -30.45 & 99.23 & 3.03 & 4.06$\times 10^{-4}$\\
MAP$_3$& 2.62 & 192.63 & -2.85 & -1.24 & -5.17 & -44.13 & -21.19 & 94.54 & 3.15 & 3.70$\times 10^{-4}$\\
MAP$_4$& 2.62 & 194.24 & -1.97 & -1.37 & -5.06 & -49.98 & -28.02 & 84.56 & 3.34 & 3.34$\times 10^{-4}$\\
MAP$_5$& 2.61 & 204.44 & -2.09 & -1.50 & -5.06 & -43.85 & -22.92 & 87.10 & 3.08 & 3.28$\times 10^{-4}$\\
  \hline
  \multicolumn{11}{c}{$M_p = 15.3M_{\oplus}$}\\
  \hline
\multirow{2}{*}{Median $\pm2\sigma$}
& \multirow{2}{*}{$2.62^{+0.01}_{-0.01}$}
& \multirow{2}{*}{$198.96^{+72.74}_{-45.78}$}
& \multirow{2}{*}{$-3.16^{+1.31}_{-5.02}$}
& \multirow{2}{*}{$-1.57^{+0.62}_{-1.04}$}
& \multirow{2}{*}{$<-4.65$}& \multirow{2}{*}{$-41.22^{+15.87}_{-15.84}$}
& \multirow{2}{*}{$-25.27^{+19.82}_{-20.11}$}
& \multirow{2}{*}{$97.54^{+39.32}_{-43.62}$}
& \multirow{2}{*}{$2.80^{+1.33}_{-0.45}$}
&...\\
  & & & & & & & & & &\\
MAP$_1$& 2.62 & 193.14 & -2.76 & -1.57 & -5.77 & -42.93 & -26.56 & 105.29 & 2.75 & 4.32$\times 10^{-4}$\\
MAP$_2$& 2.62 & 212.68 & -2.53 & -1.30 & -4.85 & -42.63 & -26.77 & 99.84 & 3.12 & 4.29$\times 10^{-4}$\\
MAP$_3$& 2.62 & 198.63 & -3.71 & -1.72 & -5.21 & -41.64 & -28.79 & 87.54 & 2.58 & 3.85$\times 10^{-4}$\\
MAP$_4$& 2.62 & 207.15 & -2.19 & -1.35 & -9.76 & -38.83 & -28.22 & 91.96 & 3.19 & 3.69$\times 10^{-4}$\\
MAP$_5$& 2.62 & 199.59 & -3.18 & -1.74 & -5.16 & -51.07 & -24.64 & 104.50 & 2.59 & 3.55$\times 10^{-4}$\\
 \hline
 \hline
\end{tabular}
    }
\end{sidewaystable}
\clearpage

\begin{sidewaystable}
    \centering
    \caption{Summary of the Maximum A Posteriori (MAP) values and median values with $\pm2\sigma$ uncertainties, or 95th percentile upper limits, from the atmospheric retrievals with super-solar haze and fixed planetary masses}
    \label{tab:retrieval results h24 fixed mp}
    \setlength{\tabcolsep}{3pt}
    \resizebox{1.0\textwidth}{!}{%
    \hspace*{-3.5cm}
    \begin{tabular}{l|ccccccccccc|cc} \hline \hline
           &  $R_p$&  $T$&  $\log \chi_{\text{CO}_2}$&  $\log \chi_{\text{CH}_\text{4}}$& $\log \chi_{\text{C}_2\text{H}_4}$&$\log \text{X}_\text{haze}$ & $\log P_\text{deck}$& $\log P_\text{base}$& $\delta_\text{NRS1}$& $\delta_\text{NRS2}$&$\delta_\text{MIRI}$ &$\mu$   &Weight\\
           Model Solutions&  $[R_{\oplus}]$&  $[\text{K}]$&  &  & & & $[\text{bar}]$& $[\text{bar}]$& $[\text{ppm}]$& $[\text{ppm}]$&$[\text{ppm}]$ & $[\text{Dalton}]$  &\\
          \hline
  \multicolumn{14}{c}{$M_p = 6.1 M_{\oplus}$}\\
         \hline
\multirow{2}{*}{Median $\pm2\sigma$}
& \multirow{2}{*}{$2.61^{+0.02}_{-0.07}$}
& \multirow{2}{*}{$131.52^{+62.44}_{-45.23}$}
& \multirow{2}{*}{$-2.79^{+1.44}_{-3.89}$}
& \multirow{2}{*}{$-1.33^{+0.75}_{-1.68}$}
& \multirow{2}{*}{$<-4.10$}& \multirow{2}{*}{$-0.19^{+1.46}_{-4.08}$}
& \multirow{2}{*}{$-6.66^{+4.90}_{-2.08}$}
& \multirow{2}{*}{$-4.44^{+4.46}_{-3.60}$}
& \multirow{2}{*}{$-55.03^{+18.72}_{-23.76}$}
& \multirow{2}{*}{$-42.44^{+25.47}_{-37.65}$}
& \multirow{2}{*}{$70.62^{+49.79}_{-65.71}$}
& \multirow{2}{*}{$3.25^{+3.59}_{-0.92}$}
&...\\
  & & & & & & & & & & & & &\\
MAP$_1$& 2.54 & 145.07 & -2.14 & -1.74 & -4.29 & 0.31 & -8.86 & -6.02 & -80.18 & -75.31 & 14.58 & 2.86 & 4.63$\times 10^{-4}$\\
MAP$_2$& 2.53 & 157.01 & -2.80 & -1.66 & -4.35 & 1.03 & -7.46 & -7.01 & -71.60 & -67.09 & 28.29 & 2.67 & 3.51$\times 10^{-4}$\\
MAP$_3$& 2.53 & 155.60 & -2.59 & -1.96 & -4.62 & 0.21 & -7.97 & -4.89 & -73.54 & -71.30 & 17.78 & 2.56 & 3.26$\times 10^{-4}$\\
MAP$_4$& 2.52 & 164.17 & -2.24 & -1.65 & -4.03 & 0.66 & -7.33 & -5.73 & -71.73 & -69.37 & 0.00 & 2.86 & 3.20$\times 10^{-4}$\\
MAP$_5$& 2.53 & 146.12 & -2.33 & -1.94 & -4.45 & 0.13 & -8.95 & -4.47 & -76.03 & -75.76 & 30.01 & 2.66 & 3.07$\times 10^{-4}$\\
  \hline
  \multicolumn{14}{c}{$M_p = 8.4 M_{\oplus}$}\\
  \hline
\multirow{2}{*}{Median $\pm2\sigma$}
& \multirow{2}{*}{$2.58^{+0.04}_{-0.05}$}
& \multirow{2}{*}{$167.50^{+49.75}_{-54.95}$}
& \multirow{2}{*}{$-3.12^{+1.46}_{-3.54}$}
& \multirow{2}{*}{$-1.90^{+1.10}_{-1.46}$}
& \multirow{2}{*}{$<-4.23$}& \multirow{2}{*}{$-0.00^{+0.89}_{-4.02}$}
& \multirow{2}{*}{$-7.29^{+5.26}_{-1.52}$}
& \multirow{2}{*}{$-4.46^{+4.35}_{-3.35}$}
& \multirow{2}{*}{$-58.81^{+23.40}_{-21.65}$}
& \multirow{2}{*}{$-49.07^{+32.63}_{-30.80}$}
& \multirow{2}{*}{$58.27^{+60.44}_{-59.57}$}
& \multirow{2}{*}{$2.61^{+2.38}_{-0.30}$}
&...\\
  & & & & & & & & & & & & &\\
MAP$_1$& 2.56 & 175.24 & -2.33 & -1.72 & -4.90 & 0.38 & -8.59 & -6.72 & -73.77 & -61.54 & 7.22 & 2.77 & 4.46$\times 10^{-4}$\\
MAP$_2$& 2.57 & 156.98 & -2.32 & -1.76 & -4.46 & 0.44 & -8.71 & -7.27 & -67.03 & -49.22 & 37.13 & 2.74 & 4.26$\times 10^{-4}$\\
MAP$_3$& 2.56 & 187.31 & -1.87 & -2.19 & -5.15 & 0.14 & -8.90 & -5.34 & -71.72 & -53.44 & 30.52 & 2.95 & 4.08$\times 10^{-4}$\\
MAP$_4$& 2.55 & 181.68 & -3.07 & -1.99 & -4.61 & 0.03 & -8.17 & -4.35 & -61.14 & -57.67 & 33.43 & 2.48 & 4.05$\times 10^{-4}$\\
MAP$_5$& 2.55 & 190.05 & -3.15 & -2.52 & -4.91 & 0.40 & -6.87 & -4.77 & -63.12 & -60.25 & 36.40 & 2.38 & 3.91$\times 10^{-4}$\\
  \hline
  \multicolumn{14}{c}{$M_p = 10.7M_{\oplus}$}\\
  \hline
\multirow{2}{*}{Median $\pm2\sigma$}
& \multirow{2}{*}{$2.61^{+0.02}_{-0.05}$}
& \multirow{2}{*}{$184.28^{+59.71}_{-56.24}$}
& \multirow{2}{*}{$-3.04^{+1.35}_{-3.84}$}
& \multirow{2}{*}{$-1.77^{+0.89}_{-1.50}$}
& \multirow{2}{*}{$<-4.45$}& \multirow{2}{*}{$-0.17^{+1.21}_{-4.20}$}
& \multirow{2}{*}{$-6.96^{+5.17}_{-1.83}$}
& \multirow{2}{*}{$-4.60^{+4.51}_{-3.42}$}
& \multirow{2}{*}{$-51.44^{+19.25}_{-20.29}$}
& \multirow{2}{*}{$-38.93^{+25.74}_{-30.09}$}
& \multirow{2}{*}{$74.11^{+50.08}_{-55.49}$}
& \multirow{2}{*}{$2.70^{+1.82}_{-0.39}$}&...\\
  & & & & & & & & & & & & &\\
MAP$_1$& 2.54 & 270.33 & -2.13 & -2.20 & -4.75 & 0.05 & -8.62 & -3.94 & -60.04 & -56.23 & 33.83 & 2.70 & 4.59$\times 10^{-4}$\\
MAP$_2$& 2.53 & 253.75 & -2.63 & -2.21 & -5.40 & 0.54 & -8.03 & -6.44 & -70.21 & -68.79 & 9.39 & 2.49 & 3.97$\times 10^{-4}$\\
MAP$_3$& 2.54 & 282.10 & -2.04 & -2.09 & -4.45 & -0.03 & -8.96 & -3.93 & -66.18 & -57.69 & 44.79 & 2.80 & 3.90$\times 10^{-4}$\\
MAP$_4$& 2.54 & 230.15 & -2.69 & -1.85 & -4.90 & 0.49 & -8.73 & -6.95 & -62.91 & -66.21 & 10.12 & 2.59 & 3.77$\times 10^{-4}$\\
MAP$_5$& 2.57 & 186.93 & -3.15 & -1.97 & -4.42 & 0.20 & -8.98 & -6.23 & -61.80 & -48.22 & 40.07 & 2.48 & 3.38$\times 10^{-4}$\\
    \hline
 \multicolumn{14}{c}{$M_p = 13. 0M_{\oplus}$}\\
   \hline
\multirow{2}{*}{Median $\pm2\sigma$}
& \multirow{2}{*}{$2.58^{+0.04}_{-0.06}$}
& \multirow{2}{*}{$229.76^{+96.62}_{-77.53}$}
& \multirow{2}{*}{$-3.21^{+1.21}_{-3.62}$}
& \multirow{2}{*}{$-2.16^{+1.06}_{-1.36}$}
& \multirow{2}{*}{$<-4.59$}& \multirow{2}{*}{$-0.05^{+0.75}_{-4.14}$}
& \multirow{2}{*}{$-7.54^{+5.44}_{-1.31}$}
& \multirow{2}{*}{$-4.48^{+4.25}_{-3.26}$}
& \multirow{2}{*}{$-54.94^{+22.08}_{-21.66}$}
& \multirow{2}{*}{$-47.80^{+33.25}_{-32.57}$}
& \multirow{2}{*}{$59.29^{+60.34}_{-62.43}$}
& \multirow{2}{*}{$2.48^{+1.17}_{-0.17}$}
&...\\
  & & & & & & & & & & & & &\\
MAP$_1$& 2.53 & 303.29 & -2.21 & -2.01 & -5.04 & 0.60 & -8.58 & -6.94 & -69.57 & -70.44 & 0.55 & 2.70 & 4.60$\times 10^{-4}$\\
MAP$_2$& 2.55 & 293.66 & -2.47 & -2.18 & -4.94 & -0.07 & -8.94 & -3.43 & -64.11 & -63.39 & 30.79 & 2.54 & 4.38$\times 10^{-4}$\\
MAP$_3$& 2.56 & 243.27 & -2.43 & -2.27 & -4.84 & 0.31 & -8.40 & -5.80 & -60.76 & -53.42 & 34.92 & 2.54 & 3.94$\times 10^{-4}$\\
MAP$_4$& 2.54 & 290.59 & -2.80 & -2.48 & -4.66 & -0.01 & -8.77 & -4.03 & -61.89 & -55.97 & 44.53 & 2.42 & 3.60$\times 10^{-4}$\\
MAP$_5$& 2.55 & 298.75 & -2.32 & -1.87 & -4.98 & 0.06 & -8.78 & -4.87 & -63.81 & -64.12 & 20.94 & 2.69 & 3.53$\times 10^{-4}$\\
  \hline
  \multicolumn{14}{c}{$M_p = 15.3M_{\oplus}$}\\
  \hline
\multirow{2}{*}{Median $\pm2\sigma$}
& \multirow{2}{*}{$2.58^{+0.05}_{-0.05}$}
& \multirow{2}{*}{$267.21^{+103.19}_{-97.01}$}
& \multirow{2}{*}{$-3.26^{+1.19}_{-3.45}$}
& \multirow{2}{*}{$-2.25^{+1.05}_{-1.26}$}
& \multirow{2}{*}{$<-4.73$}& \multirow{2}{*}{$-0.07^{+0.68}_{-4.01}$}
& \multirow{2}{*}{$-7.62^{+5.41}_{-1.24}$}
& \multirow{2}{*}{$-4.53^{+4.23}_{-3.07}$}
& \multirow{2}{*}{$-52.07^{+20.68}_{-20.34}$}
& \multirow{2}{*}{$-46.08^{+33.11}_{-30.64}$}
& \multirow{2}{*}{$62.45^{+59.57}_{-59.47}$}
& \multirow{2}{*}{$2.45^{+0.94}_{-0.14}$}
&...\\
  & & & & & & & & & & & & &\\
MAP$_1$& 2.56 & 280.25 & -3.09 & -2.36 & -5.41 & 0.20 & -8.52 & -5.76 & -57.37 & -52.99 & 41.45 & 2.40 & 4.56$\times 10^{-4}$\\
MAP$_2$& 2.56 & 286.28 & -2.59 & -1.97 & -5.58 & 0.59 & -8.78 & -7.59 & -66.19 & -57.23 & 41.53 & 2.56 & 3.00$\times 10^{-4}$\\
MAP$_3$& 2.54 & 320.30 & -3.27 & -2.08 & -5.17 & 0.48 & -8.73 & -6.93 & -59.07 & -71.58 & 24.23 & 2.44 & 2.91$\times 10^{-4}$\\
MAP$_4$& 2.55 & 303.06 & -3.68 & -2.76 & -5.19 & 0.21 & -8.46 & -5.57 & -64.95 & -56.43 & 49.88 & 2.34 & 2.87$\times 10^{-4}$\\
MAP$_5$& 2.55 & 311.47 & -2.68 & -2.34 & -6.32 & 0.12 & -8.69 & -4.78 & -57.28 & -67.64 & 35.30 & 2.46 & 2.84$\times 10^{-4}$\\
 \hline
 \hline
\end{tabular}
    }
\end{sidewaystable}
\clearpage

\begin{sidewaystable}
    \centering
    \caption{Summary of the Maximum A Posteriori (MAP) values and median values with $\pm2\sigma$ uncertainties, or 95th percentile upper limits, from the atmospheric retrievals with Titan-like haze and fixed planetary masses}
    \label{tab:retrieval results k84 fixed mp}
    \setlength{\tabcolsep}{3pt}
    \resizebox{1.0\textwidth}{!}{%
    \hspace*{-3.5cm}
    \begin{tabular}{l|ccccccccccc|cc} \hline \hline
           &  $R_p$&  $T$&  $\log \chi_{\text{CO}_2}$&  $\log \chi_{\text{CH}_\text{4}}$& $\log \chi_{\text{C}_2\text{H}_4}$&$\log \text{X}_\text{haze}$ & $\log P_\text{deck}$& $\log P_\text{base}$& $\delta_\text{NRS1}$& $\delta_\text{NRS2}$&$\delta_\text{MIRI}$ &$\mu$   &Weight\\
           Model Solutions&  $[R_{\oplus}]$&  $[\text{K}]$&  &  & & & $[\text{bar}]$& $[\text{bar}]$& $[\text{ppm}]$& $[\text{ppm}]$&$[\text{ppm}]$ & $[\text{Dalton}]$  &\\
          \hline
  \multicolumn{14}{c}{$M_p = 6.1 M_{\oplus}$}\\
         \hline
\multirow{2}{*}{Median $\pm2\sigma$}
& \multirow{2}{*}{$2.61^{+0.01}_{-0.05}$}
& \multirow{2}{*}{$131.85^{+66.25}_{-48.83}$}
& \multirow{2}{*}{$-2.46^{+1.11}_{-4.22}$}
& \multirow{2}{*}{$-1.06^{+0.54}_{-1.22}$}
& \multirow{2}{*}{$<-4.09$}& \multirow{2}{*}{$-1.07^{+2.66}_{-3.44}$}
& \multirow{2}{*}{$-5.65^{+4.10}_{-2.95}$}
& \multirow{2}{*}{$-4.37^{+4.57}_{-3.93}$}
& \multirow{2}{*}{$-51.16^{+16.18}_{-20.96}$}
& \multirow{2}{*}{$-35.74^{+20.48}_{-27.58}$}
& \multirow{2}{*}{$81.99^{+41.90}_{-50.40}$}
& \multirow{2}{*}{$4.00^{+3.22}_{-1.60}$}
&...\\
  & & & & & & & & & & & & &\\
MAP$_1$& 2.53 & 160.24 & -2.30 & -1.50 & -4.23 & 0.32 & -8.27 & -5.64 & -80.65 & -67.96 & 26.45 & 2.95 & 4.69$\times 10^{-4}$\\
MAP$_2$& 2.56 & 137.59 & -2.33 & -1.60 & -4.14 & 0.05 & -8.42 & -5.64 & -68.66 & -52.52 & 52.84 & 2.85 & 3.76$\times 10^{-4}$\\
MAP$_3$& 2.53 & 162.41 & -2.53 & -1.45 & -4.46 & 0.67 & -7.04 & -5.98 & -77.18 & -65.58 & 37.45 & 2.92 & 3.18$\times 10^{-4}$\\
MAP$_4$& 2.55 & 189.17 & -1.87 & -1.05 & -4.24 & 0.21 & -7.66 & -4.45 & -71.99 & -64.66 & 45.90 & 4.09 & 3.09$\times 10^{-4}$\\
MAP$_5$& 2.55 & 132.76 & -2.16 & -2.08 & -4.52 & 0.48 & -8.95 & -7.40 & -76.19 & -69.21 & 18.18 & 2.71 & 3.02$\times 10^{-4}$\\
  \hline
  \multicolumn{14}{c}{$M_p = 8.4 M_{\oplus}$}\\
  \hline
\multirow{2}{*}{Median $\pm2\sigma$}
& \multirow{2}{*}{$2.61^{+0.01}_{-0.05}$}
& \multirow{2}{*}{$155.08^{+52.85}_{-48.49}$}
& \multirow{2}{*}{$-2.86^{+1.31}_{-4.02}$}
& \multirow{2}{*}{$-1.50^{+0.77}_{-1.71}$}
& \multirow{2}{*}{$<-4.33$}& \multirow{2}{*}{$-0.43^{+1.89}_{-3.97}$}
& \multirow{2}{*}{$-6.23^{+4.63}_{-2.50}$}
& \multirow{2}{*}{$-4.21^{+4.42}_{-3.96}$}
& \multirow{2}{*}{$-51.26^{+18.32}_{-24.68}$}
& \multirow{2}{*}{$-36.74^{+24.34}_{-33.54}$}
& \multirow{2}{*}{$77.92^{+45.78}_{-56.14}$}
& \multirow{2}{*}{$2.98^{+2.42}_{-0.66}$}
&...\\
  & & & & & & & & & & & & &\\
MAP$_1$& 2.55 & 179.33 & -2.38 & -2.03 & -4.48 & 0.23 & -8.72 & -6.09 & -73.88 & -62.04 & 44.37 & 2.61 & 4.70$\times 10^{-4}$\\
MAP$_2$& 2.56 & 197.89 & -2.15 & -1.72 & -4.60 & 0.01 & -7.85 & -3.31 & -70.30 & -61.24 & 51.62 & 2.86 & 3.63$\times 10^{-4}$\\
MAP$_3$& 2.56 & 186.03 & -2.35 & -1.75 & -4.64 & 0.15 & -7.83 & -5.60 & -61.85 & -58.23 & 40.52 & 2.74 & 3.44$\times 10^{-4}$\\
MAP$_4$& 2.55 & 184.92 & -2.78 & -1.94 & -4.85 & 0.09 & -8.43 & -5.05 & -78.67 & -74.94 & 40.31 & 2.53 & 3.25$\times 10^{-4}$\\
MAP$_5$& 2.57 & 184.25 & -1.96 & -1.77 & -5.27 & -0.21 & -8.88 & -4.15 & -66.78 & -49.90 & 44.64 & 3.00 & 2.68$\times 10^{-4}$\\
  \hline
  \multicolumn{14}{c}{$M_p = 10.7M_{\oplus}$}\\
  \hline
\multirow{2}{*}{Median $\pm2\sigma$}
& \multirow{2}{*}{$2.61^{+0.01}_{-0.06}$}
& \multirow{2}{*}{$177.50^{+56.34}_{-51.74}$}
& \multirow{2}{*}{$-2.96^{+1.23}_{-3.77}$}
& \multirow{2}{*}{$-1.67^{+0.80}_{-1.43}$}
& \multirow{2}{*}{$<-4.46$}& \multirow{2}{*}{$-0.40^{+1.79}_{-3.97}$}
& \multirow{2}{*}{$-6.46^{+4.80}_{-2.28}$}
& \multirow{2}{*}{$-4.34^{+4.51}_{-3.84}$}
& \multirow{2}{*}{$-49.80^{+18.65}_{-24.13}$}
& \multirow{2}{*}{$-35.37^{+23.13}_{-31.94}$}
& \multirow{2}{*}{$79.54^{+45.84}_{-53.17}$}
& \multirow{2}{*}{$2.76^{+1.72}_{-0.44}$}
&...\\
  & & & & & & & & & & & & &\\
MAP$_1$& 2.55 & 231.84 & -2.54 & -2.01 & -4.30 & 0.01 & -8.84 & -5.31 & -61.56 & -55.60 & 56.08 & 2.56 & 4.57$\times 10^{-4}$\\
MAP$_2$& 2.57 & 194.50 & -2.98 & -1.88 & -5.33 & 0.52 & -7.76 & -6.40 & -74.29 & -56.78 & 34.29 & 2.53 & 4.48$\times 10^{-4}$\\
MAP$_3$& 2.56 & 205.43 & -2.93 & -2.54 & -4.87 & 0.03 & -8.83 & -5.18 & -64.53 & -58.86 & 32.13 & 2.40 & 4.20$\times 10^{-4}$\\
MAP$_4$& 2.57 & 211.89 & -2.47 & -2.02 & -4.67 & -0.16 & -8.86 & -4.40 & -51.99 & -50.15 & 48.77 & 2.58 & 4.00$\times 10^{-4}$\\
MAP$_5$& 2.57 & 198.29 & -3.05 & -1.80 & -4.58 & 0.25 & -8.18 & -6.42 & -55.49 & -49.02 & 65.04 & 2.56 & 3.76$\times 10^{-4}$\\
    \hline
 \multicolumn{14}{c}{$M_p = 13. 0M_{\oplus}$}\\
   \hline
\multirow{2}{*}{Median $\pm2\sigma$}
& \multirow{2}{*}{$2.61^{+0.01}_{-0.06}$}
& \multirow{2}{*}{$195.23^{+80.87}_{-53.82}$}
& \multirow{2}{*}{$-3.04^{+1.21}_{-4.03}$}
& \multirow{2}{*}{$-1.75^{+0.78}_{-1.37}$}
& \multirow{2}{*}{$<-4.61$}& \multirow{2}{*}{$-0.42^{+2.11}_{-4.00}$}
& \multirow{2}{*}{$-6.32^{+4.75}_{-2.41}$}
& \multirow{2}{*}{$-4.43^{+4.58}_{-3.88}$}
& \multirow{2}{*}{$-47.59^{+17.66}_{-20.57}$}
& \multirow{2}{*}{$-32.68^{+22.97}_{-29.68}$}
& \multirow{2}{*}{$83.49^{+45.50}_{-52.13}$}
& \multirow{2}{*}{$2.68^{+1.39}_{-0.35}$}
&...\\
  & & & & & & & & & & & & &\\
MAP$_1$& 2.55 & 266.89 & -2.36 & -2.28 & -5.26 & 0.29 & -8.72 & -6.23 & -66.71 & -59.20 & 61.72 & 2.56 & 4.66$\times 10^{-4}$\\
MAP$_2$& 2.55 & 262.24 & -2.79 & -1.99 & -5.49 & 0.31 & -7.83 & -5.45 & -67.61 & -58.30 & 47.85 & 2.52 & 3.95$\times 10^{-4}$\\
MAP$_3$& 2.56 & 240.56 & -3.09 & -2.55 & -5.28 & 0.29 & -8.69 & -6.82 & -61.60 & -58.06 & 32.30 & 2.38 & 2.88$\times 10^{-4}$\\
MAP$_4$& 2.57 & 242.77 & -3.12 & -2.16 & -4.98 & -0.10 & -8.45 & -3.55 & -61.19 & -47.71 & 49.21 & 2.43 & 2.83$\times 10^{-4}$\\
MAP$_5$& 2.56 & 248.61 & -2.54 & -2.47 & -6.03 & 0.33 & -7.80 & -5.85 & -58.88 & -43.76 & 38.00 & 2.47 & 2.64$\times 10^{-4}$\\
  \hline
  \multicolumn{14}{c}{$M_p = 15.3M_{\oplus}$}\\
  \hline
\multirow{2}{*}{Median $\pm2\sigma$}
& \multirow{2}{*}{$2.61^{+0.01}_{-0.06}$}
& \multirow{2}{*}{$223.51^{+109.30}_{-64.40}$}
& \multirow{2}{*}{$-3.11^{+1.19}_{-3.82}$}
& \multirow{2}{*}{$-1.90^{+0.83}_{-1.26}$}
& \multirow{2}{*}{$<-4.73$}& \multirow{2}{*}{$-0.22^{+1.50}_{-4.21}$}
& \multirow{2}{*}{$-6.75^{+5.10}_{-2.04}$}
& \multirow{2}{*}{$-4.65^{+4.71}_{-3.54}$}
& \multirow{2}{*}{$-47.37^{+18.89}_{-19.84}$}
& \multirow{2}{*}{$-33.70^{+24.94}_{-30.44}$}
& \multirow{2}{*}{$80.94^{+47.74}_{-55.21}$}
& \multirow{2}{*}{$2.59^{+1.13}_{-0.26}$}
&...\\
  & & & & & & & & & & & & &\\
MAP$_1$& 2.57 & 286.87 & -2.42 & -2.12 & -5.78 & 0.37 & -8.87 & -7.36 & -62.72 & -48.22 & 59.32 & 2.57 & 4.60$\times 10^{-4}$\\
MAP$_2$& 2.55 & 304.36 & -2.76 & -2.21 & -4.98 & 0.63 & -7.89 & -6.72 & -64.97 & -53.72 & 55.70 & 2.47 & 4.20$\times 10^{-4}$\\
MAP$_3$& 2.57 & 269.62 & -2.38 & -2.43 & -4.89 & 0.31 & -8.24 & -5.63 & -63.61 & -46.37 & 48.62 & 2.53 & 2.97$\times 10^{-4}$\\
MAP$_4$& 2.55 & 302.87 & -2.96 & -2.30 & -5.40 & 0.77 & -8.99 & -8.37 & -69.20 & -53.26 & 27.85 & 2.42 & 2.75$\times 10^{-4}$\\
MAP$_5$& 2.57 & 275.80 & -2.53 & -2.08 & -5.72 & 0.03 & -8.32 & -4.05 & -53.06 & -46.17 & 59.10 & 2.55 & 2.74$\times 10^{-4}$\\
 \hline
 \hline
\end{tabular}
    }
\end{sidewaystable}
\clearpage

\end{document}